\documentclass[review, 3p]{elsarticle}

\usepackage{graphicx, color, amsthm, amssymb}
\usepackage{txfonts, longtable, dcolumn, ctable, lscape, multirow, lineno, bbding, pifont, afterpage}
\usepackage{natbib}
\usepackage{microtype}
\bibpunct{(}{)}{;}{a}{}{,}
\journal{Icarus}

\begin{document}

\begin{frontmatter}

\ead{hanus.home@gmail.com}
%%\fntext[label2]{}
\cortext[cor1]{Corresponding author. Tel: +33 (0)4 92 00 19 44. Fax: +33 (0)4 92 00 31 21.}


%\title{Thermophysical properties of asteroids from WISE data. I.~Near-Earth asteroids (1036)~Ganymed, (1627)~Ivar, (1865)~Cerberus and (1980)~Tezcatlipoca, and Jovian Trojan asteroid (624)~Hektor}
\title{Thermophysical modeling of asteroids from WISE thermal infrared data -- Significance of the shape model and the pole orientation uncertainties}
%\titlerunning{Thermophysical modeling of asteroids from WISE thermal data}

\author[1,2]{J.~Hanu\v s\corref{cor1}}
\author[1]{M.~Delbo'}
\author[2]{J.~\v Durech}
\author[3,4,1]{V.~Al\' i-Lagoa}

\address[1]{UNS-CNRS-Observatoire de la C\^ote d'Azur, BP 4229, 06304 Nice Cedex 4, France}
\address[2]{Astronomical Institute, Faculty of Mathematics and Physics,
Charles University, V Hole\v sovi\v ck\' ach 2, 180 00 Prague, Czech Republic}
\address[3]{Instituto de Astrof\'isica de Canarias, E-38205 La Laguna, Tenerife, Spain}
\address[4]{Universidad de La Laguna, Dept. Astrof\'isica, E-38206 La Laguna, Tenerife, Spain}

\begin{abstract}
In the analysis of thermal infrared data of asteroids by means of thermophysical models (TPMs) it is a common practice to neglect the uncertainty of the shape model and the rotational state, which are taken as an input for the model. Here, we present a novel method of investigating the importance of the shape model and the pole orientation uncertainties in the thermophysical modeling -- the varied shape TPM (VS-TPM). Our method uses optical photometric data to generate various shape models that map the uncertainty in the shape and the rotational state. The TPM procedure is then run for all these shape models. We apply the implementation of the classical TPM as well as our VS-TPM to the convex shape models of several asteroids together with their thermal infrared data acquired by the NASA's Wide-field Infrared Survey Explorer (WISE) and compare the results. These show that the uncertainties of the shape model and the pole orientation can be very important (e.g., for the determination of the thermal inertia) and should be considered in the thermophysical analyses. We present thermophysical properties for six asteroids -- (624)~Hektor, (771)~Libera, (1036)~Ganymed, (1472)~Muonio, (1627)~Ivar, and (2606)~Odessa.
\end{abstract}


\end{frontmatter}

\section{Introduction}\label{introduction}

The physical characterization of asteroids has seen an enormous boost in recent years thanks to the data of the NASA Wide-field Infrared Survey Explorer (WISE) mission \citep{Wright2010, Mainzer2011a}. 
In particular, sizes and albedos have been determined from simple thermal models \citep[e.g., NEATM of][]{Harris1998} for more than 150,\,000 asteroids and for different populations thereof, including main belt asteroids 
(MBAs) -- \citet{Masiero2011, Masiero2012}, 
Hildas -- \citet{Grav2012a}, 
near-Earth objects (NEOs) -- \citet{Mainzer2014, Mainzer2011c}, 
and Trojans -- \citet{Grav2011, Grav2012b}. This  
resulted in a database of impressive quality in terms of number of observed bodies and sensitivity as compared to previous surveys: $\sim$2\,200 asteroid albedos and sizes from the SIMPS based on IRAS observations \citep{Tedesco2002}, or $\sim$5\,000 albedos and sizes based on AKARI data \citep{Usui2011}.

Simple thermal models assume spherical, non-rotating asteroids. When shape models, spin vectors, and other physical parameters are known, it is possible to apply more sophisticated methods to infer thermophysical properties such as thermal inertia and/or surface roughness. So far, such thermophysical parameters were derived for less than about 50 asteroids due to the lack of both physical properties (i.e., shape models, spin axis orientations) as well as available thermal infrared observations. However, the number of shape models has grown dramatically in the past decade and the newly obtained thermal infrared data from WISE now open the opportunity to greatly increase the number of thermal inertia determinations.

Thermal inertia, defined by $\Gamma= (\rho\kappa C)^{1/2}$, where $\rho$ is the density of the surface regolith, $\kappa$ its thermal conductivity, and $C$ its heat capacity, measures the resistance of a material to temperature change, and thus controls the temperature distribution of the surface of an atmosphere-less body.
As it affects the symmetry of the temperature distribution on asteroids, the thermal inertia controls the strength of the Yarkovsky effect, which is the rate of change in the semi-major axis of the orbit of an asteroid ($\mathrm{d}a/\mathrm{d}t$) due to the recoil force of the thermal photons \citep[][ for instance]{Bottke2006}. 
%
%When compared with the densities of meteorites -- a partial sample of the building blocks of asteroids that survive the passage through the Earth's atmosphere -- one can deduce the nature of asteroid interiors \citep{Consolmagno2008}.
%
Thermal inertia is also a sensitive indicator of the nature of the surface regolith as its value is affected by the cohesion of the material in the soil \citep[i.e., between one and several tens of millimeters of the surface layer see, e.g.,][]{Mellon2000,Jakosky1986}. 
Knowledge of the grain size of asteroid regolith is of paramount importance for future landing and/or sample-return missions \citep[such as OSIRIS--REx and Hayabusa--2, two different sample return missions to carbonaceous asteroids,][]{Lauretta2012,Okada2014}. 
Thermal inertia is strongly affected by the porosity of the material \citep{Zimbelman1986}. For a given surface composition, the higher the porosity, the lower the values of both $\kappa$ and $\Gamma$. See \citet{Vernazza2012} for a discussion of the effect of porosity on asteroids surfaces. 

In order to derive the thermal inertia and other physical parameters of asteroids, such as the diameter and the albedo, a thermophysical model \citep[hereafter TPM,][]{Lagerros1996, Lagerros1997, Lagerros1998} is typically used to analyze thermal infrared data. A TPM calculates thermal infrared fluxes given a set of physical parameters (size $D$, thermal inertia $\Gamma$, Bond albedo $A$, surface roughness $\overline\theta$) whose values are adjusted to provide the best fit between the model and the observed fluxes, by minimizing a figure of merit (chi-square). Classically, a TPM is used with an a-priori knowledge of the shape and the rotational state of the asteroid, which are taken as fixed quantities. Typically, shapes are based on radar imaging \citep[e.g., asteroids 2010~EV$_5$, or (101\,955)~Bennu,][]{AliLagoa2014, Emery2014} or on convex inversion of photometric ligthcurves \citep[e.g., asteroids (25143)~Itokawa, or (1620)~Geographos,][]{Muller2014, Rozitis2014}.

So far, shape uncertainties have never been properly accounted for in TPM analyses, but there is growing evidence that large chi-squared values are obtained when the shape model used for the thermophysical modeling differs significantly from the asteroid's true shape. In particular, \citet{Rozitis2014} have shown that if the length of the shape model's rotation axis, as determined from radar observation, is not properly estimated, this can lead to large chi-squared values and bias the value of the thermal inertia \citep[see also the case for the NEA (101955) Bennu by][]{Emery2014,Muller2012}.

Here we study how the uncertainties of the shape model and the pole orientation influence the TPM results: we perform TPM analysis of WISE thermal infrared data for a selected number of asteroids that 
have known shapes and spin poles from the Database of Asteroid Models from Inversion Techniques \citep[DAMIT\footnote{\texttt{http://astro.troja.mff.cuni.cz/projects/asteroids3D}},][]{Durech2010}.
First, using new optical lightcurve data, we derive revised shape models that we use for TPM analysis. We show that the goodness of fit of the TPM (e.g., $\chi^2$) changes when the revised shape is used instead of the DAMIT one, indicating that WISE data are sensitive to shape features. We also note that the revised shape models can, in general, improve the TPM solution.

Next, we introduce a novel method, that we call the \textit{varied shape TPM} (or VS-TPM for short), that allows us to analyze the stability of the TPM solution against the variations of the shape model and the pole orientation (Sect.~\ref{sec:varied_shape_TPM}).  We show that our method also allows us to find a shape model that improve significantly the fit of thermal infrared data compared to a classical TPM based on fixed shape and pole orientation derived uniquely from optical light curve inversion (nominal shape).

The VS-TPM consists of the following steps: we bootstrap the optical photometric data and use the technique of convex inversion of \citet{Kaasalainen2001a,Kaasalainen2001b} to determine a set of slightly different shape solutions that fit the available optical disk-integrated photometry equally well. For each shape and pole solution (varied shapes), we use a TPM to analyze thermal infrared data (e.g. WISE, IRAS) and we notice that the goodness of the fit to thermal infrared data can be significantly shape model dependent, and in most of the cases the VS-TPM allows one to find a solution that fits the thermal infrared data better than the original shape solution.

In this paper, we first describe and apply in Sect.~\ref{sec:TPM} the classical TPM scheme with an a priori convex shape model to the WISE data for nine selected asteroids. In Sect.~\ref{sec:varied_shape_TPM}, we present the VS-TPM method and use it to show how the uncertainties in the shape model and its pole orientation influence the thermophysical fit. We discuss and conclude our work in Sects.~\ref{sec:discussion}~and~\ref{sec:conclusions}.
Application of the varied shape TPM scheme to a few hundred of main-belt asteroids will then be a subject of our forthcoming work.

\section{Data}\label{sec:data}

\subsection{Visible light, disk-integrated photometry}\label{sec:photometry}

%%%
%\onecolumn
%\scriptsize{
\begin{table*}[b]
\caption{\label{tab:models}List of asteroid models derived by the lightcurve inversion method that we use for the TPM modeling. For each asteroid, the table gives the ecliptic coordinates $\lambda$ and $\beta$ of the pole solutions, the sidereal rotational period $P$, the number of dense lightcurves $N_{\mathrm{lc}}$ observed during $N_{\mathrm{app}}$ apparitions, the number of sparse data points from USNO-Flagstaff $N_{\mathrm{689}}$ and Catalina Sky Survey $N_{\mathrm{703}}$, and the reference.}
\begin{tabular}{r@{\,\,\,}l rrrr D{.}{.}{6} ccccc}
\hline 
\multicolumn{2}{c} {Asteroid} & \multicolumn{1}{c} {$\lambda_1$} & \multicolumn{1}{c} {$\beta_1$} & \multicolumn{1}{c} {$\lambda_2$} & \multicolumn{1}{c} {$\beta_2$} & \multicolumn{1}{c} {$P$} & $N_{\mathrm{lc}}$ & $N_{\mathrm{app}}$  & $N_{\mathrm{689}}$ & $N_{\mathrm{703}}$ & Reference \\
\multicolumn{2}{l} { } & [deg] & [deg] & [deg] & [deg] & \multicolumn{1}{c} {[hours]} &  &  &  &  & \\
\hline\hline
624 & Hektor & 331 & $-$32 &  &  & 6.92051 & 17 & 8 &  &  & \citet{Kaasalainen2002c} \\
 &  & 333 & $-$32 &  &  & 6.920509 & 19 & 9 & 201 & 56 & This work \\
771 & Libera & 64 & $-$78 &  &  & 5.89042 & 20 & 5 &  &  & \citet{Marciniak2009b} \\
%825 & Tanina & 46 & 48 & 231 & 60 & 6.93981 & 2 & 1 & 114 & 40 & \citet{Hanus2011} \\
832 & Karin & 242 & 46 & 59 & 44 & 18.3512 & 13 & 6 & 84 & 39 & \citet{Hanus2011} \\
%852 & Wladilena & 46 & $-$53 & 181 & $-$48 & 4.613298 & 30 & 7 & 138 & 101 & \citet{Hanus2013a} \\
1036 & Ganymed & 214 & $-$73 &  &  & 10.313 & 21 & 1 &  &  & \citet{Kaasalainen2002b} \\
 &  & 190 & $-$78 &  &  & 10.31284 & 177 & 4 & 155 & 20 & This work \\
1472 & Muonio & 42 & 62 & 249 & 61 & 8.70543 & 6 & 1 & 99 & 93 & \citet{Hanus2013a} \\
1627 & Ivar & 338 & 40 &  &  & 4.79517 & 56 & 4 &  &  & \citet{Kaasalainen2004b} \\
 &  & 334 & 39 &  &  & 4.79517 & 83 & 4 & 68 & 152 & This work \\
1865 & Cerberus & 292 & $-$72 &  &  & 6.803284 & 28 & 8 &  &  & \citet{Durech2012b} \\
 &  & 311 & $-$78 &  &  & 6.803286 & 47 & 8 &  & 62 & This work \\
1980 & Tezcatlipoca & 334 & $-$65 &  &  & 7.25226 & 48 & 5 &  &  & \citet{Kaasalainen2004b} \\
 &  & 324 & $-$69 &  &  & 7.25226 & 49 & 6 & 29 & 35 & This work \\
2606 & Odessa & 25 & $-$81 & 283 & $-$88 & 8.2444 & 3 & 1 & 25 & 129 & \citet{Hanus2013a} \\
\hline
\end{tabular}
%*Preferred solution by \citet{Marchis2014}.
\end{table*}
%\tablefoot{
%}
%}
\onecolumn
%\scriptsize{
%\begin{tabular}{r@{\,\,\,}l lc| r@{\,\,\,}l lc} \hline
% \multicolumn{2}{c} {Asteroid} & Date & Observer &  \multicolumn{2}{c}{Asteroid} & Date & Observer\\ \hline\hline
\begin{longtable}{r@{\,\,\,}l lccc}
\caption{\label{tab:observations}Optical data used for the shape model determinations.}\\
\hline
 \multicolumn{2}{c} {Asteroid} & Date & $N_\mathrm{lc}$ & Observer & Observatory (MPC code) \\ \hline\hline

\endfirsthead
\caption{continued.}\\

\hline
 \multicolumn{2}{c} {Asteroid} & Date & $N_\mathrm{lc}$ & Observer & Observatory (MPC code) \\
\hline\hline
\endhead
\hline
\endfoot
624 & Hektor & 1957 4 -- 1957 5 & 4 & \citet{Dunlap1969} & \\
 &  & 1965--02--04.3 & 1 & \citet{Dunlap1969} & \\
 &  & 1967--03--07.4 & 1 & \citet{Dunlap1969} & \\
 &  & 1968 4 -- 1968 5 & 2 & \citet{Dunlap1969} & \\
 &  & 1977 2 -- 1977 2 & 2 & \citet{Hartmann1978} & \\
 &  & 1984 10 -- 1984 10 & 3 & \citet{Detal1994} & \\
 &  & 1990 3 -- 1990 3 & 2 & \citet{Dahlgren1991} & \\
 &  & 1991 4 -- 1991 4 & 2 & \citet{Hainaut1995a} & \\
 &  & 2008 10 -- 2008 10 & 2 & \citet{Stephens2009} & \\
771 & Libera & 1984 5 -- 1984 5 & 3 & \citet{Binzel1987a} & \\
 &  & 1999 9 -- 1999 9 & 2 & \citet{Marciniak2009b} & \\
 &  & 1999 9 -- 1999 9 & 2 & \citet{Warner2000} & \\
 &  & 2005 2 -- 2005 3 & 3 & \citet{Marciniak2009b} & \\
 &  & 2006 5 -- 2006 6 & 4 & \citet{Marciniak2009b} & \\
 &  & 2008 10 -- 2009 3 & 6 & \citet{Marciniak2009b} & \\
%825 & Tanina & 1992 1 -- 1992 2 & 2 & \citet{Wisniewski1997} & \\
832 & Karin & 1984 10 -- 1984 10 & 2 & \citet{Binzel1987a} & \\
 &  & 2003 8 -- 2003 9 & 8 & \citet{Yoshida2004} & \\
 &  & 2004 9 -- 2004 9 & 3 & \citet{Ito2007} & \\
%852 & Wladilena & 1977 2 14.3 & 1 & \citet{Tedesco1979} & \\
% &  & 1982 9 -- 1982 10 & 3 & \citet{Harris1999} & \\
% &  & 1982--10--17.9 & 1 & \citet{DiMartino1984} & \\
% &  & 1993 11 -- 1993 11 & 2 & \citet{DeAngelis1995} & \\
% &  & 2003 2 -- 2003 2 & 3 & \citet{Hanus2013a} & \\
% &  & 2007 5 -- 2007 5 & 2 & \citet{Hanus2013a} & \\
% &  & 2008 8 -- 2009 1 & 7 & \citet{Hanus2013a} & \\
% &  & 2010 2 -- 2010 5 & 8 & \citet{Hanus2013a} & \\
% &  & 2010 3 -- 2010 3 & 3 & \citet{Polishook2012b} & Wise Observatory, Mitzpeh Ramon (097) \\
1036 & Ganymed & 1985 7 -- 1985 11 & 6 & \citet{Lupishko1987a} &  \\
 & & 1985 7 -- 1985 12 & 25 & \citet{Hahn1989a} &  \\
 & & 1989 4 -- 1989 7 & 11 & \citet{Chernova1995a} &  \\
 & & 2008 12 -- 2009 4 & 24 & \citet{Skiff2012a} &  Lowell Observatory \\
 & & 2011 5 -- 2012 1 & 8 & \citet{Pilcher2012b} & Multiple observatories \\
 & & 2011 5 -- 2011 12 & 103 & \citet{Velichko2013} & Multiple observatories \\ %why Benishek in lc description?
% & & 2011 7 -- 2011 12 & Mottola and Hellmich &  \\ %nasrat :-)
1472 & Muonio & 2008 9 -- 2008 9 & 3 & \citet{Stephens2009b} & \\
 & & 2008 10 -- 2008 10 & 3 & \citet{Hanus2013a} & \\
1627 & Ivar & 1985 6 -- 1985 10 & 26 & \citet{Hahn1989a} &  \\
 & & 1990 5 -- 1990 8 & 18 & \citet{Chernova1995a} &  \\
 & & 1990 5 -- 1990 5 & 2 & \citet{Velichko1990a} &  \\ %also in Chernova, remove there?
 & & 1990 5 -- 1990 5 & 2 & \citet{Hoffmann1990a} &  \\
 & & 1995 2 -- 1995 3 & 6 & \citet{Pravec1996a} & Ond\v rejov Observatory (557) \\
 & & 2008 9 -- 2009 2 & 29 & \citet{Skiff2012a} & Lowell Observatory \\
% & & 2008 9 -- 2009 2 & 16 & Brian Warner & ??? \\
1865 & Cerberus & 1980 11 -- 1980 11 & 2 & \citet{Harris1989a} & \\
 &  & 1989 11 -- 1989 11 & 2 & \citet{Wisniewski1997} & \\
 &  & 1998 10 -- 1998 10 & 2 & \citet{Sarneczky1999} & \\
 &  & 1999--09--25.0 & 1 & \citet{Szabo2001} & \\
 &  & 1999 11 -- 1999 11 & 2 & \citet{Durech2012b} & \\
 &  & 2000--7--09.1 & 1 & \citet{Szabo2001} & \\
 &  & 2008 9 -- 2008 11 & 25 & \citet{Durech2012b} & \\
 &  & 2008 10 -- 2008 11 & 5 & \citet{Skiff2012a} & \\
 &  & 2009 9 -- 2009 10 & 4 & \citet{Durech2012b} & \\
 &  & 2010--08--06.9 & 1 & \citet{Durech2012b} & \\
1980 & Tezcatlipoca & 1988 6 -- 1988 6 & 2 & \citet{Wisniewski1997} & \\
 &  & 1992--05--22.0 & 1 & \citet{Kaasalainen2004b} & \\
 &  & 1995--10--27.1 & 1 & \citet{Kaasalainen2004b} & \\
 &  & 1996 2 -- 1996 2 & 3 & \citet{Kaasalainen2004b} & \\
 &  & 1997 6 -- 1997 1 & 41 & \citet{Kaasalainen2004b} & \\
 &  & 2009--08--20.1 & 1 & \citet{Skiff2012a} & \\
2606 & Odessa & 2008 2 -- 2008 2 & 2 & \citet{Higgins2008b} & \\
 &  & 2008--03--1.5 & 1 & \citet{Hanus2013a} & \\ \hline
\end{longtable}
%\end{table*}
%\tablefoot{
%\tablefoottext{1}{Asteroid Photometric Catalogue, \texttt{http://asteroid.astro.helsinki.fi/}} %APC
%\tablefoottext{2}{On line at \texttt{http://www.david-higgins.com/Astronomy/asteroid/lightcurves.htm}} %Higgins
%\tablefoottext{3}{Downloaded from MPC, \texttt{http://www.minorplanetcenter.net/light\_curve}} %Skiff
%\tablefoottext{4}{On line at \texttt{http://www....}} %Oey
%\tablefoottext{b}{On line at \texttt{http://www.antelopehillsobservatory.org/}} %Koff
%\tablefoottext{b}{On line at \texttt{http://aslc-nm.org/Pilcher.html}} %Pilcher
%\tablefoottext{c}{Observations, reductions, and calibration methods are described in \citet{Polishook2008, Polishook2009}} %Polishook
%}
%}
%%%

We make use of optical photometric data for two reasons: (i)~to revise/improve several shape models, and (ii)~to bootstrap the photometry to derive various shape models for the VS-TPM.

We use two different types of reflected disk-integrated photometry: (i)~dense-in-time photometry, which is typically acquired by individual observers and densely covers a time interval of several hours, and (ii)~sparse-in-time photometry, which is a usual by-product of astrometric surveys and consists of a few hundred individual calibrated measurements during $\sim$15 years.
%\md{CAN YOU GIVE ONE OR TWO CITATIONS HERE ?}

Shape models adopted from the literature (stored in the DAMIT) are usually based on the dense-in-time photometry. The dense-in-time photometry is from two main sources:
(i)~the Asteroid Photometric Catalogue \citep[APC\footnote{\texttt{http://asteroid.astro.helsinki.fi/}}, ][]{Piironen2001}, and
(ii)~the data from individual observers provided by the Minor Planet Center\footnote{\texttt{http://www.minorplanetcenter.net/light\_curve2/light\_curve.php}} in the Asteroid Lightcurve Data Exchange Format \citep[ALCDEF\footnote{\texttt{http://www.minorplanet.info/alcdef.html}},][]{Warner2009}.
Several observers send us data directly or on request.
Our revised shape model determinations are based on the combined dense- and sparse-in-time data sets. We download the sparse-in-time photometric data (typically acquired by astrometric surveys) from the AstDyS site (Asteroids -- Dynamic Site\footnote{\texttt{http://hamilton.dm.unipi.it/}}) and process them similarly as in \citet{Hanus2011}: we compute the geometry of the observations, light-time correct the epochs, transform magnitudes to intensities, and exclude clear outliers. We use the sparse photometric data from the USNO--Flagstaff station (IAU code 689) and the Catalina Sky Survey Observatory \citep[IAU code 703,][]{Larson2003}.

In Tab.~\ref{tab:models}, we list for each studied asteroid information about the visible photometry used for the shape model determination, namely the number of dense-in-time lightcurves, the number of apparitions covered by dense-in-time observations and the number of sparse-in-time measurements from both astrometric surveys. The references to the photometric observations are presented in Tab.~\ref{tab:observations}.

\subsection{WISE thermal infrared fluxes, disk-integrated photometry}\label{sec:thermal_data}

We make use of the thermal infrared data of asteroids acquired by the WISE satellite, in particular the results of the NEOWISE project, which focuses on the solar system bodies \citep[see, e.g., ][]{Mainzer2011a}. The thermal infrared data are downloaded from the WISE All-Sky Single Exposure L1b Working Database via the IRSA/IPAC archive \footnote{\texttt{http://irsa.ipac.caltech.edu/Missions/wise.html}}. 

In this study, we consider only thermal infrared data from filters W3 and W4 (isophotal wavelengths at 12 and 22 $\mu$m) from the fully cryogenic phase of the mission. While W3 and W4 data are thermal-emission dominated, the fluxes in filters W1 and W2 (isophotal wavelength at 3.4 and 4.6 $\mu$m) usually at least partially consist of reflected sunlight, which cannot be properly modeled in our purely thermophysical model and these filters are therefore not considered here.

The data selection and suitability criteria applied in this work follow those of \citet{AliLagoa2014} for asteroid (341843) 2008 EV$_5$. In turn, these are based on a combination of criteria from \citet{Mainzer2011b}, \citet{Masiero2011}, and \citet{Grav2012a}: we implement the correction to the red and blue calibrator discrepancy in W3 and W4 (Cutri et al. 2012), and we use a cone search radius of $1''$ centered on the MPC ephemeris of the object in our queries. We only consider data with artefact flags p, P, and 0, and quality flags A, B, and C. A ``0'' artifact-flag entry indicates no artifact detection, whereas p and P indicate possible contamination by a latent image. Nonetheless, \citet{Masiero2011} found the pipeline criteria for P and p flags to be overly conservative, in particular, that one safely retrieves 20\% more data by allowing P-flagged. The quality flags A, B, and C, correspond to the following signal-to-noise ratios ($S/N$): $S/N > 10$ (A), $3 < S/N < 10$ (B), and $2 < S/N < 3$ (C). We require the IRSA/IPAC modified Julian date to be within four seconds of the time specified by the MPC. A positive match from the WISE Source Catalog within $6''$ of any MPC-reported detection indicates that there is an inertial source at a distance smaller than the point-spread function width of band W1. We consider that these data are contaminated if the inertial source fluxes are greater than 5\% of the asteroid flux and we remove them. Finally, we do not use data that are partially saturated to any extent.

Additionally, we find indications that the error bars of the WISE data may be underestimated, namely the fact that those measurements taken 11 seconds apart are not usually compatible within their error bars. These ``double`` measurements occur when the asteroid appears in the overlapping area of two consecutive frames. We identify $\sim$400 such double detections in the WISE data set for asteroids with convex shape models, and we find that the reported uncertainties of the thermal fluxes are underestimated by a factor of $\sim$1.4 for W3 data and of $\sim$1.3 for W4 data (described in more details in Appendix A). The change in the thermal flux of an asteroid during this short time interval is much smaller than the relative error of the fluxes and it cannot be caused by the orbital and rotational evolution either, therefore we decide to increase the error bars of the data by these factors.

Furthemore, \citet{Jarrett2011} presented a study of the WISE-Spitzer flux cross-calibration of a number of calibration stars and one galaxy situated near the poles of the ecliptic and found that (i)~the photometry for individual objects is stable for the whole cryogenic phase within less than 1\%, but (ii)~there is an rms scatter around the zero level of 4.5\% and 5.7\% of the WISE zero magnitudes in filters W3 and W4 when several objects are examined (the offset is different for each individual star). Accounting for these uncertainties in the case of asteroids is questionable, because the cross-calibration is studied for stars. Asteroids, however, have important differences compared to said calibration stars: they have widely different surface temperatures $\sim$4\,000 vs. $\sim$300 K; asteroids can present rotational variability of a factor of two or more in their fluxes due to their shapes; also the typical magnitudes of asteroids in filters W3 and W4 are much brighter than those of the stars examined by \citet{Jarrett2011}. In this context, it is reasonable to expect an unknown and different offset for each asteroid. It may even change within the dataset of the same object, due to its intrinsic time variability. Thus it can only be accounted for in our modeling by a random relative offset of the fluxes between the W3 and W4 bands. Our method to deal with this source of error and its effect on the physical parameters of asteroids derived my means of TPMs is described in Appendix A. In short, we find that the error of the cross-calibration uncertainty does not affect the results significantly.

\section{Thermophysical modeling with fixed shape models}\label{sec:TPM}

In order to study the importance of shape uncertainties in the thermophysical modeling, first we study nine selected asteroids for various reasons, for example, high quality shapes or wise data, or recent additional optical data are available. We then show that by applying the TPM on a revision of the shape model derived from the inversion of new lightcurve data in addition to the one present in the DAMIT, we obtain a different goodness-of-fit. In general, our results improved, but the revised best-fitting parameter values are consistent within the error bars. On the other hand, the $\chi^2_\mathrm{red}$ still stay above 1 in most cases.

\subsection{TPM method}\label{sec:TPM_method}

\begin{figure}[!htbp]
\begin{center}
\resizebox{\hsize}{!}{\includegraphics{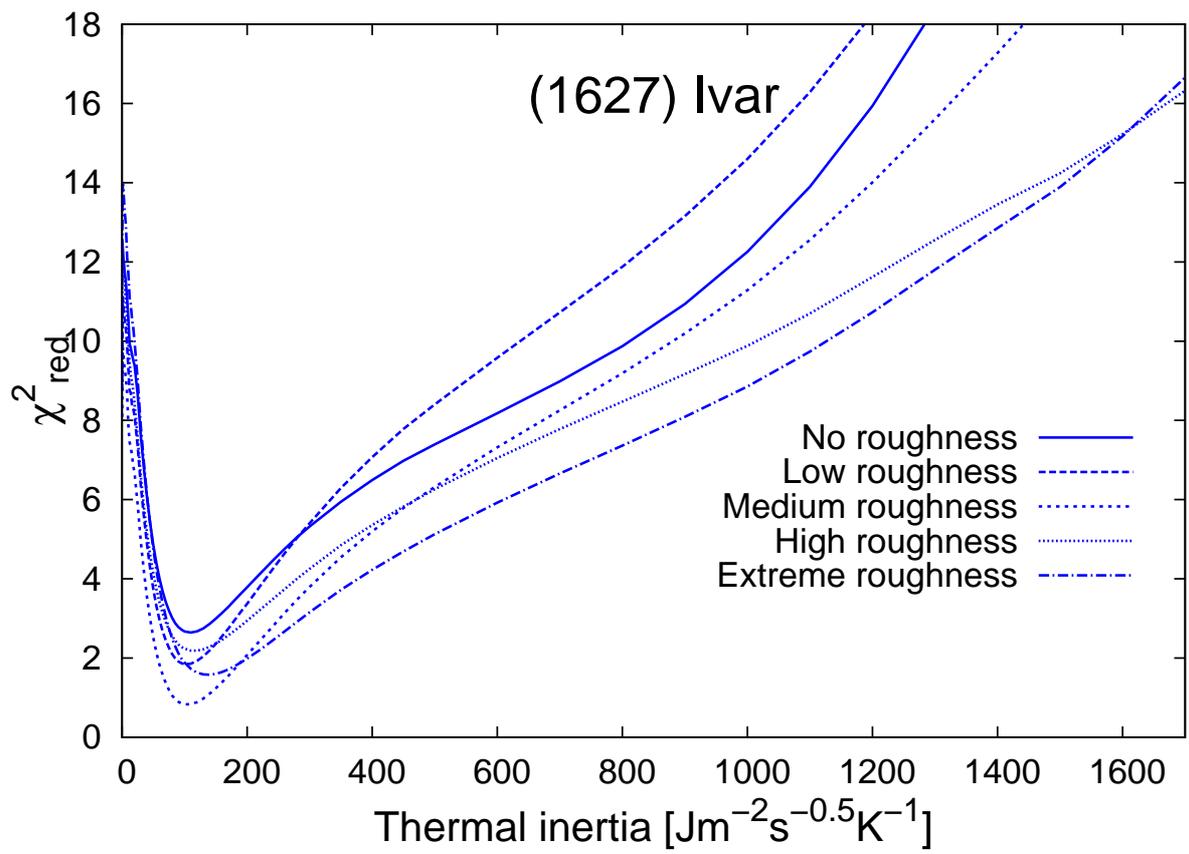}}\\
\end{center}
\caption{\label{img:tivschi}Dependence of the $\chi^2$ values of the TPM fits on the thermal inertia $\Gamma$ for a rotational phase $\phi_0=0^{\circ}$ and five different surface roughness values $\overline\theta$ for the revised shape model of asteroid (1627)~Ivar.}
\end{figure}

We use a thermophysical model implementation of \citet{Delbo2007a, Delbo2004} that is based on TPM developed by \citet{Lagerros1996, Lagerros1997, Lagerros1998, Spencer1989, Spencer1990, Emery1998}. A TPM allows thermal infrared fluxes to be calculated at different wavelengths and at a number of epochs taking into account the shape of an asteroid, its spatial orientation, and a number of physical parameters such as the size of the body, the albedo $A$, the macroscopic surface roughness $\overline \theta$ \citep[Hapke's mean surface slope,][]{Hapke1984}, and the thermal inertia $\Gamma$. The values of the parameters are determined by minimizing the difference between the observed fluxes $f_i$ and the modeled fluxes $s^2F_i$, where we consider the scale factor $s$ for the asteroid size, and $i$ corresponds to individual observations. To find the optimal set of parameter values, we minimize the metric

\begin{equation}\label{eq:chi2}
\chi^2 = \sum_i\frac{(s^2F_i-f_i)^2}{\sigma^2_i},
\end{equation}
where $\sigma_i$ represent the errors of fluxes $f_i$. 

The shape is represented by a convex polyhedron with triangular facets (models from the DAMIT database) or by a set of surfaces and normals \citep[for revised models, the so called Gaussian image, see][]{Kaasalainen2001a}. According to our tests, both representations produce similar fluxes and thus are equivalent. In the case of new models, we favor the latter representation, which is a natural outcome of the convex inversion, because it is less computationally demanding. The shape has an initial size, and the parameter $s$, adjusted, to FIT the data, is a factor that either scales linearly all vectors of the vertices of the polyhedron, or quadratically all the surfaces of the Gaussian image. Given the best-fitting value of $s$, we determine the {\em volume equivalent diameter} $D$ (i.e., the diameter of a sphere with the same volume as the scaled shape model), which is a quantity we present throughout this work and that we use to determine the body's albedo.
For the WISE bands, the bolometric emissivity can be set to a constant value of 0.9, that is the typical average spectral emissivity in the range of thermal infrared wavelengths used in thermal models. At the relevant temperatures for asteroids, the wavelength range that contributes most to this average is between 8 to 40 microns. This wavelength range contains absorptions features due  Si--O stretch and bend in silicate materials. The spectral emissivities are generally between 0.8 and 1.0 \citep[e.g.,][]{Salisbury1991,Christensen2000}. As the shapes modeled in this work are convex, there is no need to take into account topographic shadowing effect and the heating due to the light reflected and emitted by facets on other facets. The effect of roughness on the thermal infrared flux is accounted for by adding a spherical-section crater to each surface element of the shape. The crater with an opening angle $\gamma_{\mathrm{c}}$ and the crater areal density with respect to the flat part of the surface element $\rho_{\mathrm{c}}$ can be varied from 0 to 90$^{\circ}$ and from 0 to 1, respectively, to cover different values of the roughness. Following the procedure of \citet{Delbo2009} and \citet{MullerPHD}, we calculate the TPM only for a set of ten roughness models, whose parameters are given in Tab.~\ref{tab:roughness}. The correspondence between the Hapke's mean surface slope $\overline \theta$ and the adopted values of $\gamma_{\mathrm{c}}$ and $\rho_{\mathrm{c}}$ is also given in Tab.~\ref{tab:roughness}. Note the degeneracy in the $\overline \theta$ parameter -- different combinations of $\gamma_{\mathrm{c}}$ and $\rho_{\mathrm{c}}$ give the same $\overline \theta$.

%%%
\begin{table}
\caption{\label{tab:roughness}Ten different values of surface roughness used in the TPM. The table gives the opening angle $\gamma_{\mathrm{c}}$, the crater areal density $\rho_{\mathrm{c}}$, the Hapke's mean surface slope $\overline \theta$, and our designation.}
\centerline{
\begin{tabular}{cccc} \hline
$\gamma_{\mathrm{c}}$ & $\rho_{\mathrm{c}}$ & $\overline \theta$ & Designation \\ \hline\hline
0  & 0.0 &  0.0 &  No roughness \\
30  & 0.3 &  3.9 &  Low roughness \\
40  & 0.7 & 12.6 & Medium roughness \\
40  & 0.9 & 16.1 & Medium roughness \\
50  & 0.5 & 12.0 & Medium roughness \\
60  & 0.9 & 26.7 & High roughness \\
70  & 0.7 &  27.3 & High roughness \\
90  & 0.5 & 38.8 & High roughness \\
90  & 0.7 & 48.4 & Extreme roughness \\
90  & 0.9 & 55.4 & Extreme roughness \\ \hline
\end{tabular}
}
\end{table}

%%%

Shadowing of crater facets on other crater facets and mutual heating is taken into account. However, contrary to the approach of \citet{Delbo2009}, heat diffusion is not explicitly calculated within craters except for (1865)~Cerberus. Instead, the analytical approximation of \citet{Lagerros1998}, valid for small solar phase angles, is used. The phase angles for some NEAs can reach in the case of WISE observations even values of $\sim$90$^{\circ}$. The Lagerros approximation cannot be used on the night side \citep{Lagerros1998}, which, however, corresponds to the half of the surface if $\alpha\sim$90$^{\circ}$. Asteroid (1865)~Cerberus is the only one from here studied objects affected by the high phase angle observations, thus we have to explicitly calculate the heat diffusion in the craters in this case. For the remaining asteroids, fluxes computed by both approaches differ by less than 1\%, which is less than are the uncertainties of the observed fluxes we use. We favor the Lagerros approximation because of a considerable reduction of the computational time (factor of $\sim$40).

Given the asteroid convex shape and its rotational state, we run the TPM model for different values of the thermal inertia $\Gamma \in (0, 2500)$  J\,m$^{-2}$\,s$^{-1/2}$\,K$^{-1}$ and ten combinations of surface roughness (Tab.~\ref{tab:roughness}). For each value of the surface roughness, we run the TPM for the thermal inertia $\Gamma=2500$ J\,m$^{-2}$\,s$^{-1/2}$\,K$^{-1}$ and the Bond albedo $A=0.08$, and get the first size estimate $D$. Prior each following TPM run (while keeping the same surface roughness), we first compute the new value of the Bond albedo $A$ from the equation \citep[see, e.g.,][]{Harris2002}

\begin{equation}\label{eq:D}
D (\mathrm{km}) = \frac{1329}{\sqrt{p_{\mathrm{V}}}}\,10^{-0.2H},
\end{equation}
where we use diameter $D$ determined in the previous TPM run, and where the visible geometric albedo $p_{\mathrm{V}}$ can be expressed via $A\approx q\,p_{\mathrm{v}}$, where $q=0.290+0.684G$ is the phase integral \citep{Bowell1989}. We adopt the values of absolute magnitudes $H$ and slopes $G$ from the Asteroid absolute magnitude and slope catalog\footnote{\texttt{http://wiki.helsinki.fi/display/PSR/Asteroid+absolu\-te+magnitude+and+slope}} \citep[AAMS,][]{Muinonen2010, Oszkiewicz2011}. Following the procedure of \citet{MullerPHD}, we then run the TPM model with the next (lower) value of $\Gamma$ from our grid all the way until $\Gamma=0$ (always recomputing the $A$ values before each following step). We start the iteration with the highest value of $\Gamma$ from our grid, because we need few steps to reach the realistic combination of $A$ and $D$. While thermal inertia values for majority of asteroids are expected to be significantly lower, sometimes even bellow 10 J\,m$^{-2}$\,s$^{-1/2}$\,K$^{-1}$, starting the iteration with highest value of thermal inertia, could less likely bias the results than starting with the lowest one (i.e., 0 J\,m$^{-2}$\,s$^{-1/2}$\,K$^{-1}$). In any case, according to our tests for several objects, starting the iteration from zero thermal inertia gives consistent TPM results for $\Gamma\gtrsim5-10$ J\,m$^{-2}$\,s$^{-1/2}$\,K$^{-1}$.

Shape models are derived by the lightcurve inversion technique, where a large parameter space is searched: sidereal rotational period, pole orientation, shape, scattering parameters. Due to the uncertainty in the period, the initial rotational phase is typically known with an accuracy of $\pm$10 degrees. As a result, the orientation of the shape model (i.e., the rotational phase $\phi_0$) is known to this level of certainty inside the interval of optical lightcurve observations used for the lightcurve inversion. When the thermal infrared data are acquired outside this time span, however, the uncertainty of $\phi_0$ propagates proportionally with respect to the time elapsed between the acquisition of the thermal and the optical observations. For more details see Appendix B. For most studied asteroids, their expected uncertainties of $\phi_0$ are $\delta\phi_0\lesssim20^{\circ}$, which implies that it is necessary to include the rotational phase $\phi_0$ into the TPM optimization because even a change of 3--4 degrees can significantly improve the fit. We decide to scan $\phi_0$ with a step of 2$^{\circ}$ within the expected range.

An example of such sequence of TPM runs for five different values of surface roughness is shown in Fig.~\ref{img:tivschi}, where we plot the dependence of the reduced $\chi^2$ values of the TPM fits on the thermal inertia for asteroid (1627)~Ivar.

After scanning the parameter space of the thermal inertia $\Gamma$, surface roughness $\overline \theta$, Bond albedo $A$ and rotational phase $\phi_0$, we find the solution with the lowest $\chi^2$ value. The $\chi^2$ metric is used to find the best-fitting solution takes into account directly only the uncertainties of the thermal infrared fluxes. It neglects uncertainties of the shape model and the pole orientation. We investigate their importance in thermophysical modeling below.

We scale the size of the shape model directly by the TPM. After that, we determine the {\em volume equivalent diameter} $D$ (i.e. the diameter of a sphere with the same volume as the scaled shape model), which is a quantity we present throughout this work.

While the $\chi^2$ metric is used to find the best-fitting solution, it is common practice in the TPM analysis also to use the $\chi^2$ statistic to estimate the goodness-of-fit \citep[see, e.g., ][]{AliLagoa2014, Emery2014}. The reduced chi-square is computed from the standard chi-square metric defined by Eq.~\ref{eq:chi2} as $\chi^2_{\mathrm{red}}=\chi^2/\nu$, where the number of degrees of freedom $\nu$ corresponds to the number of data points minus the number of free parameters. All solutions within $\chi^2_{\mathrm{red}}<(1+\sigma)$, where $\sigma=\sqrt{2\nu}/\nu$, are considered indistinguishable \citep[e.g.,][]{Press1986} and are used to estimate the uncertainties of the fitted parameters. However, this approach is, in principle, reliable only if $\chi^2_{\mathrm{red}} \approx 1$ and in the (ideal) case of normally distributed and independent errors of the thermal infrared fluxes. 
Values of $\chi^2_{\mathrm{red}}\gtrsim1$ indicate that the model fit has not fully captured the data, but such solutions can still produce realistic results if care is exercised in the analysis. Values of $\chi^2_{\mathrm{red}}\gg1$ indicate a poor fit that should be rejected \citep[e.g., in ][the authors rejected solutions with $\chi^2_{\mathrm{red}}\sim8$]{Delbo2009}.
An alternative is to estimate the fitted parameter uncertainties by means of an empirical approach: the error bars include all parameter values corresponding to all solutions with $\chi^2_{\mathrm{red}}<\chi^2_{\mathrm{min}}*(1+\sigma)$. Such uncertainties often span well the minima in the thermal inertia parameter space for our typical $\chi^2_{\mathrm{red}}$ values of 1--3, and we always check the appearance of the minima individually to ensure that this is the case.

\subsection{TPM with fixed shape models}\label{sec:TPM_classical}

\begin{figure*}[!htbp]
\begin{center}
\resizebox{\hsize}{!}{\includegraphics{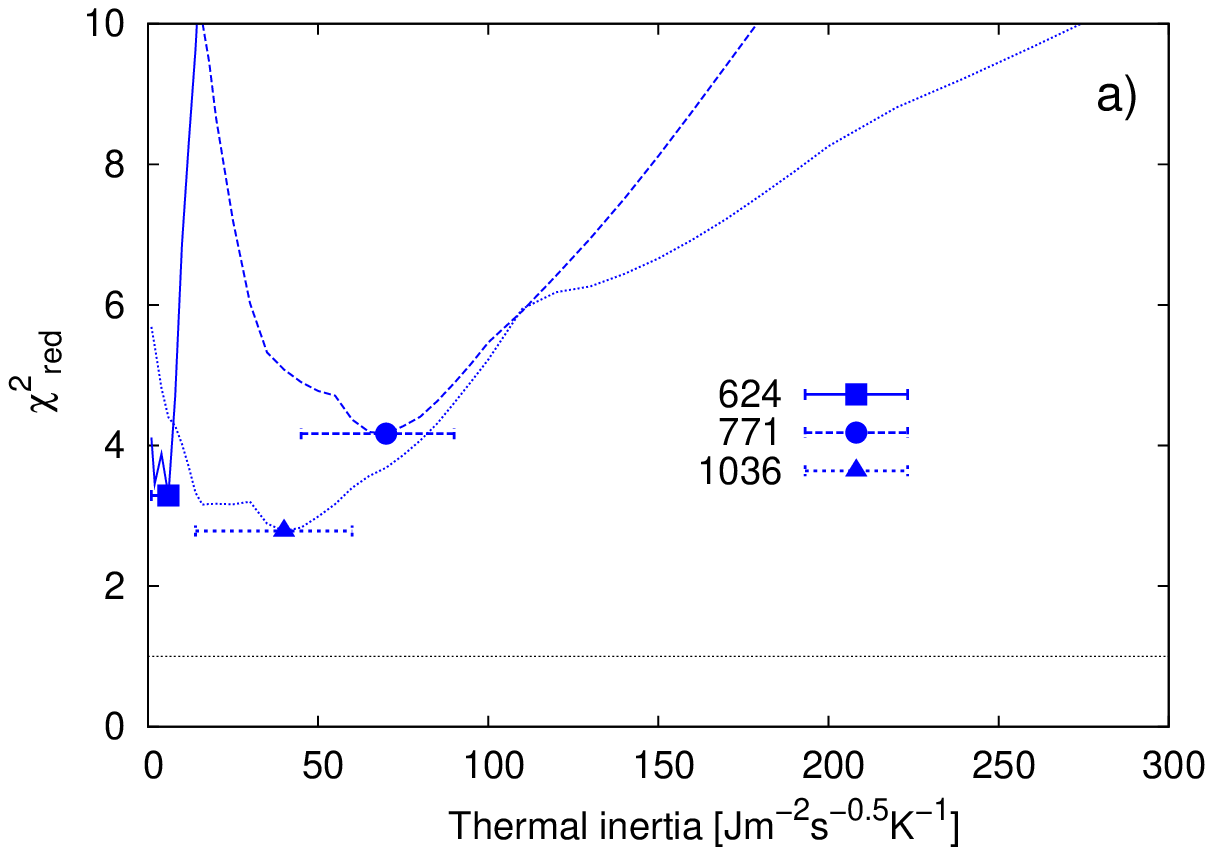}\includegraphics{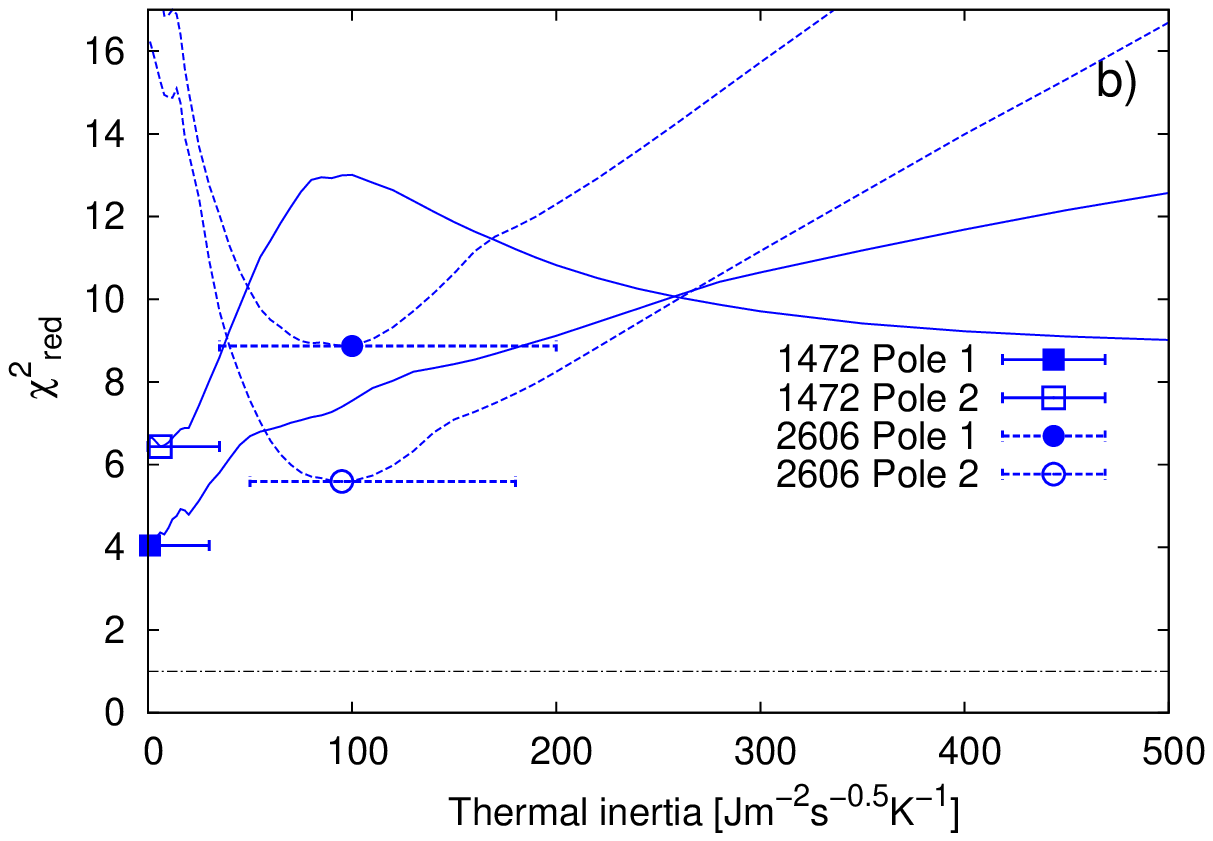}}\\
\resizebox{\hsize}{!}{\includegraphics{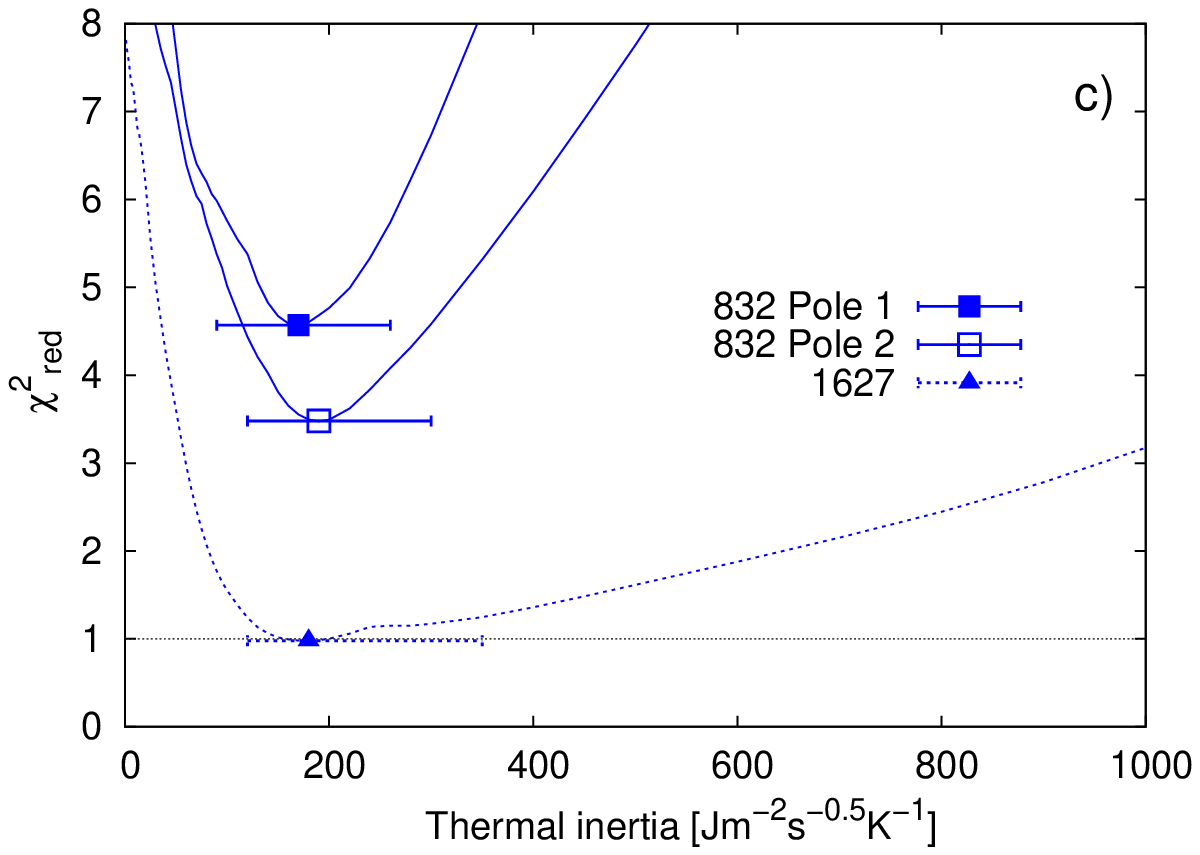}\includegraphics{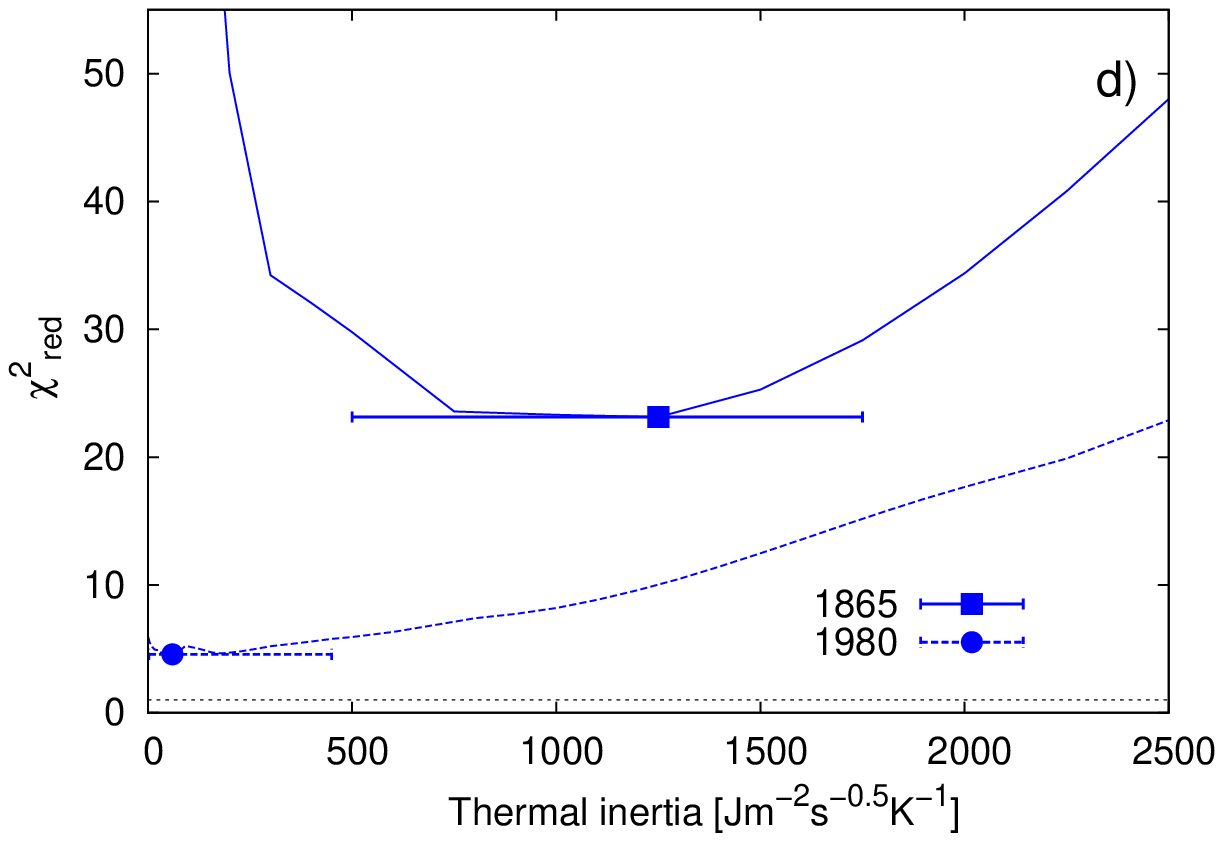}}\\
\end{center}
\caption{\label{img:gamma_vs_chi2_all}Thermal inertia $\Gamma$ fits for the best-fitting surface roughness and rotational phase for nine studied asteroids (DAMIT shape models). For asteroids (832)~Karin, (1472)~Muonio and (2606)~Odessa, we plot the dependence for both ambiguous pole solutions. The dashed horizontal line indicates $\chi^2_{\mathrm{red}}=1$. }
\end{figure*}

%\onecolumn
\begin{landscape}
\LTcapwidth=\textwidth
\begin{longtable}{r@{\,\,\,}l ccccc ccc cc c cc c}
\caption{\label{tab:TI}Thermophysical characteristics of asteroids derived by the classical TPM with the DAMIT (D in the ``pole`` column) and revised (R) shape models, as well as by the VS-TPM (VS). We provide the asteroid number and name, the pole solution, the number of IRAS thermal IR measurements $N_\mathrm{IRAS}$, the number of WISE thermal IR measurements in filters W3 $N_\mathrm{W3}$ and W4 $N_\mathrm{W4}$, volume equivalent diameter $D$, thermal inertia $\Gamma$, visual geometric albedo $p_\mathrm{V}$, Hapke's mean surface slope $\overline \theta$, rotational phase $\phi_0$, reduced chi-square of the best fit $\chi^2_{\mathrm{red}}$, absolute magnitude $H$ and slope $G$ \citep[AAMS,][]{Muinonen2010, Oszkiewicz2011}, taxonomical class and average heliocentric distance $r$ of the asteroids when observed by WISE. For the taxonomy, we primarilly show SMASS II class \citep{Bus2002}, if not available, then the Tholen class \citep{Tholen1984,Tholen1989} or taxonomy from \citet{DeMeo2013}.}\\
\hline 
\multicolumn{2}{c} {Asteroid} & \multicolumn{1}{c} {Pole} & \multicolumn{1}{c} {$N_\mathrm{IRAS}$} & \multicolumn{1}{c} {$N_\mathrm{W3}$} & \multicolumn{1}{c} {$N_\mathrm{W4}$} & \multicolumn{1}{c} {$D$} & \multicolumn{1}{c} {$\Gamma$} & \multicolumn{1}{c} {$p_\mathrm{V}$} & \multicolumn{1}{c} {$\overline \theta$} & \multicolumn{1}{c} {$\phi_0$} & \multicolumn{1}{c} {$\chi^2_{\mathrm{red}}$} & \multicolumn{1}{c} {$H$} & \multicolumn{1}{c} {$G$} & \multicolumn{1}{c} {TAX}  & \multicolumn{1}{c} {$r$} \\
\multicolumn{2}{l} { } &  &  &  &  & \multicolumn{1}{c} {[km]} & \multicolumn{1}{c} {[J\,m$^{-2}$\,s$^{-1/2}$\,K$^{-1}$]} &  &  & [deg] &  & \multicolumn{1}{c} {[mag]} &  &  & [AU] \\
\hline\hline

\endfirsthead
\caption{continued.}\\

\hline 
\multicolumn{2}{c} {Asteroid} & \multicolumn{1}{c} {Pole} & \multicolumn{1}{c} {$N_\mathrm{IRAS}$} & \multicolumn{1}{c} {$N_\mathrm{W3}$} & \multicolumn{1}{c} {$N_\mathrm{W4}$} & \multicolumn{1}{c} {$D$} & \multicolumn{1}{c} {$\Gamma$} & \multicolumn{1}{c} {$p_\mathrm{V}$} & \multicolumn{1}{c} {$\overline \theta$} & \multicolumn{1}{c} {$\phi_0$} & \multicolumn{1}{c} {$\chi^2_{\mathrm{red}}$} & \multicolumn{1}{c} {$H$} & \multicolumn{1}{c} {$G$} & \multicolumn{1}{c} {TAX}  & \multicolumn{1}{c} {$r$} \\
\multicolumn{2}{l} { } &  &  &  &  & \multicolumn{1}{c} {[km]} & \multicolumn{1}{c} {[J\,m$^{-2}$\,s$^{-1/2}$\,K$^{-1}$]} &  &  & [deg] &  & \multicolumn{1}{c} {[mag]} &  &  & [AU] \\
\hline\hline

\endhead
\hline
\endfoot

  624 &	         Hektor &	 1D &	   &	 11 &	 11 &	 181$^{+1}_{-10}$ &	 6$^{+1}_{-5}$ &	 0.05$^{+0.01}_{-0.01}$ &	26.7 &	 354 &	 3.3 &	 7.3 &	 0.33 &	   D &	 5.3 \\
      &	                &	 1R &	    &	    &	    &	 175$^{+6}_{-21}$ &	 6$^{+2}_{-6}$ &	 0.056$^{+0.017}_{-0.003}$ &	26.7 &	   2 &	 4.5 &	  &	 &	  &	 \\
      &	                &	 1VS &	    &	    &	    &	 186$^{+1}_{-34}$ &	 6$^{+4}_{-6}$ &	 0.058$^{+0.017}_{-0.007}$ &	26.7 &	     &	 2.4 &	     &	      &	     &	     \\
  771 &	         Libera &	 1D &	  6 &	 31 &	 31 &	 30.5$^{+1.3}_{-1.3}$ &	 70$^{+20}_{-25}$ &	 0.14$^{+0.01}_{-0.01}$ &	27.3 &	   2 &	 4.2 &	 10.3 &	 0.32 &	   X &	 2.8 \\
      &	                &	 1VS &	    &	    &	    &	 32$^{+2}_{-5}$ &	 65$^{+85}_{-35}$ &	 0.13$^{+0.05}_{-0.02}$ &	26.7 &	     &	 3.4 &	      &	      &	     &	     \\
  832 &	          Karin &	 1D &	   &	 12 &	 12 &	 16.4$^{+0.5}_{-1.3}$ &	 170$^{+90}_{-80}$ &	 0.21$^{+0.04}_{-0.01}$ &	38.8 &	  12 &	 4.6 &	 11.1 &	 0.16 &	   S &	 2.9 \\
      &	                &	 2D &	   &	    &	    &	 16.8$^{+0.5}_{-1.5}$ &	 190$^{+110}_{-70}$ &	 0.20$^{+0.04}_{-0.01}$ &	38.8 &	   6 &	 3.5 &	  &	  &	    &	  \\
      &	                &	 1VS &	    &	    &	    &	 16.6$^{+0.8}_{-1.5}$ &	 90$^{+150}_{-90}$ &	 0.22$^{+0.04}_{-0.04}$ &	26.7 &	     &	 1.9 &	      &	      &	     &	     \\
      &	                &	 2VS &	    &	    &	    &	 17$^{+2}_{-2}$ &	 65$^{+215}_{-65}$ &	 0.23$^{+0.05}_{-0.06}$ &	16.1 &	     &	 1.6 &	      &	      &	     &	     \\
 1036 &	        Ganymed &	 1D &	  2 &	 22 &	 22 &	 35.5$^{+1.9}_{-0.8}$ &	 40$^{+20}_{-26}$ &	 0.26$^{+0.01}_{-0.03}$ &	26.7 &	 140 &	 2.8 &	 9.2 &	 0.31 &	   S &	 3.9 \\
      &	                &	 1R &	    &	    &	    &	 36.0$^{+1.0}_{-0.5}$ &	 35$^{+10}_{-5}$ &	 0.253$^{+0.008}_{-0.011}$ &	26.7 &	  12 &	 1.2 &	  &	 &	  &	 \\
      &	                &	 1VS &	    &	    &	    &	 37$^{+2}_{-4}$ &	 35$^{+65}_{-29}$ &	 0.25$^{+0.05}_{-0.03}$ &	26.7 &	     &	 1.1 &	     &	      &	     &	     \\
 1472 &	         Muonio &	 1D &	   &	 10 &	  9 &	 8.6$^{+0.9}_{-0.8}$ &	 1$^{+29}_{-1}$ &	 0.24$^{+0.05}_{-0.05}$ &	3.9 &	 358 &	 4.0 &	 12.4 &	 0.35 &	   - &	 2.7 \\
      &	                &	 2D &	   &	    &	    &	 8.1$^{+1.1}_{-0.9}$ &	 6$^{+29}_{-6}$ &	 0.27$^{+0.07}_{-0.06}$ &	12.6 &	  12 &	 6.4 &	  &	  &	    &	  \\
      &	                &	 1VS &	    &	    &	    &	 9.1$^{+1.1}_{-1.5}$ &	 25$^{+65}_{-25}$ &	 0.23$^{+0.09}_{-0.06}$ &	16.1 &	     &	 1.3 &	      &	      &	      &	     \\
      &	                &	 2VS &	    &	    &	    &	 9.2$^{+0.3}_{-3}$ &	 0$^{+40}_{-0}$ &	 0.23$^{+0.09}_{-0.06}$ &	0.0 &	     &	 2.2 &	      &	      &	      &	     \\
 1627 &	           Ivar &	 1D &	   &	 13 &	 13 &	 7.4$^{+0.2}_{-0.2}$ &	 180$^{+170}_{-60}$ &	 0.26$^{+0.02}_{-0.02}$ &	16.1 &	   4 &	 1.0 &	 12.6 &	 0.33 &	   S &	 2.1 \\
      &	                &	 1R &	    &	    &	    &	 7.4$^{+0.1}_{-0.2}$ &	 100$^{+30}_{-20}$ &	 0.257$^{+0.015}_{-0.005}$ &	12.0 &	  10 &	 0.8 &	  &	 &	  &	 \\
      &	                &	 1VS &	    &	    &	    &	 8.0$^{+0.3}_{-0.9}$ &	 100$^{+120}_{-40}$ &	 0.255$^{+0.02}_{-0.014}$ &	12.0 &	     &	 0.8 &	      &	      &	     &	     \\
 1865 &	       Cerberus* &	 1D &	   &	 10 &	  9 &	 $\sim$1.3 &	 $\sim$1250 &	 $\sim$0.17 &	16.1 &	 350 &	 23.1 &	 16.6 &	 0.37 &	   S &	 1.1 \\
      &	               * &	 1R &	    &	    &	    &	 $\sim$1.2 &	 $\sim$400 &	 $\sim$0.215 &	55.4 &	 352 &	 16.2 &	  &	 &	  &	 \\
      &	               * &	 1VS &	    &	    &	    &	 $\sim$1.2 &	 $>$300 &	 $\sim$0.30 &	55.4 &	     &	 6.3 &	      &	      &	     &	     \\
 1980 &	   Tezcatlipoca &	 1D &	   &	 17 &	 15 &	 5.0$^{+1.1}_{-0.6}$ &	 60$^{+390}_{-58}$ &	 0.23$^{+0.07}_{-0.08}$ &	16.1 &	 356 &	 4.6 &	 13.6 &	 0.18 &	  Sl &	 2.3 \\
      &	                &	 1R &	    &	    &	    &	 4.6$^{+1.1}_{-0.4}$ &	 60$^{+440}_{-42}$ &	 0.29$^{+0.05}_{-0.11}$    &	16.1 &	 350 &	 3.5 &	  &	 &	  &	 \\
      &	                &	 1VS &	    &	    &	    &	 5.4$^{+0.7}_{-1.3}$ &	 220$^{+380}_{-204}$ &	 0.22$^{+0.13}_{-0.07}$ &	48.4 &	     &	 2.9 &	      &	      &	     &	     \\
 2606 &	         Odessa &	 1D &	   &	  9 &	  9 &	 17$^{+3}_{-4}$ &	 100$^{+100}_{-65}$ &	 0.10$^{+0.06}_{-0.02}$ &	12.0 &	 348 &	 8.9 &	 11.7 &	 0.21 &	   X &	 3.5 \\
      &	                &	 2D &	   &	    &	    &	 16$^{+2}_{-3}$ &	 95$^{+85}_{-45}$ &	 0.12$^{+0.05}_{-0.03}$ &	12.6 &	 348 &	 5.6 &	  &	  &	    &	  \\
      &	                &	 1VS &	    &	    &	    &	 20$^{+1}_{-5}$ &	 100$^{+100}_{-55}$ &	 0.09$^{+0.05}_{-0.02}$ &	12.0 &	     &	 1.6 &	      &	      &	     &	     \\
      &	                &	 2VS &	    &	    &	    &	 17$^{+2}_{-3}$ &	 90$^{+80}_{-40}$ &	 0.13$^{+0.04}_{-0.04}$ &	12.0 &	     &	 1.5 &	      &	      &	     &	     \\

\hline
%\multicolumn{16}{l} {*Formally non-acceptable TPM fit}

%\tablefoot{
%\hspace{2cm}*Poor TPM solution.
%}
\end{longtable}
*Poor TPM fit
%For each asteroid, the table gives the  .\\
%\tablefoottext{1}{1.}
%\tablefoottext{2}{2.} }
%\footnotetext[2]{Multiple pole solution with similar $\beta$, period value is secure.}
\end{landscape}

We take shape models of four near-Earth asteroids (1036)~Ganymed, (1627)~Ivar, (1865)~Cerberus and (1980)~Tezcatlipoca, of four main-belt asteroids (771)~Libera, (832)~Karin, (1472)~Muonio and (2606)~Odessa, and of the Jovian Trojan asteroid (624)~Hektor from the publicly available DAMIT database. In Tab.~\ref{tab:models}, we list rotational parameters, information about the photometry used for the shape model determination, and the references to the original publications. These solutions have well defined sidereal rotational period values (the accuracy corresponds to the last decimal place in period values in Tab.~\ref{tab:models}), and usually two symmetric pole orientations with similar ecliptic latitudes $\beta$ and difference in ecliptic longitude $\lambda$ of $\sim180^{\circ}$ (the so-called pole ambiguity, a typical example is asteroid (832)~Karin). However, due to the more various observing geometries of NEAs than of the main-belt asteroids or Jovian Trojans, the pole ambiguity is not present in our sample of NEAs. Moreover, \citet{Marchis2014} removed the pole ambiguity for asteroid (624)~Hektor thanks to the disk-resolved images, thus we use only the preferred solution (see Tab~\ref{tab:models}) for the TPM.

We run the TPM for each studied asteroid and for various initial values of thermophysical parameters as described in Sect.~\ref{sec:TPM_method}. In Tab.~\ref{tab:TI}, we list thermophysical parameters of the best-fitting solutions -- $\Gamma$, $D$, $\overline\theta$ and the geometric visible albedo $p_\mathrm{V}$ (instead of the Bond albedo $A$).

We obtain best-fitting TPM solution with $\chi^2_{\mathrm{red}}\sim1$ only for asteroid (1627)~Ivar. This solution can be considered formally acceptable.

For asteroids (624)~Hektor, (771)~Libera, (832)~Karin, (1036)~Ganymed, (1472)~Muonio, (1980)~Tezcatlipoca and (2606)~Odessa, we obtain TPM fits with higher $\chi^2_{\mathrm{red}}$ values in the range of 3--9, which means that our model is not fully reproducing the thermal observations. The extreme case is asteroid (1865)~Cerberus with $\chi^2_{\mathrm{red}}>20$, which means that we failed to constrain any of its properties.

The best-fitting thermophysical properties and their uncertainty estimates are included in Tab.~\ref{tab:TI}.

The convergence of the solution in the thermal inertia parameter space for all studied asteroids is shown in Figs.~\ref{img:gamma_vs_chi2_all}a,b,c,d, where the $\chi^2$ curves correspond to the best-fitting surface roughness and rotational phase for each value thermal inertia. We show both pole solutions when they are available. This figure shows that there is a more or less prominent minimum $\chi^2_{\mathrm{red}}$ for most asteroids, and therefore the thermal inertia is usually well constrained (e.g., (1036)~Ganymed, (1472)~Muonio or (2606)~Odessa). The thermal inertia for asteroid (624)~Hektor is the best constrained one ($\Gamma<30$ J\,m$^{-2}$\,s$^{-1/2}$\,K$^{-1}$), but in general, the range of acceptable thermal inertia values is broader (see~Tab.~\ref{tab:TI}). For asteroid (1980)~Tezcatlipoca, the thermal inertia is poorly constrained providing only an upper limit of $\lesssim$500~J\,m$^{-2}$\,s$^{-1/2}$\,K$^{-1}$.

The high $\chi^2_{\mathrm{red}}$ values could be, in principle, explained by:
(i)~shape model uncertainties (i.e., convex shape, pole orientation) dominate over the flux uncertainties, 
(ii)~systematic uncertainties due to various assumptions of our TPM model are biasing modeled fluxes (for example, a finite grid of searched parameters, or the usage of a specific roughness model), or finally
(iii)~an underestimation of the uncertainties of the WISE thermal infrared measurements (including systematics in calibration and outliers). 
These are discussed in the following subsections. In Sect.\ref{sec:varied_shape_TPM}, we present and apply the varied-shape TPM method, which shows that accounting for the uncertainties in the shape model and the pole orientation is essential for the determination of reliable thermophysical properties by TPM.

TPM modeling of asteroid (1036)~Ganymed is unique because the rotational phase $\phi_0$ is not constrained at all. The shape model is based on photometric data from only one apparition in 1985, so the uncertainty of the sidereal rotational period of 0.001 hour for the WISE observations (year 2010) corresponds to a rotational uncertainty up to $\sim540$ degrees.

In all three cases with ambiguous pole solutions, the TPM produce different fits (more than one in the best-fitting $\chi^2_{\mathrm{red}}$ values), which suggests that one pole solution is preferred, however, the thermal inertia values are consistent.

The fits to the thermal infrared data for selected/all asteroids are shown in the Supplementary material.

\subsection{Revised shape models}\label{sec:revised_models}

\begin{figure*}[!htbp]
\begin{center}
\resizebox{\hsize}{!}{\includegraphics{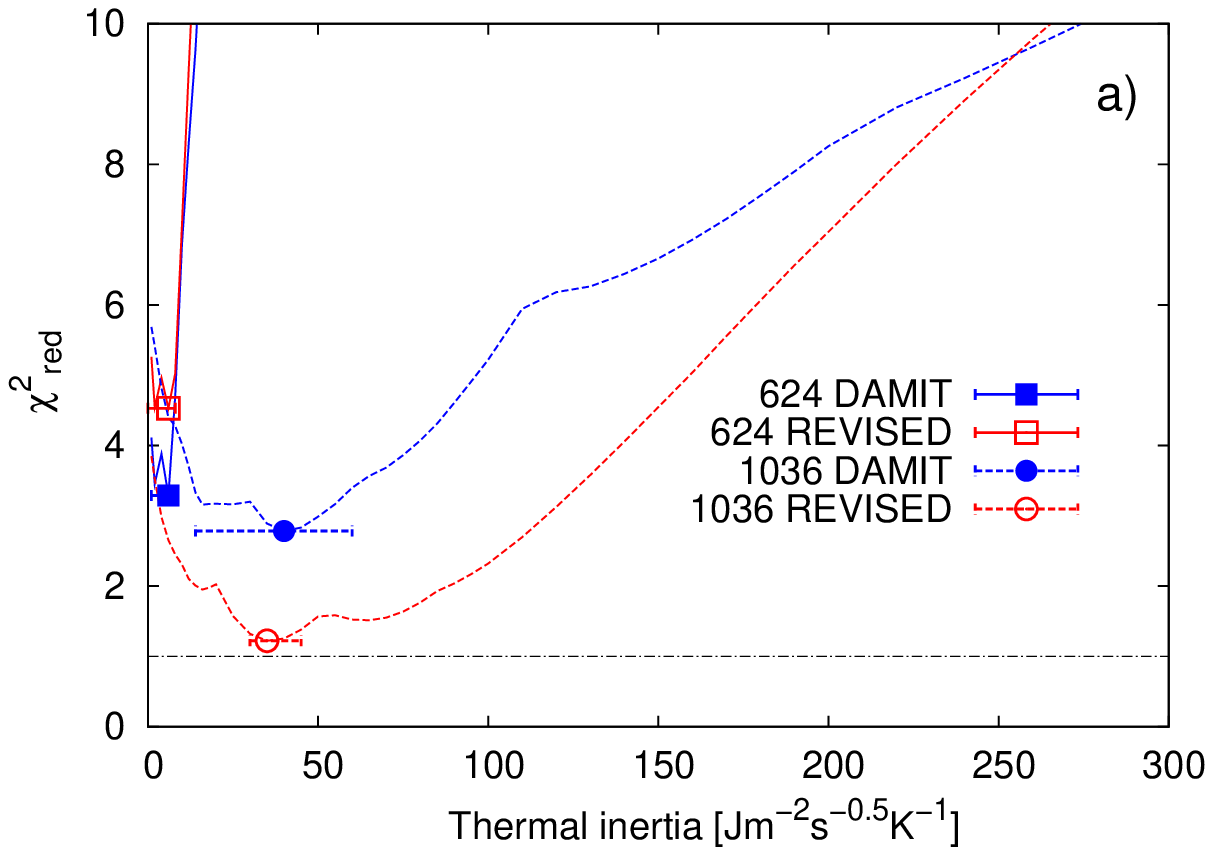}\includegraphics{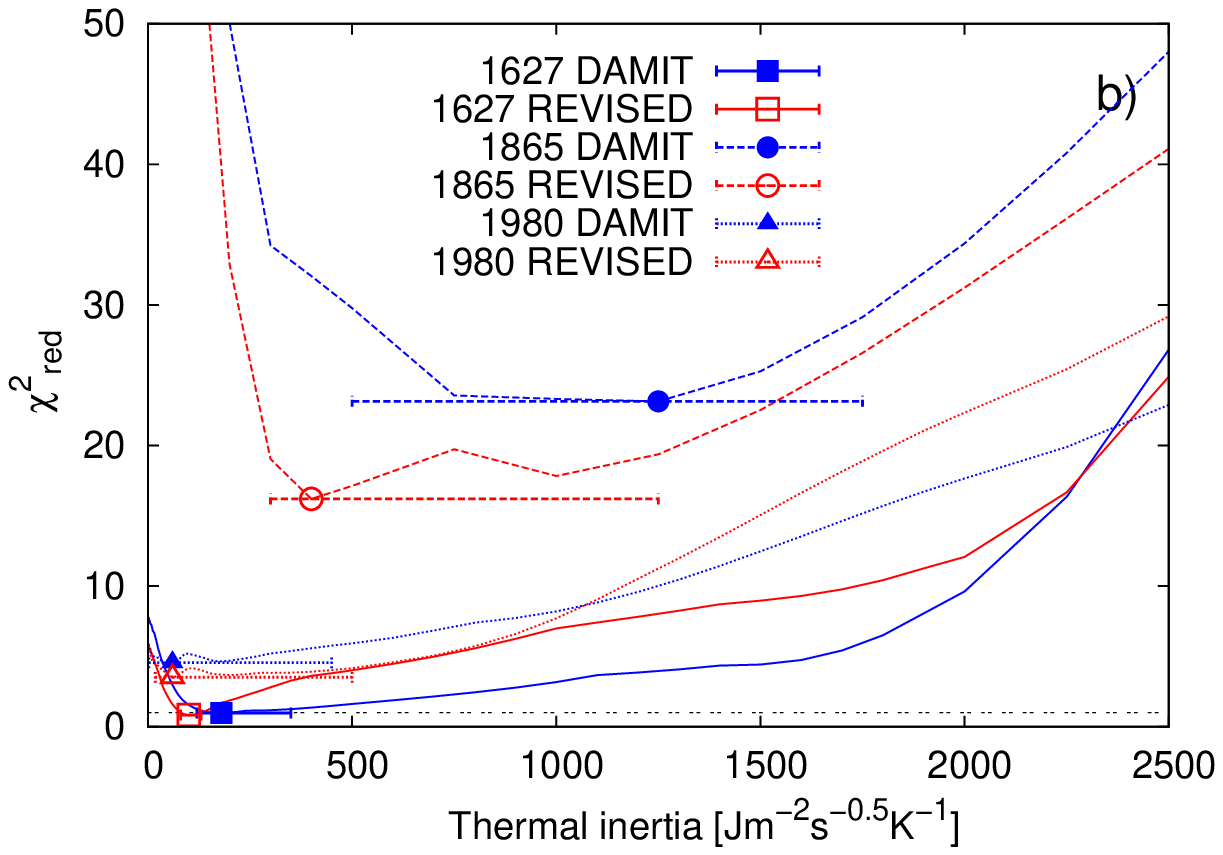}}\\
\end{center}
\caption{\label{img:gamma_vs_chi2_rev}Thermal inertia $\Gamma$ fits for the best-fitting surface roughness and rotational phase for DAMIT (blue) and revised (red) shape models of asteroids  (624)~Hektor, (1036)~Ganymed, (1627)~Ivar, (1865)~Cerberus and (1980)~Tezcatlipoca. The dashed horizontal line indicates $\chi^2_{\mathrm{red}}=1$. A color version of the figure is available in the electronic version of the journal.}
\end{figure*}

For some asteroids new optical lightcurves were acquired after their shape models were determined and included in the DAMIT. Thus we have the opportunity to revise the shape models of asteroids (624)~Hektor, (1036)~Ganymed, (1627)~Ivar, (1865)~Cerberus and (1980)~Tezcatlipoca. All revised shape models, derived following the procedure of \citet{Hanus2011, Hanus2013a}, are based on combined dense- and sparse-in-time photometric data sets. Rotational parameters and information about the photometry are included in Tab.~\ref{tab:models}.

The typical uncertainty of the pole orientation, which depends on the amount and quality of available photometric data, is in the ecliptic coordinate frame (5--10$^{\circ}$)/$\cos \beta$ in longitude $\lambda$ and 5--10$^{\circ}$ in latitude $\beta$ (see also Fig.~\ref{img:shapes_poles}, where we map a typical pole uncertainty for all studied asteroids). The same applies to the DAMIT models we already used in Sect.~\ref{sec:TPM_classical}.

We run the TPM with the revised shape models as fixed inputs as in the previous section and we present the derived thermophysical parameters in Tab.~\ref{tab:TI}. The best fitting values of TPM fits do not change significantly but the quality of the fits is generally improved (see the comparison in Figs.~\ref{img:gamma_vs_chi2_rev}a,b). The only exception is (624)~Hektor, for which the TPM fit with the revised shape model is slightly worse, but consistent in the best-fitting values of thermophysical properties.

These results indicate that the introduction of revised shape models affects the goodness-of-fit. For asteroid (1627)~Ivar, the TPM with the revised shape model reduced the uncertainty of the thermal inertia value by a factor of $\sim$3. In turn, we do not significantly lower the $\chi^2_{\mathrm{red}}$ values for asteroids (624)~Hektor, (1865)~Cerberus and (1980)~Tezcatlipoca, thus we still do not ideally reproduce the thermal infrared data. On the other hand, the TPM fit of asteroid (1036)~Ganymed has now best-fitting $\chi^2_{\mathrm{red}}$ value close to one.

One question that still remains is whether some improvements in the fitting technique could be the main cause of our $\chi^2_\mathrm{red}$ values being higher than one. In the next subsection we investigate this and other related issues.

\subsection{Additional model uncertainties}\label{sec:systematics}

Because we run TPM with thermophysical parameters (thermal inertia $\Gamma$, surface roughness $\overline \theta$, rotational phase $\phi_0$)  from a finite grid of values, our best-fitting solution could not correspond to a real global minimum in the searched parameter space; however, it should be close. By using a finer grid of parameter values, we should be in principle able to find a solution with a lower $\chi^2$. By varying $\Gamma$ and $\phi_0$, we are able to reduce the $\chi^2$ for all studied asteroids typically by only few percent, which is a rather marginal improvement. This means that our grid in these parameters is sufficiently fine.

To deal with the surface roughness, we use the Lagerros analytical crater approximation, which is computationally less demanding than performing a full heat diffusion computation within the craters. The differences in the fluxes produced by these two models are usually considerably smaller than 1\%, which are always lower than the uncertainties of the observed fluxes by a factor of at least five. As a result, the quality of the fit should be affected only marginally.

On the other hand, we use only ten different values for the surface roughness $\overline \theta$ in the modeling (see Tab.~\ref{tab:roughness}). In the practical application, we rather use the opening angle $\gamma_{\mathrm{c}}$ and the crater areal density $\rho_{\mathrm{c}}$ than the Hapke's mean slope $\overline \theta$. By using more combination of $\gamma_{\mathrm{c}}$ and $\rho_{\mathrm{c}}$, we are able to reduce the $\chi^2$ values of the TPM fits by few percent, which is again a marginal improvement of the fit.

We also do not account for the uncertainty in the input $H$~and $G$ values. According to our tests (TPM with different $H$ values), a change of $\pm0.5$ mag in $H$ is compensated by the change in Bond albedo $A$. However, the size remains similar (see Eq.~\ref{eq:D}), as well as the best-fitting values of thermal inertia, surface roughness, or indeed the resulting minimum $\chi^2_{\mathrm{red}}$.

%Including more values for the surface roughness helps to reduce the chi-square, however, each value puts considerable demands on the CPU needed for the computation. Because we do not receive different thermophysical parameters, including only few additional values for the surface roughness parametrization should be a reasonable compromise -- we decease the $\chi^2$ by reasonably increasing the computational time. However, we should keep in mind that a small improvement of the fit of the order of $\sim$10\% in $\chi^2$ could be still possible.

Our results call for additional investigation of the role of the shape model uncertainties in the TPM, which we perform by using our novel VS-TPM method in the next section.

\section{Varied shape TPM}\label{sec:varied_shape_TPM}

\subsection{VS-TPM -- method}\label{sec:varied_shape_TPM_method}

\begin{figure}[!htbp]
\begin{center}
 \resizebox{0.99\hsize}{!}{\includegraphics{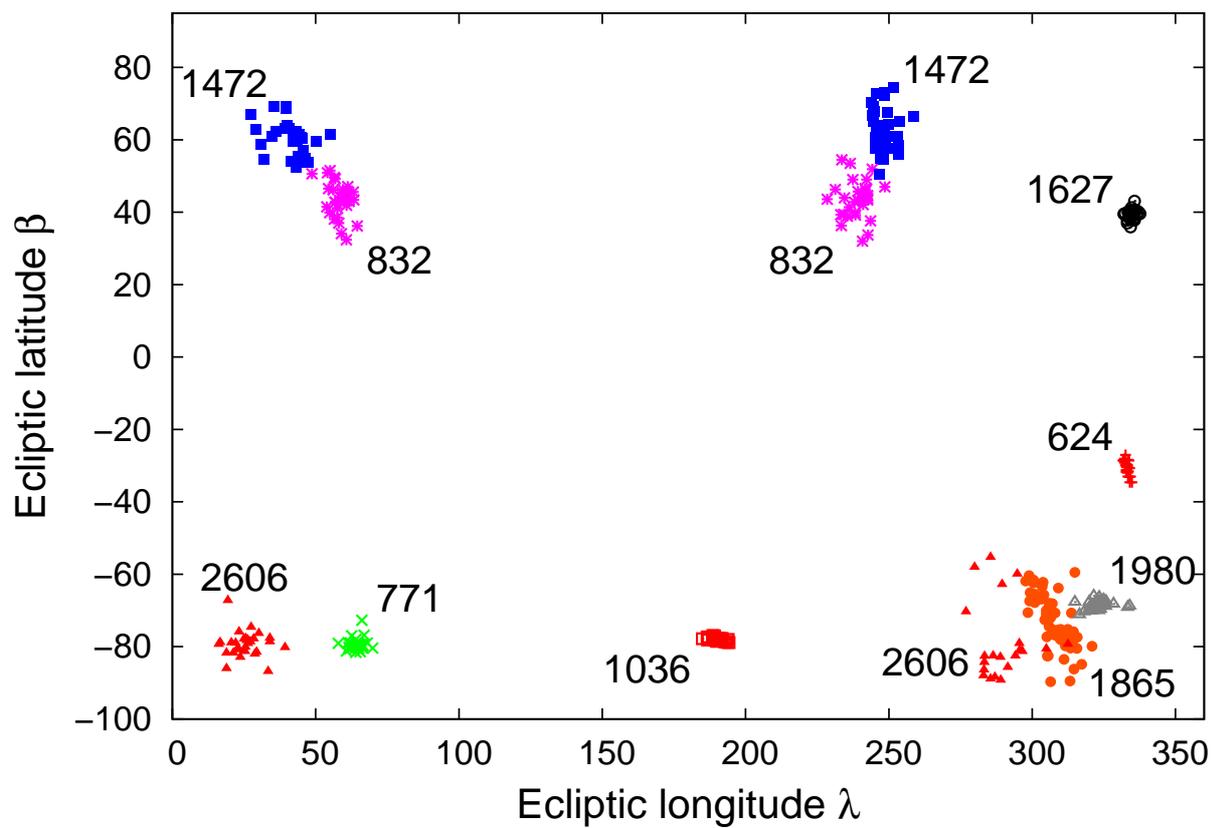}}\\
\end{center}
 \caption{\label{img:shapes_poles}Orientations of the spin axes of all varied shape models for here studied asteroids. A color version of the figure is available in the electronic version of the journal.}
\end{figure}

\begin{figure}[!htbp]
\begin{center}
 \resizebox{0.48\hsize}{!}{\includegraphics{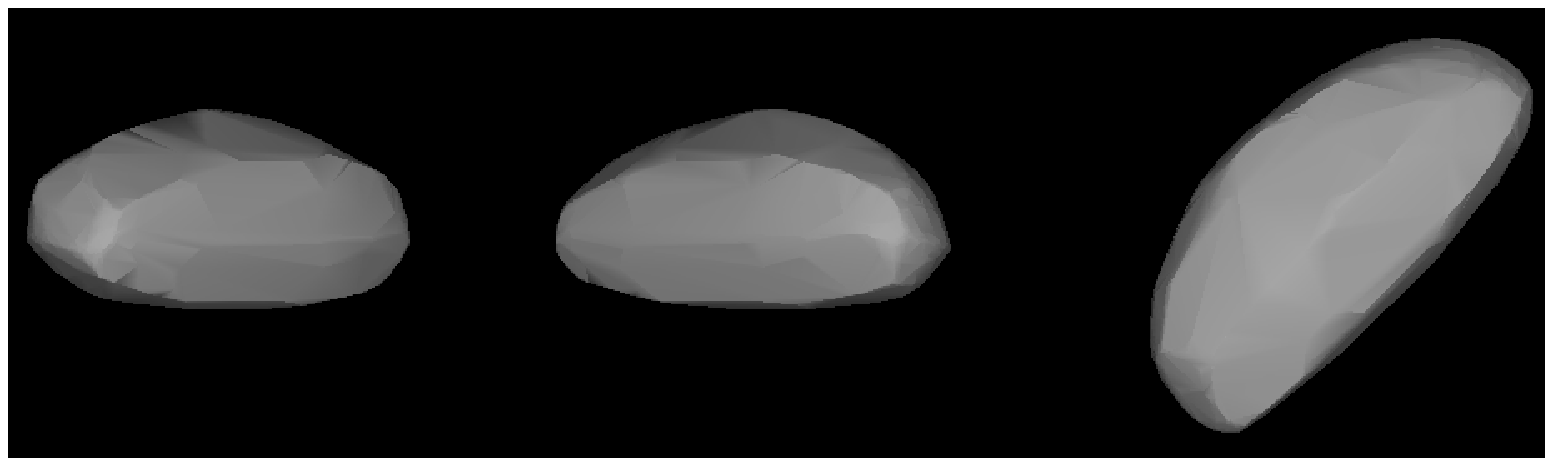}}\,\resizebox{0.495\hsize}{!}{\includegraphics{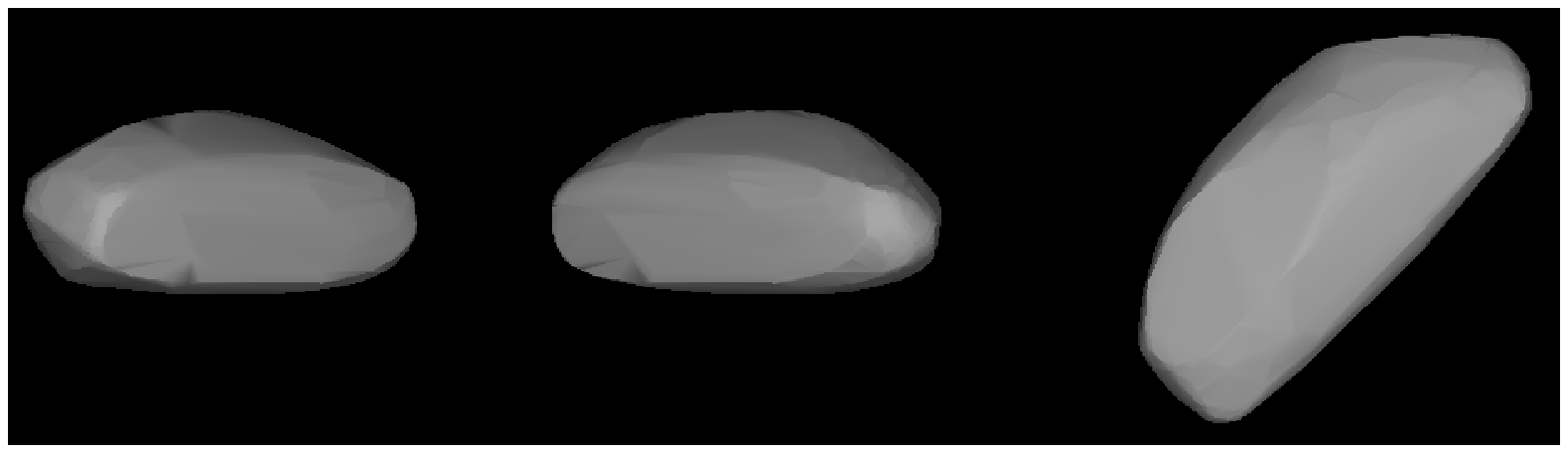}}\\
 \resizebox{0.48\hsize}{!}{\includegraphics{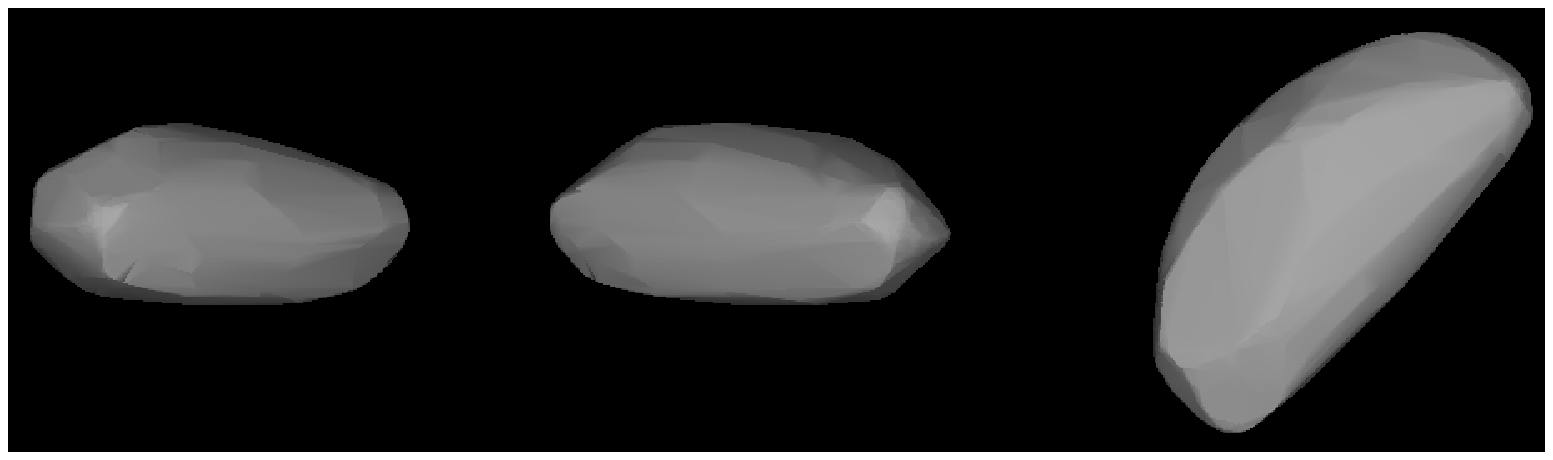}}\,\resizebox{0.495\hsize}{!}{\includegraphics{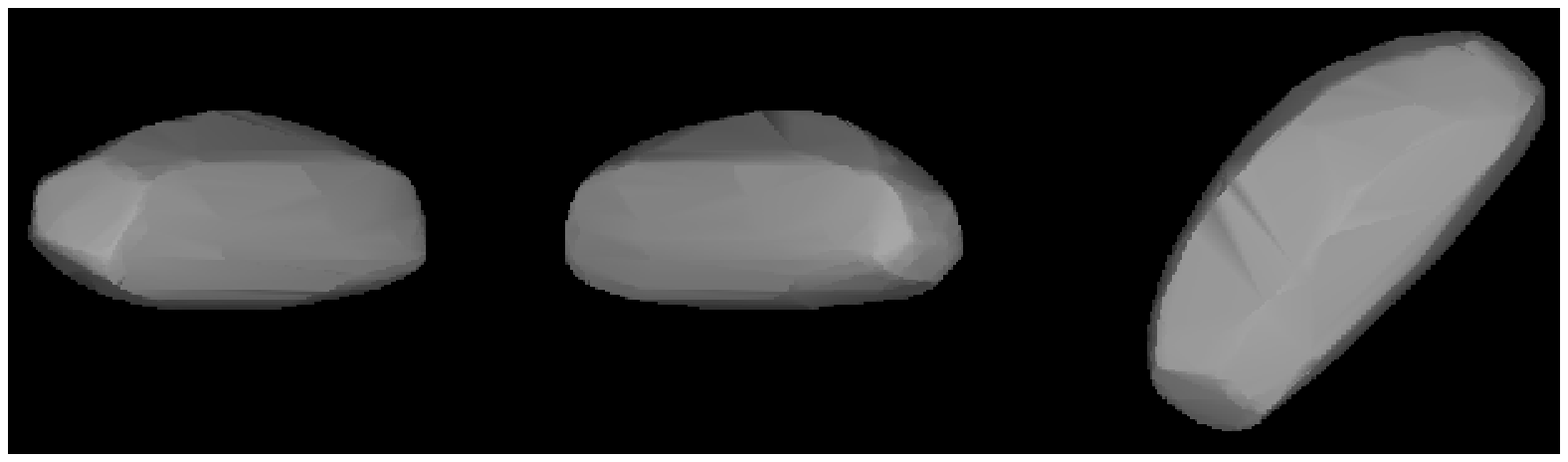}}\\
 \resizebox{0.48\hsize}{!}{\includegraphics{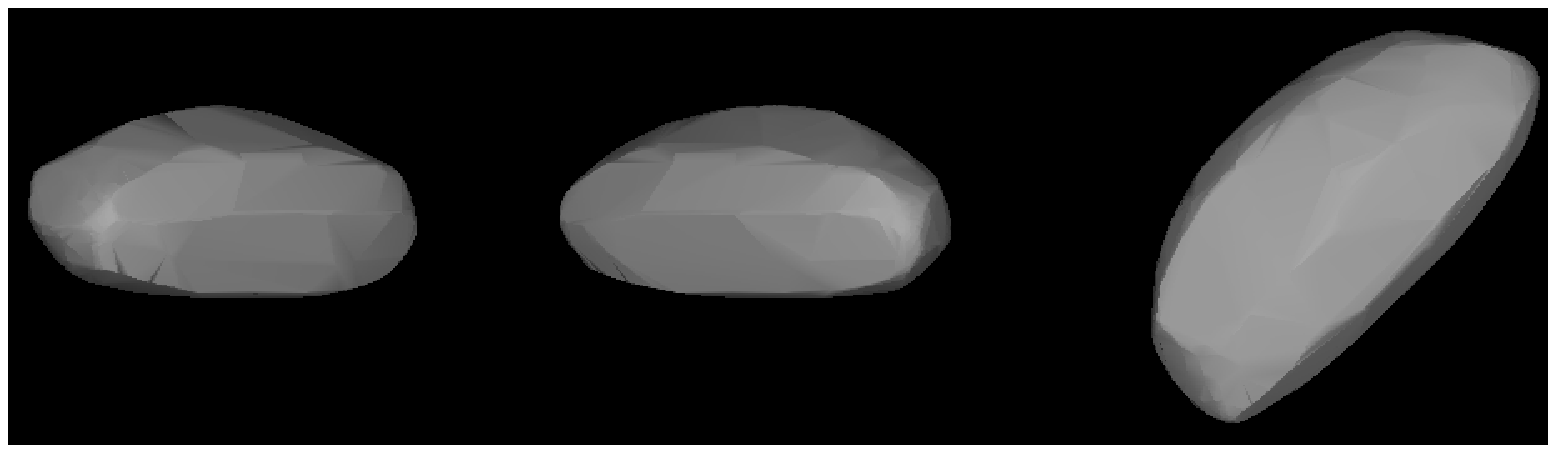}}\,\resizebox{0.495\hsize}{!}{\includegraphics{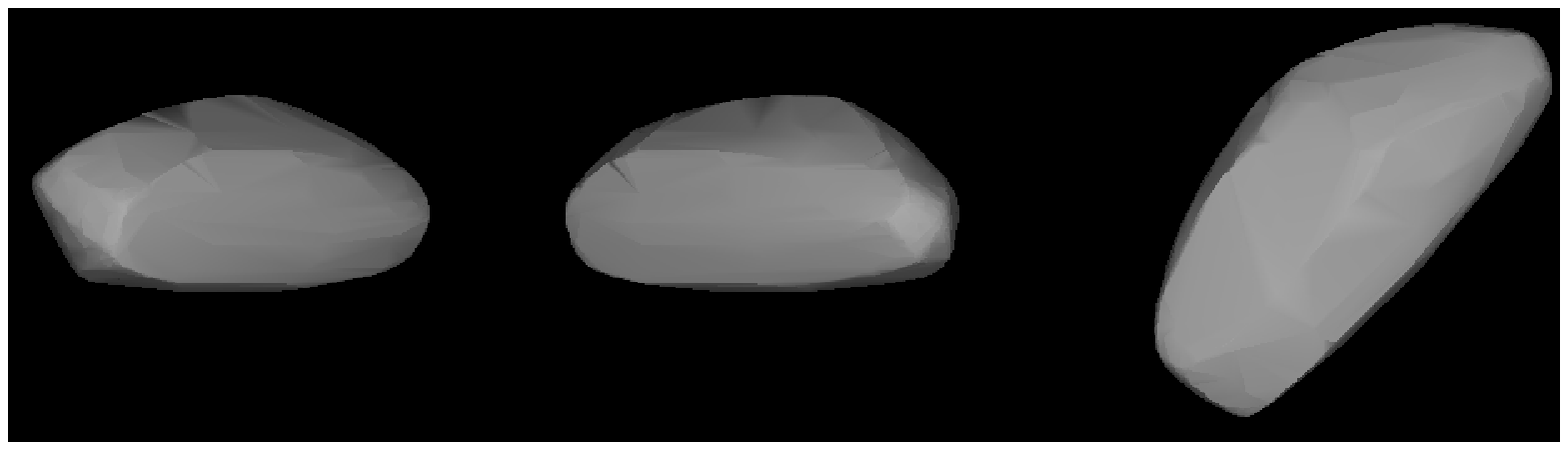}}\\
\end{center}
 \caption{\label{img:shapes_1627}Six shape models of asteroid (1627)~Ivar derived by the lightcurve inversion method by bootstrapping the optical photometric data sets. First two images are equator-on views rotated by 90$^{\circ}$ and the third one is a pole-on view.}
\end{figure}

It is known that the disk-integrated photometry is not significantly influenced by the non-convexities and shape features, and that the convex inversion is usually stable and produce similar results even if we reasonably vary the photometric data \citep[see e.g.,][]{Kaasalainen2001b}. As a result, most of the available shape models are convex. In the case of very precise thermal measurements (such as from the WISE satellite), the shape features together with the pole uncertainty could play a more significant role for the reliable determination of thermophysical properties, because they could considerably influence the temperature profile on the surface and thus the thermal infrared fluxes. To investigate this behavior, we introduce here a procedure that varies both the shape model and the pole orientation, while keeping the shape convex. Each varied shape model can be then used as input for the TPM modeling. We call this method {\em varied shape TPM (VS-TPM)}.

Instead of varying the shape models directly \footnote{For example by deforming the polyhedron, by introducing concavities, or by more sophisticated methods such as transforming the polyhedron into the Gaussian image, varying the facets and transforming it back, or by representing the shape by spherical harmonic series and varying the coefficients.}, we bootstrap the photometric data sets and use the convex inversion method to derive varied shape models. The advantage of this approach is that it is based only on the data and it also maps uncertainties in the pole direction. 

The whole procedure of investigating the stability of the TPM on the shape model variations can be divided into the following steps:
\begin{itemize}
 \item[1.] Bootstrapping of the photometric data. We randomly select a number of dense lightcurves from the original photometric data set equal to the number it contains, i.e., if the original data set contains ten individual lightcurves, we randomly select ten lightcurves from that data set. As a result, we can have some lightcurve multiple times in the new data set, and some of the lightcurves can be missing. This, in principle, corresponds to a different weighing of the lightcurves. For the sparse data, the modification procedure is similar: we randomly choose individual measurements from the pool of original data, while keeping the same number of observations in the whole sparse-in-time lightcurve. Data from different astrometric sources are treated separately. Generally speaking, we use the bootstrap method \citep{Press1986}.
 \item[2.] Shape model determination. We use the randomly selected photometric data set to derive the varied shape model by the lightcurve inversion. The original rotational state solution serves as an initial guess for the optimized parameters. The rotational state is usually close to the original one as can be seen in Fig.~\ref{img:shapes_poles}, where we show the orientations of the spin axes of varied shape models of all nine asteroids studied here. The difference  in the rotational state corresponds to the expected uncertainty, which is typically 5--10$^{\circ}$ in the pole orientation. The shape appearance is similar, as can be seen in Fig.~\ref{img:shapes_1627}, where we show several varied shape models of asteroid (1627)~Ivar based on modified photometric data sets.
 \item[3.] TPM modeling. We perform the TPM optimization scheme for each varied shape and its rotational state  the same way as for the original shape model.
 \item[4.] We repeat steps 1--3 to obtain a desired statistical sample. We run VS-TPM for each pole solution individually.
\end{itemize}

All the varied shape models fit the visible lightcurves equally well. Usually, the dense lightcurves are not weighed when the shape models are computed by the lightcurve inversion, but it is obvious that individual lightcurves do not have equal quality. On the other hand, the sparse data set is weighed with respect to the dense data to penalize its lower photometric accuracy. Our method not only naturally varies the lightcurve weights, but also selects subsets of the photometric data by ignoring some lightcurves, and thus produces sets of slightly different shape models whose rotational parameters sample the underlying uncertainties of the optical data.

\subsection{VS-TPM -- application to nine asteroids}\label{sec:TPM_stability}

\begin{figure}[!htbp]
\begin{center}
\resizebox{\hsize}{!}{\includegraphics{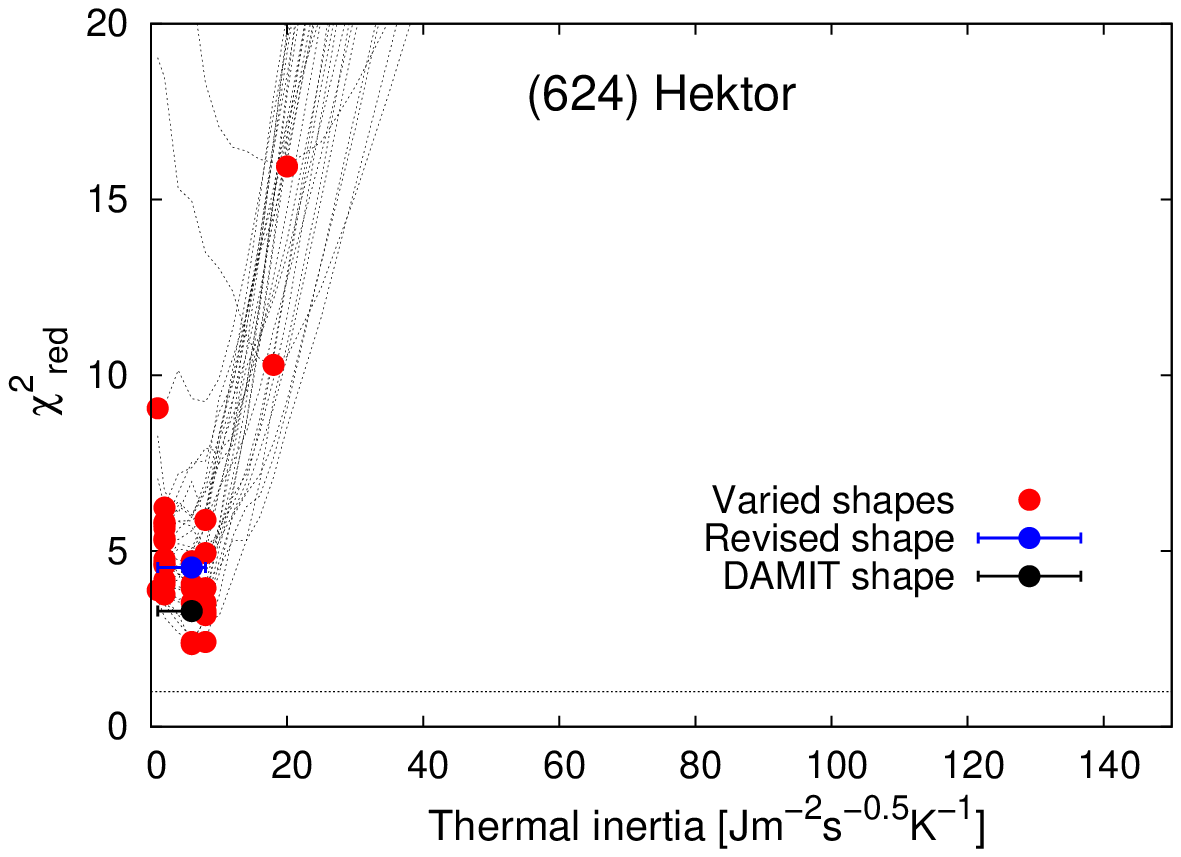}\includegraphics{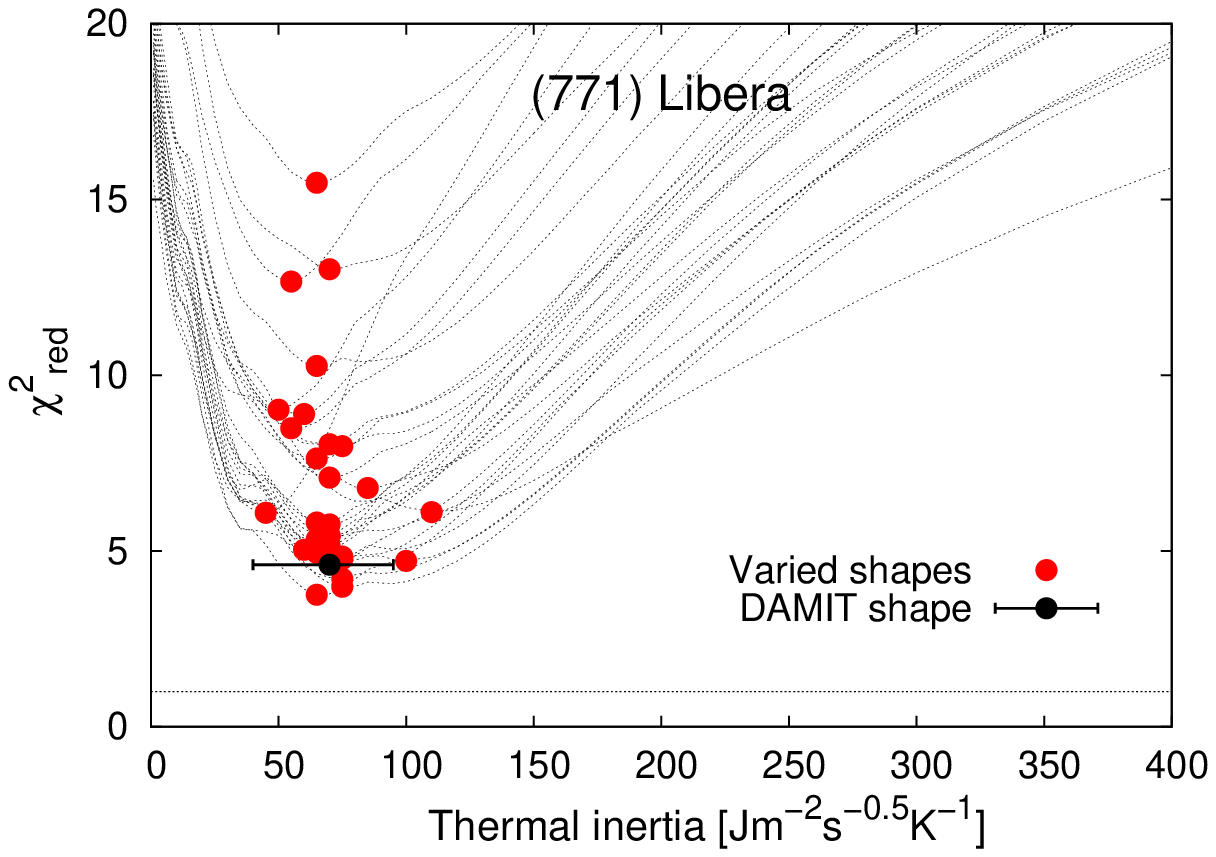}}\\
\resizebox{\hsize}{!}{\includegraphics{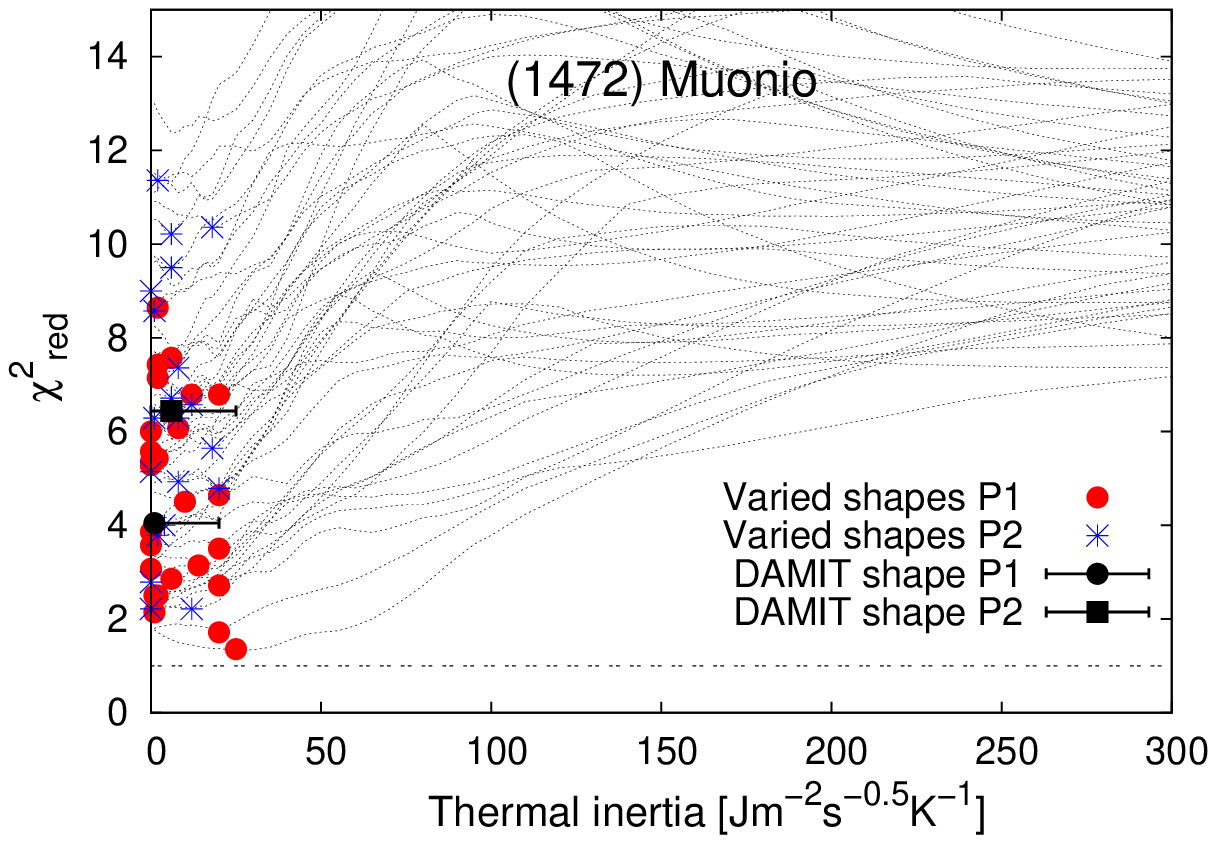}\includegraphics{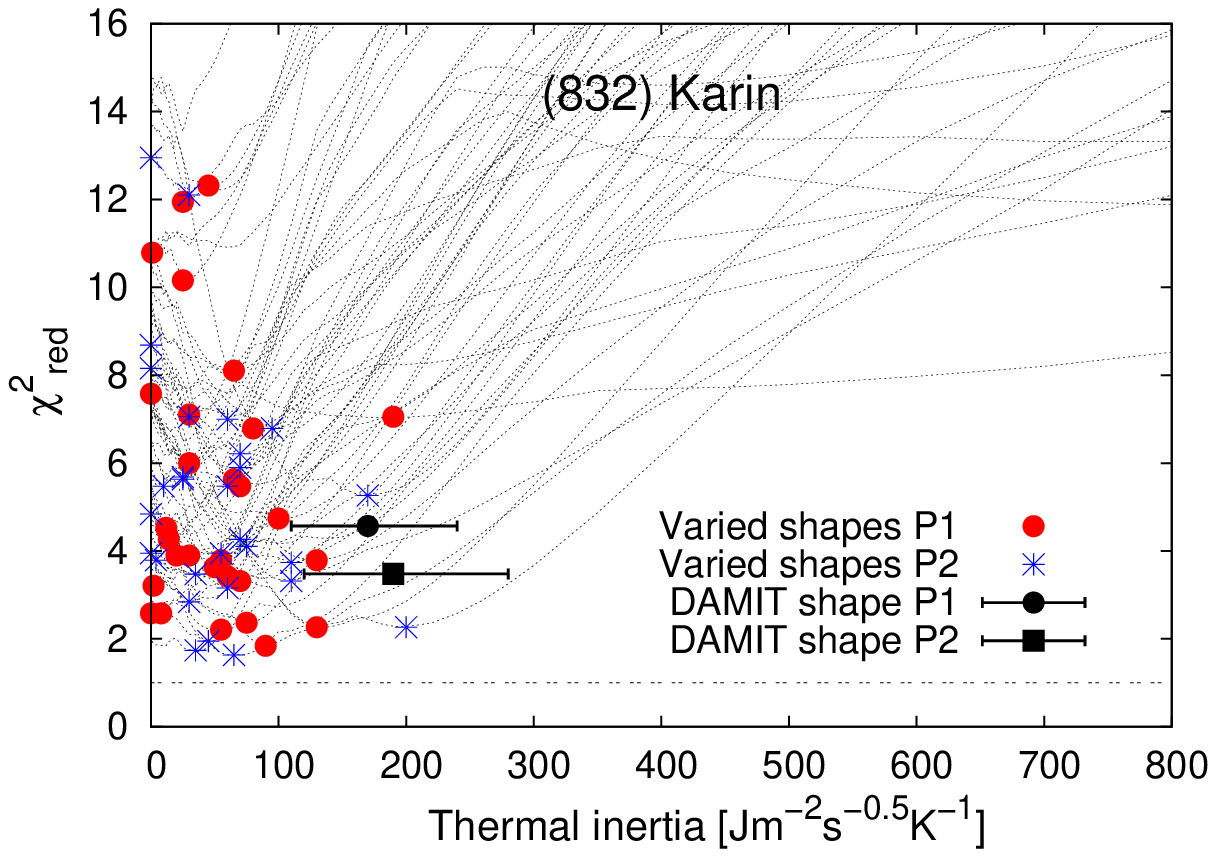}}\\
\resizebox{0.5\hsize}{!}{\includegraphics{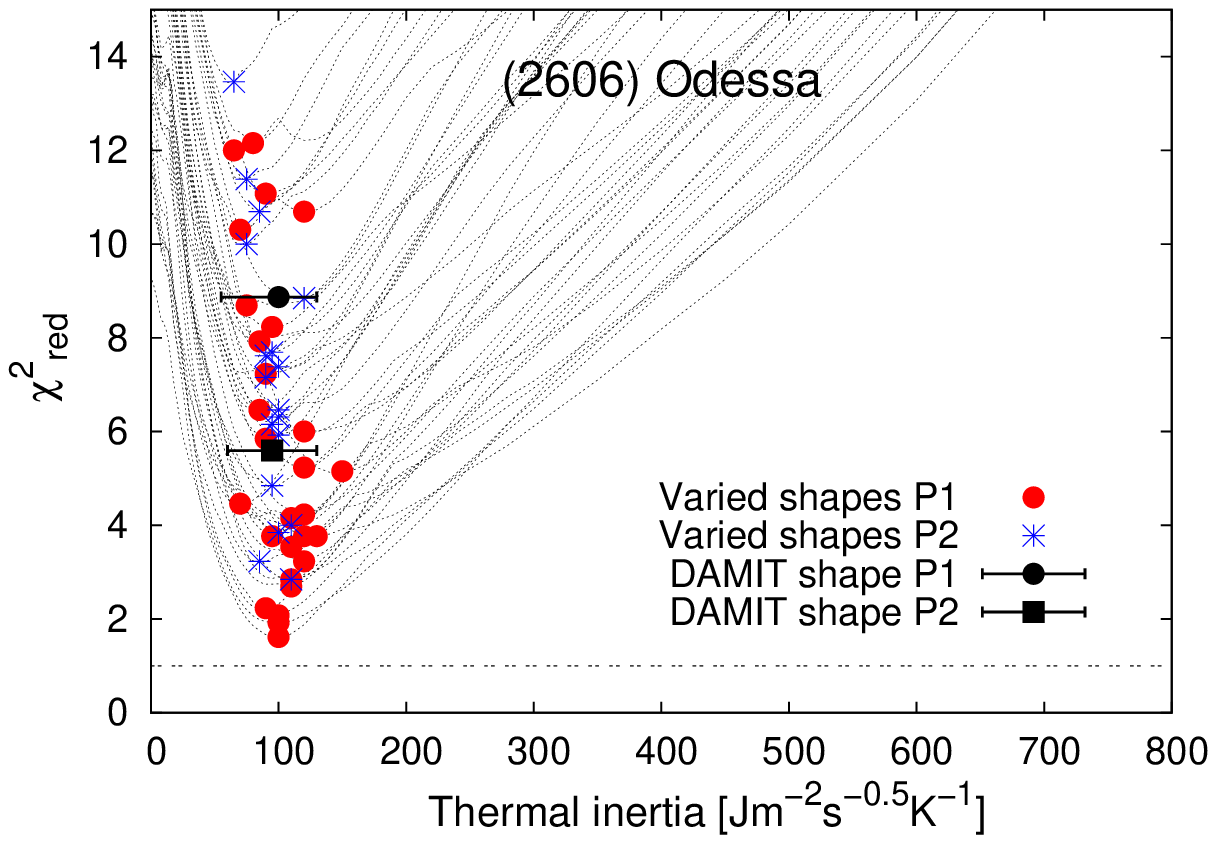}}\\
\end{center}
\caption{\label{img:MBAs_TI}Distribution of the best-fitted thermal inertia values (red full circles) determined for the varied shape models of the Jovian Trojan (624)~Hektor and main-belt asteroids (771)~Libera, (832)~Karin, (1472)~Muonio and (2606)~Odessa, together with the convergence in the thermal inertia (black lines). The second ambiguous pole solution is represented by blue asterisks. We also show the thermal inertia solutions with the corresponding error bars derived by the classical TPM for both revised (blue) and the DAMIT (black) shape models. The dashed horizontal line indicates $\chi^2_{\mathrm{red}}=1$. Note the different scales for the thermal inertia. A color version of the figure is available in the electronic version of the journal.}
\end{figure}

\begin{figure*}[!htbp]
\begin{center}
\resizebox{\hsize}{!}{\includegraphics{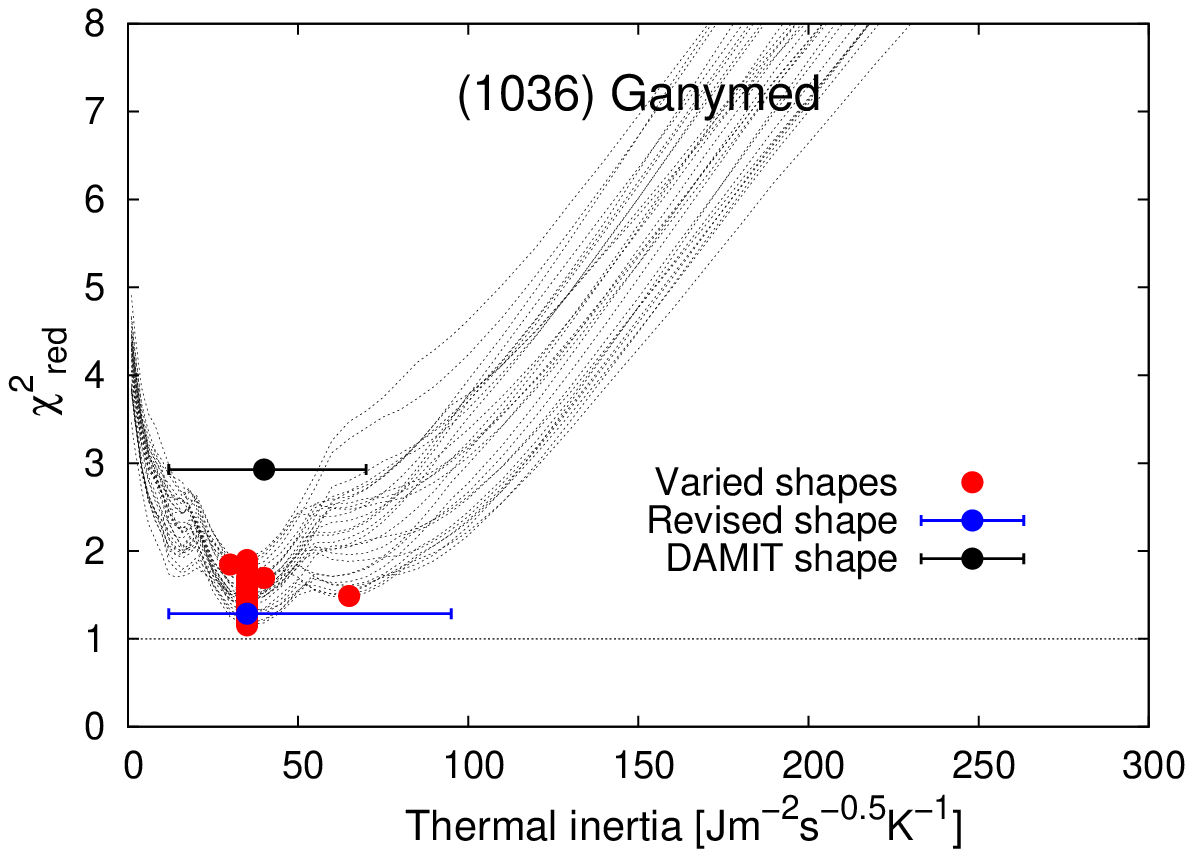}\includegraphics{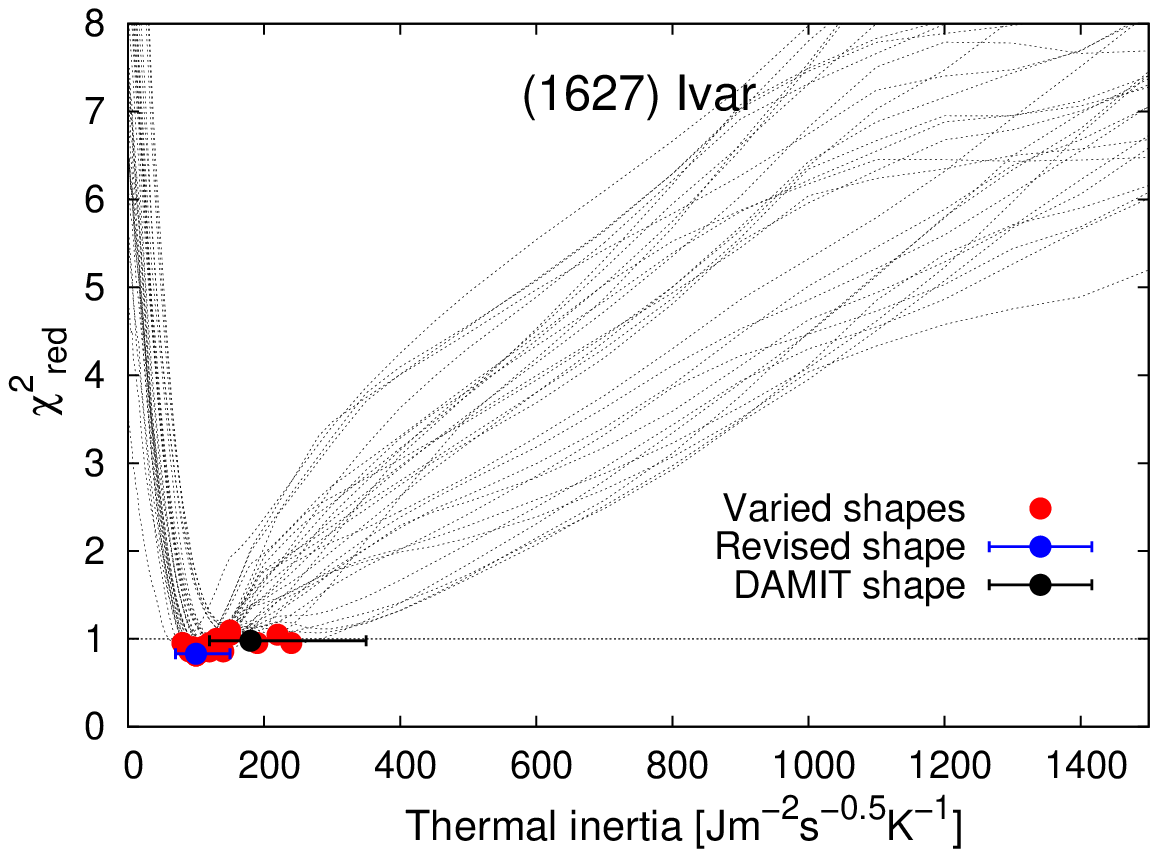}}\\
\resizebox{\hsize}{!}{\includegraphics{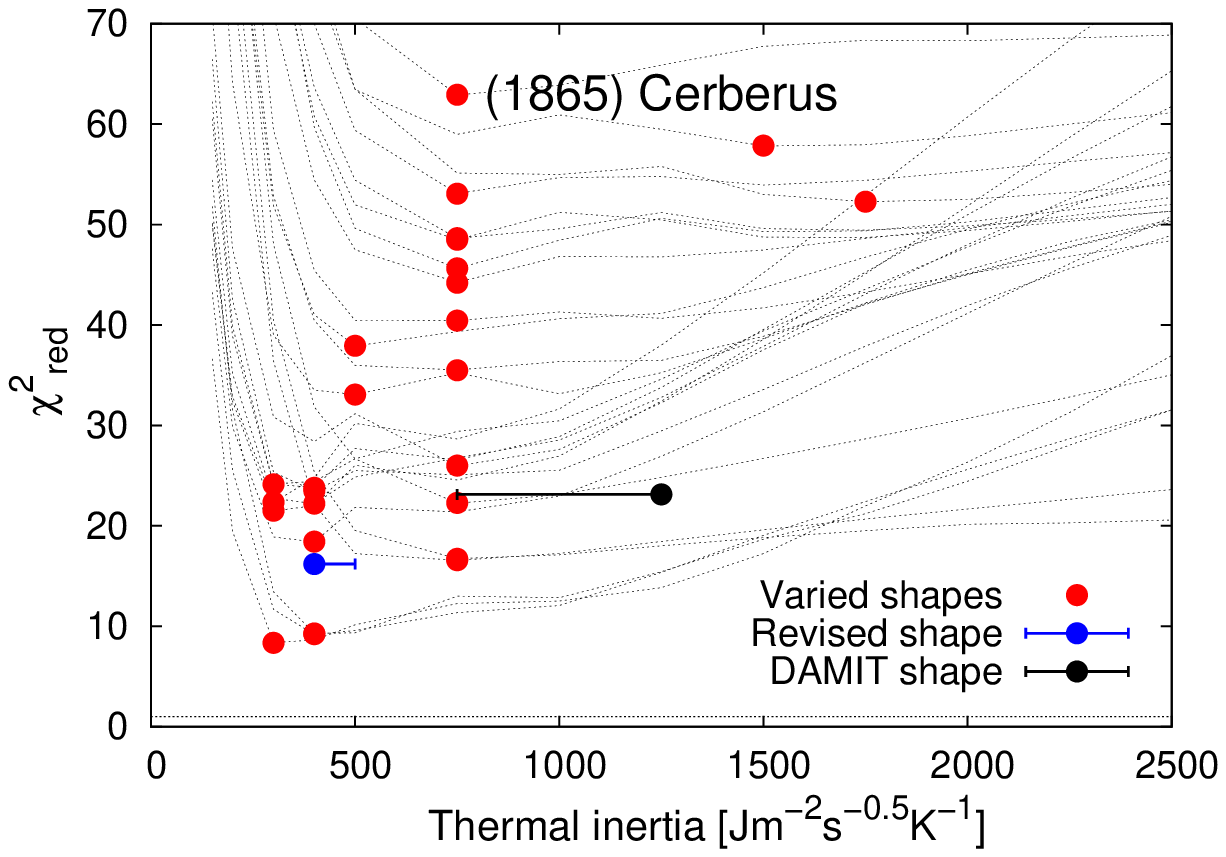}\includegraphics{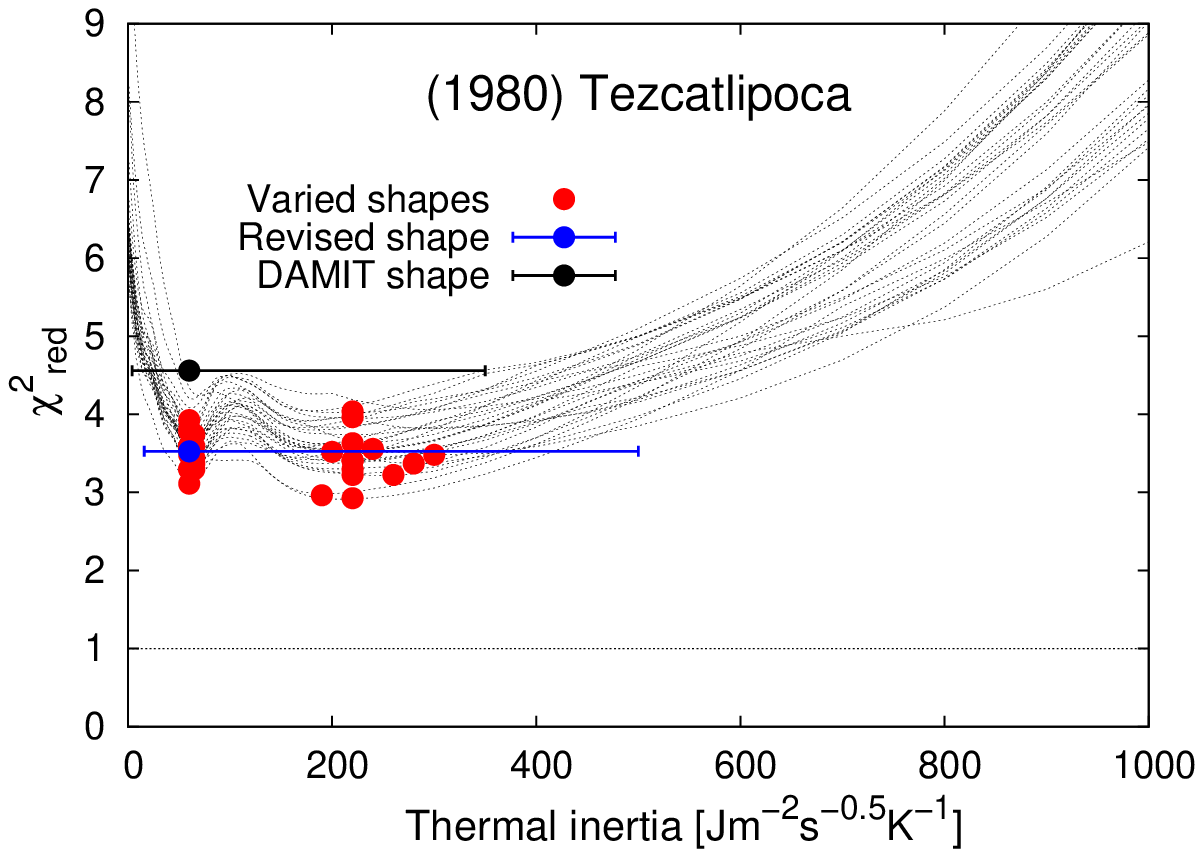}}\\
\end{center}
\caption{\label{img:NEAs_TI}Distribution of the best-fitted thermal inertia values (red full circles) determined for the varied shape models of four near-Earth asteroids (1036)~Ganymed, (1627)~Ivar, (1865)~Cerberus and (1980)~Tezcatlipoca, together with the convergence in the thermal inertia (black lines). We also show the thermal inertia solutions with the corresponding error bars derived by the classical TPM for both the revised (blue) and the DAMIT (black) shape models. The dashed horizontal line indicates $\chi^2_{\mathrm{red}}=1$. Note the different scales for the thermal inertia. A color version of the figure is available in the electronic version of the journal.}
\end{figure*}

In Sects.~\ref{sec:TPM_classical}~and~\ref{sec:revised_models}, we already illustrate for several asteroids that the accuracy of the shape and its rotational state should be considered in the TPM modeling. In particular, in the case of (1865)~Cerberus, the classical TPM approach leads to slightly different fits when the revised or the DAMIT shapes are used, indicating that the solution does not appear stable against variations of the shape and the pole orientation. Additionally the high $\chi^2_{\mathrm{red}}$ values of the TPM fits for asteroids (1472)~Muonio, (1865)~Cerberus, (1980)~Tezcatlipoca and (2606)~Odessa indicate that the current shape models are not able to reproduce carefully the observed thermal fluxes. We thus apply our novel method VS-TPM to investigate the stability of the TPM solution against the shape and pole uncertainties and to estimate the contribution of the shape model uncertainty to the $\chi^2_{\mathrm{red}}$ for all nine studied asteroids.

%In the following, we focus on the stability of the TPM solution with respect to only variations of the revised shape models. %The comparison between the results of the TPM with the original models (i.e., based on the non-varied photometric data sets) and the varied shape models is reasonable only for the revised shape models, where we have control over the model determination scheme. 

We create for each studied asteroid a sample of 29 varied shape models and their corresponding rotational states following the scheme described in Sect.~\ref{sec:varied_shape_TPM_method}. Together with the original shape model, we have a statistical sample of 30 slightly different shape models for each asteroid. In Fig.~\ref{img:shapes_poles}, we show the distribution of spin vectors of the varied shape models for all nine studied asteroids. They are consistent within their expected uncertainties. The typical dispersion is $\sim$10--20$^{\circ}$. 

Thereafter, we run the TPM for each varied shape model and pole solution and show the resulting thermophysical parameters in Tab~\ref{tab:TI}. In Figs.~\ref{img:MBAs_TI} and ~\ref{img:NEAs_TI}, we plot the dependence of the $\chi^2_{\mathrm{red}}$ versus thermal inertias (gray dashed lines) for all varied shape models and highlight each minimum: red full circles or blue asterisks, corresponding to different ambiguous pole solutions. We also include the solutions from Sect.~\ref{sec:TPM_classical} with their uncertainties for both revised (if available) and DAMIT shape models.

The $\chi^2_{\mathrm{red}}$ of the solutions of the varied shape TPM reach values $\sim$1 in five out of nine cases, which shows that we are reliably fitting the thermal infrared data and that the VS-TPM supersedes the classical approach. This is because varied shape models, even though they are indistinguishable in terms of reproducing the optical photometry, are clearly different in the thermal infrared. Note that this method is not an optimization of the shape, rather it is a way to map the uncertainties in the TPM analyses related to the uncertainties of the shape and the pole orientation given by the optical photometry. A technique that truly optimizes the shape (still as convex), the rotational state and the thermophysical properties simultaneously from optical and the thermal infrared photometry is under development \citep[multiple source data inversion method,][]{Durech2014}.

We observe various behaviors for the distribution of the best TPM solutions in our sample of asteroids. In the cases of asteroids (624)~Hektor, (1036)~Ganymed and (1627)~Ivar, the chi-square curves and their corresponding minima present qualitatively similar behavior and lead to statistically indistinguishable solutions. This suggests that the TPM results are not strongly dependent on the varied shape models, likely because the optical photometric data sets for these asteroids are particularly large and, therefore, their shapes are well constrained. Not surprisingly, the range of acceptable thermal inertia values is larger than suggested by the classical TPM analysis.

The most intriguing result is the fact that we also obtain TPM fits with a large range of $\chi^2_{\mathrm{red}}$ values for most of the studied asteroids. This suggests that TPM modeling with individual varied shape models differ considerably, and sometimes the best fitting solutions have $\chi^2_{\mathrm{red}}$ values that differ by a factor of 5 while keeping thermal inertia determinations consistent, as can be seen in Fig.~\ref{img:MBAs_TI} for asteroids (771)~Libera or (1472)~Muonio. The most extreme case is asteroid (2606)~Odessa, with $\chi^2_{\mathrm{red}}$ between 2 and 12. Contrary to the classical TPM approach, the application of the VS-TPM has enabled us to obtain satisfactory fits (with $\chi^2_{\mathrm{red}}\sim1$) for asteroids (1472)~Muonio and (2606)~Odessa. 

On the other hand, the best thermal inertia solutions within the varied shape models of asteroid (832)~Karin range from zero to two hundred, some of which are not consistent with the values determined by the classical approach. This is a warning that fixing a single shape model with the classical TPM and neglecting the uncertainties in the shape and the pole orientation can strongly bias the solution. In turn, the VS-TPM cannot fit the thermal infrared data of asteroid (1865)~Cerberus. The distribution of the best fitting thermal inertia solutions (red full circles in Fig.~\ref{img:NEAs_TI}) as well as their corresponding $\chi^2_{\mathrm{red}}$ for (1865)~Cerberus span the largest range of values. This shows a very strong dependence on the shape model and rotational state, probably enhanced by the high phase angle of the observations ($\alpha\sim90$) and the extreme elongation of the shape.

%\textbf{The thermal inertia of (771)~Libera and (1980)~Tezcatlipoca shows no sensitivitity to the uncertainty of the shape model, are $\sim$3--4, the overall stability of the TPM solution gives us confidence that the thermophysical parameters (although not constrained much for Tezcatlipoca) are reliable.}

To sum up, derived thermal inertia values of asteroids (624)~Hektor, (771)~Libera, (1036)~Ganymed, (1472)~Muonio, (1627)~Ivar and (2606)~Odessa seem to be well constrained, contrary to those of (832)~Karin and (1980)~Tezcatlipoca. The TPM solution for (1865)~Cerberus is still poor. Contrary to the classical TPM, the VS-TPM approach does not always help reject one of the ambiguous pole solutions. In particular, there is not a preferred solution for asteroids (832)~Karin, (1472)~Muonio and (2606)~Odessa.

%To estimate thermophysical properties and their corresponding uncertainties based on the VS-TPM, we consider the confidence interval given by the 21 ($\sim68\%$) TPM solutions with the lowest $\chi^2_{\mathrm{red}}$. For each case, we calculate the uncertainties of the parameters by the standard 1$\sigma$ approach if $\chi^2_{\mathrm{red}}\sim1$, or we consider in other cases the range of solutions with $\chi^2_{\mathrm{red}}<1+\chi^2_{\mathrm{min}}$. By finding the maximum and minimum values allowed by the individual uncertainties within these 21 cases, we obtain the final uncertainties quoted in Tab.~\ref{tab:TI}. Note that the solution for (1865)~Cerberus is rather poor, thus we do not report the uncertainties, however, we give a lower limit for the thermal inertia.

To estimate thermophysical properties and their corresponding uncertainties based on the VS-TPM, we consider the confidence interval given by the 21 ($\sim68\%$) TPM solutions with the lowest best-fitting $\chi^2_{\mathrm{red}}$. This criterion allows us to exclude the most extreme bootstrapped shapes from the analysis, and, simultaneously, to account for the TPM parameter ranges spanned by the varied shape models. We calculate the uncertainties of the fitted parameters by the empirical approach the same way as in Sect.~\ref{sec:TPM_classical} for each single varied shape: $\chi^2_{\mathrm{red}}<\chi^2_{\mathrm{min}}*(1+\sqrt{2\nu}/\nu)$. By finding the maximum and minimum values allowed by the individual uncertainties within these 21 cases, we obtain the final uncertainties quoted in Tab.~\ref{tab:TI}. Note that the TPM fit for (1865)~Cerberus is poor, thus we do not report the uncertainties; however, we are still able to estimate a lower limit of the thermal inertia.

%Although the thermal infrared data sets are significantly smaller compared to the optical data sets, we also apply the bootstrap method to them. In general, the TPM solutions for varied shape models and bootstrapped thermal data are consistent with the VS-TPM solutions with the original thermal data sets. However, as expected, we end up with less constrained thermophysical properties because of the significant loss of information associated with the reduction of the number of data points. We conclude that due to the low amount of thermal data, their bootstrapping is not as robust as in the case of optical lightcurves and thus we do not discuss it further.

\subsection{Individual cases}\label{sec:TPM_idividual}

\paragraph{(624) Hektor}

The low thermal inertia of this D-type asteroids is not surprising, it is slightly smaller than the one of another Trojan asteroid (617)~Patroclus \citep[20$\pm$15 J\,m$^{-2}$\,s$^{-1/2}$\,K$^{-1}$,][]{Muller2010}, or other Jovian Trojans \citep[62$\pm$37 J\,m$^{-2}$\,s$^{-1/2}$\,K$^{-1}$ for (1173)~Anchises, 7$\pm$7 J\,m$^{-2}$\,s$^{-1/2}$\,K$^{-1}$ for (2363)~Cebriones, and 15$\pm$15 J\,m$^{-2}$\,s$^{-1/2}$\,K$^{-1}$ for (3063)~Makhaon,][]{Horner2012,Fernandez2003}. 

Note that Hektor is a binary minor body consisting of a primary component and a moonlet with a size ratio of XXX. Thus, infrared fluxes are dominated by the thermal emission of the primary and the moonlet can be neglected in the TPM analysis.  

Our volume equivalent diameter of $D=186^{+1}_{-34}$~km is 1$\sigma$ larger than the diameter reported by \cite{Grav2012b} based on the NEATM, but lower than the diameter based on the AKARI data \citep[230~km, based on a standard thermal model,][]{Usui2011}. Moreover, \citet{Marchis2014} used their own convex shape model to fit the contour profiles of adaptive optics images acquired by Keck telescope and derived a volume-equivalent diameter of $D=270\pm22$~km. We also perform the TPM analysis of the WISE data with the convex shape model of \citet{Marchis2014} as input and still obtain a significantly smaller size. 
\cite{Marchis2014} give some evidence that the primary component of (624)~Hektor has a bi-lobed shape. If this is the case, as the convex shape model usually corresponds to the convex hull of the real shape, the diameter estimated from TPM could be, in principle, both over- or underestimated depending on the observing/viewing geometry. In our particular case, the observing configuration during the (almost) maximum elongation is such that the projected area of the convex model would be bigger and flatter than that of the bi-lobed shape model. Given the low phase angle of the observations the model temperatures of the convex shape are expected to be higher than those of a non convex shape. Consequently, the best fitting diameter becomes smaller to compensate for the higher temperatures on the visible hemisphere. While we caution that more accurate adaptive optics images, and/or occultation data are likely necessary to investigate the shape of this body, we point out that the value of  $p_\mathrm{V}=0.058^{+0.017}_{-0.007}$ corresponding to $D\sim190$ km is in very good agreement with the average albedo of large trojans of 0.07$\pm$0.03 reported by \citet{Grav2011}. On the other hand, the value of $p_\mathrm{V}=0.029$, corresponding to $D\sim270$ km is significantly lower.

Jovian Trojan (624)~Hektor is another asteroid with an emission band at $\sim10$ $\mu$m and low thermal inertia \citep[such as (24)~Themis, (1)~Ceres, or (617)~Patroclus,][]{Vernazza2012}. Such low thermal inertia could be consistent with a very fine and porous surface material (fluffy material). 

\paragraph{(1036) Ganymed}

The photometric data set for this asteroid is exceptionally large and the TPM produces thermal inertia of $\Gamma=35^{+65}_{-29}$, which is comparable to the value for the Moon's fine regolith \citep{Spencer1989}. Such low value of thermal inertia is not common among NEAs (indeed, Ganymed is a NEAs with the second lowest $\Gamma$ so far), for which most determinations are of the order of some hundreds J\,m$^{-2}$\,s$^{-1/2}$\,K$^{-1}$ \citep[see, e.g.,][]{Delbo2007a}.

However, we note that the thermal conductivity in the regolith is temperature dependent \citep{Keihm1984}, and so is thermal inertia. Under the assumption that heat is transported in the regolith mainly by radiative conduction between grains, the thermal conductivity is proportional to $T^3$, with $T$ being the temperature of the regolith grains \citep{Kuhrt1989, Jakosky1986}. In this case $\Gamma \propto T^{3/2}$ and, because $\Gamma \propto \sqrt{\kappa}$ then $\Gamma \propto r^{-3/4}$, where $r$ is the heliocentric distance of the body. As Ganymed was observed at $r\sim$ 3.7 AU we expect that its thermal inertia value at 1 AU would be around three times higher, namely $\sim$100-150 J\,m$^{-2}$\,s$^{-1/2}$\,K$^{-1}$. Accurate modeling of the heat transfer in the regolith \citep{Gundlach2013} shows that the heat conduction between touching regolith grains is also important and the temperature dependence of the conductivity might depart from the pure radiative term. Using the model of \citet{Gundlach2013} and assuming S-type like thermal properties we can estimate that the thermal inertia can increase by a factor between 1.3 and 2.7 as Ganymed moves from 3.7 to 1 AU depending on the average grain size and regolith packing fraction. We conclude that the thermal inertia value at 1 AU of Ganymed is between 45 and 100 J\,m$^{-2}$\,s$^{-1/2}$\,K$^{-1}$, which is still among the lowest values measured for NEAs \citep[the lowest value so far is $\Gamma=36^{30}_{-20}$ J\,m$^{-2}$\,s$^{-1/2}$\,K$^{-1}$ for asteroid 1950 DA,][]{Rozitis2014b}. 

Another explanation for the low thermal inertia could be the unusually large size of this NEA: $D=37^{+2}_{-4}$~km. This is significantly larger than the other NEAs with thermal inertia determinations (usually of the order of 1 km). This is also consistent with the size vs. thermal inertia dependence proposed by \citet{Delbo2007a}. 

The lower-than-average value for the thermal inertia of Ganymed is likely due to the presence of a finer than average surface regolith, implying that this asteroid has a surface that has been exposed longer than the average NEA to the micrometeorite bombardment, which is one of the mechanisms that has been claimed to be capable of comminuting rocks on asteroids \citep{Horz1997}. A more convincing explanation is thermal cracking: it has been recently shown that thermal fragmentation induced by the diurnal temperature variations breaks up rocks larger than a few centimeters more quickly than do micrometeoroid impacts on NEAs \citep{Delbo2014}. Because the process of thermal fragmentation is strongly dependent on the value of of the diurnal temperature excursion, which increases with decreasing perihelion distance, we expect NEAs with low perihelion distances to have finer regolith and therefore lower thermal inertias. We note that this is the case for Ganymed, which is one of the most heated (``hot'') NEAs according to \citet{Marchi2009} (see their Fig~4). Analysis of the orbital history of Ganymed \citep{Marchi2009} shows that this asteroid has 98\% probability to have had a perihelion distance smaller or equal to 0.1 AU (!) in the past for at least 4,000 years, which implies that thermal cracking was very efficient for this asteroid. 

The geometric visible albedo is $p_\mathrm{V}=0.25^{+0.05}_{-0.03}$, which is a typical value for an S-type asteroid. Based on the value of $\Gamma$, we expect that the regolith grain sizes are comparable to those on the Moon.

\section{Discussion}\label{sec:discussion}

%\textbf{The most challenging issue encountered in our study is to reliably estimate the errors of derived thermophysical parameters.} DISSCUSSION OF THE ERRORS. EFFECTS OF SYSTEMATICS.

%are dependent on the errors of the data we use. Even though we increase the nominal errors of the infrared fluxes according to systematics found by \citet{Jarrett2011}, these values have their own uncertainties. We also neglect the systematic errors of our TPM model, for example, those errors related to the particular choice of the roughness model or the assumption of the Lambertian emission. On the other hand, the model errors of typically few percent should be always lower than the errors of the fluxes, which are $\gtrsim5\%$ considering the \citet{Jarrett2011} systematics. 

Asteroids shape models are determined from optical lightcurves that usually cover only several apparitions, and therefore a limited number of observing geometries. The shape model then predicts well the lightcurves for the apparitions covered by the observations but could be less accurate for other apparitions, especially when the geometry of observations is significantly different. In such cases, parts of the surface that are exposed to the observer could not be realistic. If the shape model is not based on dense-in-time photometry from the same apparition as the WISE data were observed in, we should expect that the computed fluxes are less accurate/realistic.

Most TPM works so far have relied on convex shape models for several reasons: 
(i)~the majority of shape model determinations are convex, 
(ii)~the high quality optical photometric data (i.e., with high photometric accuracy and/or acquired at high phase angles) necessary for a non-convex shape model determination is available for only very few asteroids, moreover, there are usually no WISE thermal infrared data for them, because filters W3 and W4 saturate for big asteroids, 
(iii)~radar observations, which are typically used for non-convex shape model determinations, are limited to only few largest and closest MBAs or close flybys of small NEAs, or 
(iv)~the convex shape model usually fits the optical data to the level of the noise, thus optimization with a non-convex model does not provide meaningful results because the disk-integrated photometry usually contains little information about the non-convexities.

On the other hand, the convex approximation could be sufficient because convex and non-convex models of asteroid (1620)~Geographos give consistent thermophysical solutions \citep{Rozitis2014}, and thermophysical properties of asteroid (21)~Lutetia based on a convex shape model \citep{Carry2012} are consistent with the findings based on the Rosetta/VIRTIS data \citep{Coradini2011}. It was also shown in \citet{Marchis2006,Hanus2013b} that the disk-resolved images acquired by the 8--10m class telescopes equipped with AO systems usually well correspond to the asteroid's 2D shape projections, thus the shape models globally well represent the real asteroid's appearance. The fact, that in some cases we are not able to reproduce well the observed fluxes by the TPM with such convex shape models (i.e., with those that are reliable global representations of the true shapes), or that individual varied shapes give significantly different $\chi^2_{\mathrm{red}}$ values of the TPM fits, provides the evidence that the observed thermal emission is indeed more sensitive than optical observations to topographic features and concavities that do not significantly alter the global shape. 

Our results show that the shape model (together with the pole orientation) is a significant limiting factor for the goodness of the TPM fit. We expect this effect to be more prominent when the shape model is based mainly on sparse-in-time opticaL data. We also show the stability of the values of the physical parameters derived from TPM with respect to changes to the shape model and the pole orientation. This raises our confidence that thermophysical properties are in general reliable.

The VS-TPM also shows that the quoted errors based on the classical TPM approach, which account only for statistical uncertainties of the observed infrared fluxes, should be interpreted as minimum estimates of the real uncertainties. The VS-TPM, which accounts for the uncertainties in the shape and pole orientation of the asteroid, is a very viable method to estimate realistic error bars of the parameter values. However, at the moment and for the foreseeable future, our shapes are and will be based on optical lightcurve inversion, and thus remain convex. Probably, $\chi^2$ values would be lowered if it was possible to derive a non-convex shape model from lightcurve inversion.

Finally, a model capable of optimizing photometric and thermal infrared data simultaneously (multi-data inversion technique) should be a promising step forward towards the improvement of (still convex) asteroid shapes using thermal infrared data together with optical observations \citep{Durech2014}.

%The determination of reliable thermophysical properties benefits from the availability of the high quality shape models. As the visible photometric data rarely allow the derivation of a non-convex shape model, the majority of available shape models will remain convex. However, additional photometric data for the improvement of the shape models (there are shape models based on small photometric data sets consisting mostly of sparse photometric measurements, such shape models are rather coarse) will be necessary in order to derive reliable thermophysical fits (stable with respect to the shape model and the pole uncertainties).

\section{Conclusions}\label{sec:conclusions}

We present a novel method to investigate the importance of the shape model and the pole orientation uncertainties in the thermophysical modeling -- the varied shape TPM (VS-TPM).

We apply the VS-TPM to nine asteroids and reveal the strong dependence of the TPM fit on the shape model and the pole orientation uncertainty for several asteroids. The best-fitting parameters are presented in Tab.~\ref{tab:TI} and discussed in Sect.~\ref{sec:TPM_stability}. From this table, one can see that the uncertainties of these properties derived by the classical TPM method are usually underestimated. In most cases, the best-fitting values of the physical parameter derived from the TPM analysis are consistent between the classical TPM and the VS-TPM. However, there are exemples where VS-TPM shows that neglecting the uncertainties in the shape model and the rotational state can result in biased values of these parameters (e.g., the thermal inertia for the asteroid (832)~Karin).

Furthermore, the VS-TPM allows us to find a significantly better TPM solution than the classical approach (e.g, (1472)~Muonio and (2606)~Odessa). 

Additionally, the obvious differences between the TPM fits found for some of the ambiguous pole solutions based on the standard approach are not that prominent after applying the VS-TPM. This suggests that one should be very cautious when making conclusions based on the classical results.

Based on our findings, we recommend to always consider the uncertainties of the shape model and its pole orientation in the thermophysical modeling. 

%\textbf{Derived thermophysical properties could be significantly biased due to a possible systematic offset in the zero magnitudes of the thermal data reported by \citet{Jarrett2011}. However, this offset is, in principle, ``random'' for each asteroid and we cannot control it. This offset is irrelevant for the our study of the shape model importance for TPM.}

With the tools developed in this work, we are now ready to exploit the WISE catalog and determine thermophysical properties, especially thermal inertias, for hundreds of asteroids with convex shape models.

\section*{Acknowledgements}
The work of JH, MD and VAL  was carried under the contract 11-BS56-008 (SHOCKS) of the French Agence National de la Recherche (ANR), and JD has been supported by grants GACR P209/10/0537  and 15-04816S of the Czech Science Foundation. VAL acknowledges support from the project AYA2011-29489-C03-02 (MEC, former Spanish Ministry of Education and Science). The computations have been done on the ``Mesocentre'' computers, hosted by the Observatoire de la C\^{o}te d'Azur, and on the computational cluster Tiger at the Astronomical Institute of Charles University in Prague (\texttt{http://sirrah.troja.mff.cuni.cz/tiger}).

This publication uses data products from NEOWISE, a project of the Jet Propulsion Laboratory/California Institute of Technology, funded by the Planetary Science Division of the NASA. We made use of the NASA/IPAC Infrared Science Archive, which is operated by the Jet Propulsion Laboratory, California Institute of Technology, under contract with the NASA.

\section*{References}
\bibliography{mybib}

\begin{thebibliography}{102}
\expandafter\ifx\csname natexlab\endcsname\relax\def\natexlab#1{#1}\fi

\bibitem[{{Al{\'i}-Lagoa} {et~al.}(2014){Al{\'i}-Lagoa}, {Lionni}, {Delbo},
  {Gundlach}, {Blum}, \& {Licandro}}]{AliLagoa2014}
{Al{\'i}-Lagoa}, V., {Lionni}, L., {Delbo}, M., {et~al.} 2014, \aap, 561, A45

\bibitem[{{Binzel}(1987)}]{Binzel1987a}
{Binzel}, R.~P. 1987, \icarus, 72, 135

\bibitem[{{Bottke} {et~al.}(2006){Bottke}, {Vokrouhlick{\'y}}, {Rubincam}, \&
  {Nesvorn{\'y}}}]{Bottke2006}
{Bottke}, J. W.~F., {Vokrouhlick{\'y}}, D., {Rubincam}, D.~P., \&
  {Nesvorn{\'y}}, D. 2006, Annual Review of Earth and Planetary Sciences, 34,
  157

\bibitem[{{Bowell} {et~al.}(1989){Bowell}, {Hapke}, {Domingue}, {Lumme},
  {Peltoniemi}, \& {Harris}}]{Bowell1989}
{Bowell}, E., {Hapke}, B., {Domingue}, D., {et~al.} 1989, in {Asteroids II},
  ed. R.~P. {Binzel}, T.~{Gehrels}, \& M.~S. {Matthews}, 524--556

\bibitem[{{Bus} \& {Binzel}(2002)}]{Bus2002}
{Bus}, S.~J. \& {Binzel}, R.~P. 2002, Icarus, 158, 146

\bibitem[{{Carry} {et~al.}(2012){Carry}, {Kaasalainen}, {Merline},
  {M{\"u}ller}, {Jorda}, {Drummond}, {Berthier}, {O'Rourke}, {{\v D}urech},
  {K{\"u}ppers}, {Conrad}, {Tamblyn}, {Dumas}, {Sierks}, \& {OSIRIS
  Team}}]{Carry2012}
{Carry}, B., {Kaasalainen}, M., {Merline}, W.~J., {et~al.} 2012, \planss, 66,
  200

\bibitem[{{Chernova} {et~al.}(1995){Chernova}, {Kiselev}, {Krugley},
  {Lupishko}, {Shevchenko}, {Velichko}, \& {Mohamed}}]{Chernova1995a}
{Chernova}, G.~P., {Kiselev}, N.~N., {Krugley}, Y.~N., {et~al.} 1995, \aj, 110,
  1875

\bibitem[{{Christensen} {et~al.}(2000){Christensen}, {Bandfield}, {Hamilton},
  {Howard}, {Lane}, {Piatek}, {Ruff}, \& {Stefanov}}]{Christensen2000}
{Christensen}, P.~R., {Bandfield}, J.~L., {Hamilton}, V.~E., {et~al.} 2000,
  \jgr, 105, 9735

\bibitem[{{Coradini} {et~al.}(2011){Coradini}, {Capaccioni}, {Erard}, {Arnold},
  {De Sanctis}, {Filacchione}, {Tosi}, {Barucci}, {Capria}, {Ammannito},
  {Grassi}, {Piccioni}, {Giuppi}, {Bellucci}, {Benkhoff}, {Bibring}, {Blanco},
  {Blecka}, {Bockelee-Morvan}, {Carraro}, {Carlson}, {Carsenty}, {Cerroni},
  {Colangeli}, {Combes}, {Combi}, {Crovisier}, {Drossart}, {Encrenaz},
  {Federico}, {Fink}, {Fonti}, {Giacomini}, {Ip}, {Jaumann}, {Kuehrt},
  {Langevin}, {Magni}, {McCord}, {Mennella}, {Mottola}, {Neukum}, {Orofino},
  {Palumbo}, {Schade}, {Schmitt}, {Taylor}, {Tiphene}, \&
  {Tozzi}}]{Coradini2011}
{Coradini}, A., {Capaccioni}, F., {Erard}, S., {et~al.} 2011, Science, 334, 492

\bibitem[{{Dahlgren} {et~al.}(1991){Dahlgren}, {Lagerkvist}, {Fitzsimmons}, \&
  {Williams}}]{Dahlgren1991}
{Dahlgren}, M., {Lagerkvist}, C.-I., {Fitzsimmons}, A., \& {Williams}, I.~P.
  1991, \mnras, 250, 115

\bibitem[{Delbo'(2004)}]{Delbo2004}
Delbo', M. 2004, PhD thesis - Freie Univesitaet Berlin, 1

\bibitem[{Delbo'(2014)}]{Delbo2014}
Delbo', M. 2014, Nature

\bibitem[{{Delbo'} {et~al.}(2007){Delbo'}, {dell'Oro}, {Harris}, {Mottola}, \&
  {Mueller}}]{Delbo2007a}
{Delbo'}, M., {dell'Oro}, A., {Harris}, A.~W., {Mottola}, S., \& {Mueller}, M.
  2007, \icarus, 190, 236

\bibitem[{{Delbo'} \& {Tanga}(2009)}]{Delbo2009}
{Delbo'}, M. \& {Tanga}, P. 2009, \planss, 57, 259

\bibitem[{{DeMeo} \& {Carry}(2013)}]{DeMeo2013}
{DeMeo}, F.~E. \& {Carry}, B. 2013, \icarus, 226, 723

\bibitem[{{Detal} {et~al.}(1994){Detal}, {Hainaut}, {Pospieszalska-Surdej},
  {Schils}, {Schober}, \& {Surdej}}]{Detal1994}
{Detal}, A., {Hainaut}, O., {Pospieszalska-Surdej}, A., {et~al.} 1994, \aap,
  281, 269

\bibitem[{{Dunlap} \& {Gehrels}(1969)}]{Dunlap1969}
{Dunlap}, J.~L. \& {Gehrels}, T. 1969, \aj, 74, 796

\bibitem[{{\v Durech} {et~al.}(2014){\v Durech}, {Hanu\v s}, {Delb\' o}, {Al\'
  i-Lagoa}, \& {Carry}}]{Durech2014}
{\v Durech}, J., {Hanu\v s}, J., {Delb\' o}, M., {Al\' i-Lagoa}, V., \&
  {Carry}, B. 2014, in AAS/Division for Planetary Sciences Meeting Abstracts,
  Vol.~46, AAS/Division for Planetary Sciences Meeting Abstracts, \#509.11

\bibitem[{{\v Durech} {et~al.}(2010){\v Durech}, {Sidorin}, \&
  {Kaasalainen}}]{Durech2010}
{\v Durech}, J., {Sidorin}, V., \& {Kaasalainen}, M. 2010, \aap, 513, A46

\bibitem[{{\v{D}urech} {et~al.}(2012){\v{D}urech}, {Vokrouhlick{\'y}},
  {Baransky}, {Breiter}, {Burkhonov}, {Cooney}, {Fuller}, {Gaftonyuk}, {Gross},
  {Inasaridze}, {Kaasalainen}, {Krugly}, {Kvaratshelia}, {Litvinenko},
  {Macomber}, {Marchis}, {Molotov}, {Oey}, {Polishook}, {Pollock}, {Pravec},
  {S{\'a}rneczky}, {Shevchenko}, {Slyusarev}, {Stephens}, {Szab{\'o}},
  {Terrell}, {Vachier}, {Vanderplate}, {Viikinkoski}, \&
  {Warner}}]{Durech2012b}
{\v{D}urech}, J., {Vokrouhlick{\'y}}, D., {Baransky}, A.~R., {et~al.} 2012,
  \aap, 547, A10

\bibitem[{{Emery} {et~al.}(2014){Emery}, {Fern{\'a}ndez}, {Kelley}, {Warden
  (n{\`e}e Crane)}, {Hergenrother}, {Lauretta}, {Drake}, {Campins}, \&
  {Ziffer}}]{Emery2014}
{Emery}, J.~P., {Fern{\'a}ndez}, Y.~R., {Kelley}, M.~S.~P., {et~al.} 2014,
  \icarus, 234, 17

\bibitem[{{Emery} {et~al.}(1998){Emery}, {Sprague}, {Witteborn}, {Colwell},
  {Kozlowski}, \& {Wooden}}]{Emery1998}
{Emery}, J.~P., {Sprague}, A.~L., {Witteborn}, F.~C., {et~al.} 1998, \icarus,
  136, 104

\bibitem[{{Fern{\'a}ndez} {et~al.}(2003){Fern{\'a}ndez}, {Sheppard}, \&
  {Jewitt}}]{Fernandez2003}
{Fern{\'a}ndez}, Y.~R., {Sheppard}, S.~S., \& {Jewitt}, D.~C. 2003, \aj, 126,
  1563

\bibitem[{{Grav} {et~al.}(2012{\natexlab{a}}){Grav}, {Mainzer}, {Bauer},
  {Masiero}, {Spahr}, {McMillan}, {Walker}, {Cutri}, {Wright}, {Eisenhardt},
  {Blauvelt}, {DeBaun}, {Elsbury}, {Gautier}, {Gomillion}, {Hand}, \&
  {Wilkins}}]{Grav2012a}
{Grav}, T., {Mainzer}, A.~K., {Bauer}, J., {et~al.} 2012{\natexlab{a}}, \apj,
  744, 197

\bibitem[{{Grav} {et~al.}(2011){Grav}, {Mainzer}, {Bauer}, {Masiero}, {Spahr},
  {McMillan}, {Walker}, {Cutri}, {Wright}, {Eisenhardt}, {Blauvelt}, {DeBaun},
  {Elsbury}, {Gautier}, {Gomillion}, {Hand}, \& {Wilkins}}]{Grav2011}
{Grav}, T., {Mainzer}, A.~K., {Bauer}, J., {et~al.} 2011, \apj, 742, 40

\bibitem[{{Grav} {et~al.}(2012{\natexlab{b}}){Grav}, {Mainzer}, {Bauer},
  {Masiero}, \& {Nugent}}]{Grav2012b}
{Grav}, T., {Mainzer}, A.~K., {Bauer}, J.~M., {Masiero}, J.~R., \& {Nugent},
  C.~R. 2012{\natexlab{b}}, \apj, 759, 49

\bibitem[{{Gundlach} \& {Blum}(2013)}]{Gundlach2013}
{Gundlach}, B. \& {Blum}, J. 2013, \icarus, 223, 479

\bibitem[{{Hahn} {et~al.}(1989){Hahn}, {Magnusson}, {Harris}, {Young},
  {Belkora}, {Fico}, {Lupishko}, {Shevchenko}, {Velichko}, {Burchi}, {Ciunci},
  {di Martino}, \& {Debehogne}}]{Hahn1989a}
{Hahn}, G., {Magnusson}, P., {Harris}, A.~W., {et~al.} 1989, \icarus, 78, 363

\bibitem[{{Hainaut-Rouelle} {et~al.}(1995){Hainaut-Rouelle}, {Hainaut}, \&
  {Detal}}]{Hainaut1995a}
{Hainaut-Rouelle}, M.-C., {Hainaut}, O.~R., \& {Detal}, A. 1995, \aaps, 112,
  125

\bibitem[{{Hanu\v{s}} {et~al.}(2011){Hanu\v{s}}, {\v{D}urech}, {Bro\v{z}},
  {Warner}, {Pilcher}, {Stephens}, {Oey}, {Bernasconi}, {Casulli}, {Behrend},
  {Polishook}, {Henych}, {Lehk{\'y}}, {Yoshida}, \& {Ito}}]{Hanus2011}
{Hanu\v{s}}, J., {\v{D}urech}, J., {Bro\v{z}}, M., {et~al.} 2011, \aap, 530,
  A134

\bibitem[{{Hanu\v{s}} {et~al.}(2013{\natexlab{a}}){Hanu\v{s}}, {Marchis}, \&
  {{\v D}urech}}]{Hanus2013b}
{Hanu\v{s}}, J., {Marchis}, F., \& {{\v D}urech}, J. 2013{\natexlab{a}},
  \icarus, 226, 1045

\bibitem[{{Hanu\v{s}} {et~al.}(2013{\natexlab{b}}){Hanu\v{s}}, {{\v D}urech},
  {Bro\v{z}}, {Marciniak}, {Warner}, {Pilcher}, {Stephens}, {Behrend}, {Carry},
  {\v{C}apek}, {Antonini}, {Audejean}, {Augustesen}, {Barbotin}, {Baudouin},
  {Bayol}, {Bernasconi}, {Borczyk}, {Bosch}, {Brochard}, {Brunetto}, {Casulli},
  {Cazenave}, {Charbonnel}, {Christophe}, {Colas}, {Coloma}, {Conjat},
  {Cooney}, {Correira}, {Cotrez}, {Coupier}, {Crippa}, {Cristofanelli},
  {Dalmas}, {Danavaro}, {Demeautis}, {Droege}, {Durkee}, {Esseiva}, {Esteban},
  {Fagas}, {Farroni}, {Fauvaud}, {Fauvaud}, {Del Freo}, {Garcia}, {Geier},
  {Godon}, {Grangeon}, {Hamanowa}, {Hamanowa}, {Heck}, {Hellmich}, {Higgins},
  {Hirsch}, {Husarik}, {Itkonen}, {Jade}, {Kami{\'n}ski}, {Kankiewicz},
  {Klotz}, {Koff}, {Kryszczy{\'n}ska}, {Kwiatkowski}, {Laffont}, {Leroy},
  {Lecacheux}, {Leonie}, {Leyrat}, {Manzini}, {Martin}, {Masi}, {Matter},
  {Micha{\l}owski}, {Micha{\l}owski}, {Micha{\l}owski}, {Michelet},
  {Michelsen}, {Morelle}, {Mottola}, {Naves}, {Nomen}, {Oey}, {Og{\l}oza},
  {Oksanen}, {Oszkiewicz}, {P{\"a}{\"a}kk{\"o}nen}, {Paiella}, {Pallares},
  {Paulo}, {Pavic}, {Payet}, {Poli{\'n}ska}, {Polishook}, {Poncy}, {Revaz},
  {Rinner}, {Rocca}, {Roche}, {Romeuf}, {Roy}, {Saguin}, {Salom}, {Sanchez},
  {Santacana}, {Santana-Ros}, {Sareyan}, {Sobkowiak}, {Sposetti}, {Starkey},
  {Stoss}, {Strajnic}, {Teng}, {Tr{\'e}gon}, {Vagnozzi}, {Velichko},
  {Waelchli}, {Wagrez}, \& {W{\"u}cher}}]{Hanus2013a}
{Hanu\v{s}}, J., {{\v D}urech}, J., {Bro\v{z}}, M., {et~al.}
  2013{\natexlab{b}}, \aap, 551, A67

\bibitem[{{Hapke}(1984)}]{Hapke1984}
{Hapke}, B. 1984, \icarus, 59, 41

\bibitem[{{Harris}(1998)}]{Harris1998}
{Harris}, A.~W. 1998, \icarus, 131, 291

\bibitem[{{Harris} \& {Lagerros}(2002)}]{Harris2002}
{Harris}, A.~W. \& {Lagerros}, J.~S.~V. 2002, Asteroids III, 205

\bibitem[{{Harris} \& {Young}(1989)}]{Harris1989a}
{Harris}, A.~W. \& {Young}, J.~W. 1989, \icarus, 81, 314

\bibitem[{{Hartmann} \& {Cruikshank}(1978)}]{Hartmann1978}
{Hartmann}, W.~K. \& {Cruikshank}, D.~P. 1978, \icarus, 36, 353

\bibitem[{{Higgins} {et~al.}(2008){Higgins}, {Pravec}, {Kusnirak}, {Hornoch},
  {Brinsfield}, {Allen}, \& {Warner}}]{Higgins2008b}
{Higgins}, D., {Pravec}, P., {Kusnirak}, P., {et~al.} 2008, Minor Planet
  Bulletin, 35, 123

\bibitem[{{Hoffmann} \& {Geyer}(1990)}]{Hoffmann1990a}
{Hoffmann}, M. \& {Geyer}, E.~H. 1990, \actaa, 40, 389

\bibitem[{{Horner} {et~al.}(2012){Horner}, {M{\"u}ller}, \&
  {Lykawka}}]{Horner2012}
{Horner}, J., {M{\"u}ller}, T.~G., \& {Lykawka}, P.~S. 2012, \mnras, 423, 2587

\bibitem[{{Horz} \& {Cintala}(1997)}]{Horz1997}
{Horz}, F. \& {Cintala}, M. 1997, Meteoritics and Planetary Science, 32, 179

\bibitem[{{Ito} \& {Yoshida}(2007)}]{Ito2007}
{Ito}, T. \& {Yoshida}, F. 2007, \pasj, 59, 269

\bibitem[{{Jakosky}(1986)}]{Jakosky1986}
{Jakosky}, B.~M. 1986, \icarus, 66, 117

\bibitem[{{Jarrett} {et~al.}(2011){Jarrett}, {Cohen}, {Masci}, {Wright},
  {Stern}, {Benford}, {Blain}, {Carey}, {Cutri}, {Eisenhardt}, {Lonsdale},
  {Mainzer}, {Marsh}, {Padgett}, {Petty}, {Ressler}, {Skrutskie}, {Stanford},
  {Surace}, {Tsai}, {Wheelock}, \& {Yan}}]{Jarrett2011}
{Jarrett}, T.~H., {Cohen}, M., {Masci}, F., {et~al.} 2011, \apj, 735, 112

\bibitem[{{Kaasalainen} {et~al.}(2004){Kaasalainen}, {Pravec}, {Krugly},
  {\v{S}arounov{\'a}}, {Torppa}, {Virtanen}, {Kaasalainen}, {Erikson},
  {Nathues}, {\v{D}urech}, {Wolf}, {Lagerros}, {Lindgren}, {Lagerkvist},
  {Koff}, {Davies}, {Mann}, {Ku\v{s}nir{\'a}k}, {Gaftonyuk}, {Shevchenko},
  {Chiorny}, \& {Belskaya}}]{Kaasalainen2004b}
{Kaasalainen}, M., {Pravec}, P., {Krugly}, Y.~N., {et~al.} 2004, \icarus, 167,
  178

\bibitem[{{Kaasalainen} \& {Torppa}(2001)}]{Kaasalainen2001a}
{Kaasalainen}, M. \& {Torppa}, J. 2001, Icarus, 153, 24

\bibitem[{{Kaasalainen} {et~al.}(2001){Kaasalainen}, {Torppa}, \&
  {Muinonen}}]{Kaasalainen2001b}
{Kaasalainen}, M., {Torppa}, J., \& {Muinonen}, K. 2001, Icarus, 153, 37

\bibitem[{{Kaasalainen} {et~al.}(2002{\natexlab{a}}){Kaasalainen}, {Torppa}, \&
  {Piironen}}]{Kaasalainen2002c}
{Kaasalainen}, M., {Torppa}, J., \& {Piironen}, J. 2002{\natexlab{a}}, \aap,
  383, L19

\bibitem[{{Kaasalainen} {et~al.}(2002{\natexlab{b}}){Kaasalainen}, {Torppa}, \&
  {Piironen}}]{Kaasalainen2002b}
{Kaasalainen}, M., {Torppa}, J., \& {Piironen}, J. 2002{\natexlab{b}}, Icarus,
  159, 369

\bibitem[{{Keihm}(1984)}]{Keihm1984}
{Keihm}, S.~J. 1984, \icarus, 60, 568

\bibitem[{{Kuhrt} \& {Giese}(1989)}]{Kuhrt1989}
{Kuhrt}, E. \& {Giese}, B. 1989, \icarus, 81, 102

\bibitem[{{Lagerros}(1996)}]{Lagerros1996}
{Lagerros}, J.~S.~V. 1996, \aap, 310, 1011

\bibitem[{{Lagerros}(1997)}]{Lagerros1997}
{Lagerros}, J.~S.~V. 1997, {\aap}, 325, 1226

\bibitem[{{Lagerros}(1998)}]{Lagerros1998}
{Lagerros}, J.~S.~V. 1998, \aap, 332, 1123

\bibitem[{{Larson} {et~al.}(2003){Larson}, {Beshore}, {Hill}, {Christensen},
  {McLean}, {Kolar}, {McNaught}, \& {Garradd}}]{Larson2003}
{Larson}, S., {Beshore}, E., {Hill}, R., {et~al.} 2003, in {Bulletin of the
  American Astronomical Society}, Vol.~35, {AAS/Division for Planetary Sciences
  Meeting Abstracts \#35}, 982

\bibitem[{{Lauretta} {et~al.}(2012){Lauretta}, {Barucci}, {Bierhaus},
  {Brucato}, {Campins}, {Christensen}, {Clark}, {Connolly}, {Dotto}, {Dworkin},
  {Emery}, {Garvin}, {Hildebrand}, {Libourel}, {Marshall}, {Michel}, {Nolan},
  {Nuth}, {Rizk}, {Sandford}, {Scheeres}, \& {Vellinga}}]{Lauretta2012}
{Lauretta}, D.~S., {Barucci}, M.~A., {Bierhaus}, E.~B., {et~al.} 2012, LPI
  Contributions, 1667, 6291

\bibitem[{{Lupishko} {et~al.}(1987){Lupishko}, {Velichko}, {Kazakov}, \&
  {Shevchenko}}]{Lupishko1987a}
{Lupishko}, D.~F., {Velichko}, F.~P., {Kazakov}, V.~V., \& {Shevchenko}, V.~G.
  1987, Kinematika i Fizika Nebesnykh Tel, 3, 92

\bibitem[{{Mainzer} {et~al.}(2011{\natexlab{a}}){Mainzer}, {Bauer}, {Grav},
  {Masiero}, {Cutri}, {Dailey}, {Eisenhardt}, {McMillan}, {Wright}, {Walker},
  {Jedicke}, {Spahr}, {Tholen}, {Alles}, {Beck}, {Brandenburg}, {Conrow},
  {Evans}, {Fowler}, {Jarrett}, {Marsh}, {Masci}, {McCallon}, {Wheelock},
  {Wittman}, {Wyatt}, {DeBaun}, {Elliott}, {Elsbury}, {Gautier}, {Gomillion},
  {Leisawitz}, {Maleszewski}, {Micheli}, \& {Wilkins}}]{Mainzer2011a}
{Mainzer}, A., {Bauer}, J., {Grav}, T., {et~al.} 2011{\natexlab{a}}, \apj, 731,
  53

\bibitem[{{Mainzer} {et~al.}(2014){Mainzer}, {Bauer}, {Grav}, {Masiero},
  {Cutri}, {Wright}, {Nugent}, {Stevenson}, {Clyne}, {Cukrov}, \&
  {Masci}}]{Mainzer2014}
{Mainzer}, A., {Bauer}, J., {Grav}, T., {et~al.} 2014, \apj, 784, 110

\bibitem[{{Mainzer} {et~al.}(2011{\natexlab{b}}){Mainzer}, {Grav}, {Bauer},
  {Masiero}, {McMillan}, {Cutri}, {Walker}, {Wright}, {Eisenhardt}, {Tholen},
  {Spahr}, {Jedicke}, {Denneau}, {DeBaun}, {Elsbury}, {Gautier}, {Gomillion},
  {Hand}, {Mo}, {Watkins}, {Wilkins}, {Bryngelson}, {Del Pino Molina}, {Desai},
  {G{\'o}mez Camus}, {Hidalgo}, {Konstantopoulos}, {Larsen}, {Maleszewski},
  {Malkan}, {Mauduit}, {Mullan}, {Olszewski}, {Pforr}, {Saro}, {Scotti}, \&
  {Wasserman}}]{Mainzer2011c}
{Mainzer}, A., {Grav}, T., {Bauer}, J., {et~al.} 2011{\natexlab{b}}, \apj, 743,
  156

\bibitem[{{Mainzer} {et~al.}(2011{\natexlab{c}}){Mainzer}, {Grav}, {Masiero},
  {Bauer}, {Wright}, {Cutri}, {McMillan}, {Cohen}, {Ressler}, \&
  {Eisenhardt}}]{Mainzer2011b}
{Mainzer}, A., {Grav}, T., {Masiero}, J., {et~al.} 2011{\natexlab{c}}, \apj,
  736, 100

\bibitem[{{Marchi} {et~al.}(2009){Marchi}, {Delbo'}, {Morbidelli}, {Paolicchi},
  \& {Lazzarin}}]{Marchi2009}
{Marchi}, S., {Delbo'}, M., {Morbidelli}, A., {Paolicchi}, P., \& {Lazzarin},
  M. 2009, \mnras, 400, 147

\bibitem[{{Marchis} {et~al.}(2014){Marchis}, {Durech}, {Castillo-Rogez},
  {Vachier}, {Cuk}, {Berthier}, {Wong}, {Kalas}, {Duchene}, {van Dam},
  {Hamanowa}, \& {Viikinkoski}}]{Marchis2014}
{Marchis}, F., {Durech}, J., {Castillo-Rogez}, J., {et~al.} 2014, \apjl, 783,
  L37

\bibitem[{{Marchis} {et~al.}(2006){Marchis}, {Kaasalainen}, {Hom}, {Berthier},
  {Enriquez}, {Hestroffer}, {Le Mignant}, \& {de Pater}}]{Marchis2006}
{Marchis}, F., {Kaasalainen}, M., {Hom}, E.~F.~Y., {et~al.} 2006, Icarus, 185,
  39

\bibitem[{{Marciniak} {et~al.}(2009){Marciniak}, {Micha{\l}owski}, {Hirsch},
  {Behrend}, {Bernasconi}, {Descamps}, {Colas}, {Sobkowiak}, {Kami{\'n}ski},
  {Kryszczy{\'n}ska}, {Kwiatkowski}, {Poli{\'n}ska}, {Rudawska}, {Fauvaud},
  {Santacana}, {Bruno}, {Fauvaud}, {Teng-Chuen-Yu}, \&
  {Peyrot}}]{Marciniak2009b}
{Marciniak}, A., {Micha{\l}owski}, T., {Hirsch}, R., {et~al.} 2009, \aap, 508,
  1503

\bibitem[{{Masiero} {et~al.}(2011){Masiero}, {Mainzer}, {Grav}, {Bauer},
  {Cutri}, {Dailey}, {Eisenhardt}, {McMillan}, {Spahr}, {Skrutskie}, {Tholen},
  {Walker}, {Wright}, {DeBaun}, {Elsbury}, {Gautier}, {Gomillion}, \&
  {Wilkins}}]{Masiero2011}
{Masiero}, J.~R., {Mainzer}, A.~K., {Grav}, T., {et~al.} 2011, \apj, 741, 68

\bibitem[{{Masiero} {et~al.}(2012){Masiero}, {Mainzer}, {Grav}, {Bauer},
  {Cutri}, {Nugent}, \& {Cabrera}}]{Masiero2012}
{Masiero}, J.~R., {Mainzer}, A.~K., {Grav}, T., {et~al.} 2012, \apjl, 759, L8

\bibitem[{{Mellon} {et~al.}(2000){Mellon}, {Jakosky}, {Kieffer}, \&
  {Christensen}}]{Mellon2000}
{Mellon}, M.~T., {Jakosky}, B.~M., {Kieffer}, H.~H., \& {Christensen}, P.~R.
  2000, \icarus, 148, 437

\bibitem[{{Mueller}(2012)}]{MullerPHD}
{Mueller}, M. 2012, ArXiv e-prints

\bibitem[{{Mueller} {et~al.}(2010){Mueller}, {Marchis}, {Emery}, {Harris},
  {Mottola}, {Hestroffer}, {Berthier}, \& {di Martino}}]{Muller2010}
{Mueller}, M., {Marchis}, F., {Emery}, J.~P., {et~al.} 2010, \icarus, 205, 505

\bibitem[{{Muinonen} {et~al.}(2010){Muinonen}, {Belskaya}, {Cellino},
  {Delb{\`o}}, {Levasseur-Regourd}, {Penttil{\"a}}, \&
  {Tedesco}}]{Muinonen2010}
{Muinonen}, K., {Belskaya}, I.~N., {Cellino}, A., {et~al.} 2010, \icarus, 209,
  542

\bibitem[{{M{\"u}ller} {et~al.}(2014){M{\"u}ller}, {Hasegawa}, \&
  {Usui}}]{Muller2014}
{M{\"u}ller}, T.~G., {Hasegawa}, S., \& {Usui}, F. 2014, \pasj, 66, 52

\bibitem[{{M{\"u}ller} {et~al.}(2012){M{\"u}ller}, {O'Rourke}, {Barucci},
  {P{\'a}l}, {Kiss}, {Zeidler}, {Altieri}, {Gonz{\'a}lez-Garc{\'i}a}, \&
  {K{\"u}ppers}}]{Muller2012}
{M{\"u}ller}, T.~G., {O'Rourke}, L., {Barucci}, A.~M., {et~al.} 2012, \aap,
  548, A36

\bibitem[{{Okada} {et~al.}(2014){Okada}, {Fukuhara}, {Tanaka}, {Taguchi},
  {Imamura}, {Arai}, {Senshu}, {Ogawa}, {Demura}, {Kitazato}, {Nakamura},
  {Sekiguchi}, {Hasegawa}, {Matsunaga}, {Wada}, {Takita}, {Sakatani},
  {Horikawa}, {Helbert}, {Mueller}, \& {Hagermann}}]{Okada2014}
{Okada}, T., {Fukuhara}, T., {Tanaka}, S., {et~al.} 2014, in Lunar and
  Planetary Inst. Technical Report, Vol.~45, Lunar and Planetary Science
  Conference, 1201

\bibitem[{{Oszkiewicz} {et~al.}(2011){Oszkiewicz}, {Muinonen}, {Bowell},
  {Trilling}, {Penttil{\"a}}, {Pieniluoma}, {Wasserman}, \&
  {Enga}}]{Oszkiewicz2011}
{Oszkiewicz}, D.~A., {Muinonen}, K., {Bowell}, E., {et~al.} 2011, Journal of
  Quantitative Spectroscopy \& Radiative Transfer, 112, 1919

\bibitem[{{Piironen} {et~al.}(2001){Piironen}, {Lagerkvist}, {Torppa},
  {Kaasalainen}, \& {Warner}}]{Piironen2001}
{Piironen}, J., {Lagerkvist}, C., {Torppa}, J., {Kaasalainen}, M., \& {Warner},
  B. 2001, in {Bulletin of the American Astronomical Society}, Vol.~33,
  {Bulletin of the American Astronomical Society}, 1562

\bibitem[{{Pilcher} {et~al.}(2012){Pilcher}, {Benishek}, {Briggs}, {Ferrero},
  {Klinglesmith}, \& {Warren}}]{Pilcher2012b}
{Pilcher}, F., {Benishek}, V., {Briggs}, J.~W., {et~al.} 2012, Minor Planet
  Bulletin, 39, 141

\bibitem[{{Pravec} {et~al.}(1996){Pravec}, {\v{S}arounov{\'a}}, \&
  {Wolf}}]{Pravec1996a}
{Pravec}, P., {\v{S}arounov{\'a}}, L., \& {Wolf}, M. 1996, \icarus, 124, 471

\bibitem[{{Press} {et~al.}(1986){Press}, {Flannery}, \&
  {Teukolsky}}]{Press1986}
{Press}, W.~H., {Flannery}, B.~P., \& {Teukolsky}, S.~A. 1986, {Numerical
  recipes. The art of scientific computing} (Cambridge: University Press, 1986)

\bibitem[{{Rozitis} \& {Green}(2014)}]{Rozitis2014}
{Rozitis}, B. \& {Green}, S.~F. 2014, \aap, 568, A43

\bibitem[{Rozitis {et~al.}(2014)Rozitis, MacLennan, \& Emery}]{Rozitis2014b}
Rozitis, B., MacLennan, E., \& Emery, J.~P. 2014, Nature, 512, 174

\bibitem[{{Salisbury} {et~al.}(1991){Salisbury}, {D'Aria}, \&
  {Jarosewich}}]{Salisbury1991}
{Salisbury}, J.~W., {D'Aria}, D.~M., \& {Jarosewich}, E. 1991, \icarus, 92, 280

\bibitem[{{S{\'a}rneczky} {et~al.}(1999){S{\'a}rneczky}, {Szab{\'o}}, \&
  {Kiss}}]{Sarneczky1999}
{S{\'a}rneczky}, K., {Szab{\'o}}, G., \& {Kiss}, L.~L. 1999, \aaps, 137, 363

\bibitem[{{Skiff} {et~al.}(2012){Skiff}, {Bowell}, {Koehn}, {Sanborn},
  {McLelland}, \& {Warner}}]{Skiff2012a}
{Skiff}, B.~A., {Bowell}, E., {Koehn}, B.~W., {et~al.} 2012, Minor Planet
  Bulletin, 39, 111

\bibitem[{{Spencer}(1990)}]{Spencer1990}
{Spencer}, J.~R. 1990, \icarus, 83, 27

\bibitem[{{Spencer} {et~al.}(1989){Spencer}, {Lebofsky}, \&
  {Sykes}}]{Spencer1989}
{Spencer}, J.~R., {Lebofsky}, L.~A., \& {Sykes}, M.~V. 1989, \icarus, 78, 337

\bibitem[{{Stephens}(2009{\natexlab{a}})}]{Stephens2009}
{Stephens}, R.~D. 2009{\natexlab{a}}, Minor Planet Bulletin, 36, 59

\bibitem[{{Stephens}(2009{\natexlab{b}})}]{Stephens2009b}
{Stephens}, R.~D. 2009{\natexlab{b}}, Minor Planet Bulletin, 36, 18

\bibitem[{{Szab{\'o}} {et~al.}(2001){Szab{\'o}}, {Cs{\'a}k}, {S{\'a}rneczky},
  \& {Kiss}}]{Szabo2001}
{Szab{\'o}}, G.~M., {Cs{\'a}k}, B., {S{\'a}rneczky}, K., \& {Kiss}, L.~L. 2001,
  \aap, 375, 285

\bibitem[{{Tedesco} {et~al.}(2002){Tedesco}, {Noah}, {Noah}, \&
  {Price}}]{Tedesco2002}
{Tedesco}, E.~F., {Noah}, P.~V., {Noah}, M., \& {Price}, S.~D. 2002,
  Astronomical Journal, 123, 1056

\bibitem[{{Tholen}(1984)}]{Tholen1984}
{Tholen}, D.~J. 1984, PhD thesis, Arizona Univ., Tucson.

\bibitem[{{Tholen}(1989)}]{Tholen1989}
{Tholen}, D.~J. 1989, in Asteroids II, ed. R.~P. {Binzel}, T.~{Gehrels}, \&
  M.~S. {Matthews}, 1139--1150

\bibitem[{{Usui} {et~al.}(2011){Usui}, {Kuroda}, {M{\"u}ller}, {Hasagawa},
  {Ishiguro}, {Ootsubo}, {Ishihara}, {Kataza}, {Takita}, {Oyabu}, {Ueno},
  {Matsuhara}, \& {Onaka}}]{Usui2011}
{Usui}, F., {Kuroda}, D., {M{\"u}ller}, T.~G., {et~al.} 2011, \pasj, 63, 1117

\bibitem[{{Velichko} {et~al.}(1990){Velichko}, {Krugly}, {Lupishko}, \&
  {Mokhamed}}]{Velichko1990a}
{Velichko}, F.~P., {Krugly}, Y.~N., {Lupishko}, D.~F., \& {Mokhamed}, R.~A.
  1990, Astronomicheskij Tsirkulyar, 1546, 39

\bibitem[{{Velichko} {et~al.}(2013){Velichko}, {Psarev}, {Kiselev}, {Zaitsev},
  {Velichko}, \& {Krymsaljuk}}]{Velichko2013}
{Velichko}, F.~P., {Psarev}, V.~A., {Kiselev}, N.~N., {et~al.} 2013, in Lunar
  and Planetary Science Conference, Vol.~44, Lunar and Planetary Science
  Conference, 2372

\bibitem[{{Vernazza} {et~al.}(2012){Vernazza}, {Delbo}, {King}, {Izawa},
  {Olofsson}, {Lamy}, {Cipriani}, {Binzel}, {Marchis}, {Mer{\'i}n}, \&
  {Tamanai}}]{Vernazza2012}
{Vernazza}, P., {Delbo}, M., {King}, P.~L., {et~al.} 2012, \icarus, 221, 1162

\bibitem[{{Warner}(2000)}]{Warner2000}
{Warner}, B. 2000, Minor Planet Bulletin, 27, 4

\bibitem[{{Warner} {et~al.}(2009){Warner}, {Harris}, \& {Pravec}}]{Warner2009}
{Warner}, B.~D., {Harris}, A.~W., \& {Pravec}, P. 2009, Icarus, 202, 134

\bibitem[{{Wisniewski} {et~al.}(1997){Wisniewski}, {Micha{\l}owski}, {Harris},
  \& {McMillan}}]{Wisniewski1997}
{Wisniewski}, W.~Z., {Micha{\l}owski}, T.~M., {Harris}, A.~W., \& {McMillan},
  R.~S. 1997, \icarus, 126, 395

\bibitem[{{Wright} {et~al.}(2010){Wright}, {Eisenhardt}, {Mainzer}, {Ressler},
  {Cutri}, {Jarrett}, {Kirkpatrick}, {Padgett}, {McMillan}, {Skrutskie},
  {Stanford}, {Cohen}, {Walker}, {Mather}, {Leisawitz}, {Gautier}, {McLean},
  {Benford}, {Lonsdale}, {Blain}, {Mendez}, {Irace}, {Duval}, {Liu}, {Royer},
  {Heinrichsen}, {Howard}, {Shannon}, {Kendall}, {Walsh}, {Larsen}, {Cardon},
  {Schick}, {Schwalm}, {Abid}, {Fabinsky}, {Naes}, \& {Tsai}}]{Wright2010}
{Wright}, E.~L., {Eisenhardt}, P.~R.~M., {Mainzer}, A.~K., {et~al.} 2010, \aj,
  140, 1868

\bibitem[{{Yoshida} {et~al.}(2004){Yoshida}, {Dermawan}, {Ito}, {Sawabe},
  {Haji}, {Saito}, {Hirai}, {Nakamura}, {Sato}, {Yanagisawa}, \&
  {Malhotra}}]{Yoshida2004}
{Yoshida}, F., {Dermawan}, B., {Ito}, T., {et~al.} 2004, \pasj, 56, 1105

\bibitem[{{Zimbelman}(1986)}]{Zimbelman1986}
{Zimbelman}, J.~R. 1986, \icarus, 68, 366

\end{thebibliography}
\bibliographystyle{aa}

\afterpage{%

\section*{Appendix A. WISE data}\label{sec:wise_errors}

\subsection*{A.1. Our estimation of the uncertainties of WISE thermal infrared data}\label{sec:WISE_errors}

\begin{figure*}[!htbp]
\begin{center}
\resizebox{\hsize}{!}{\includegraphics{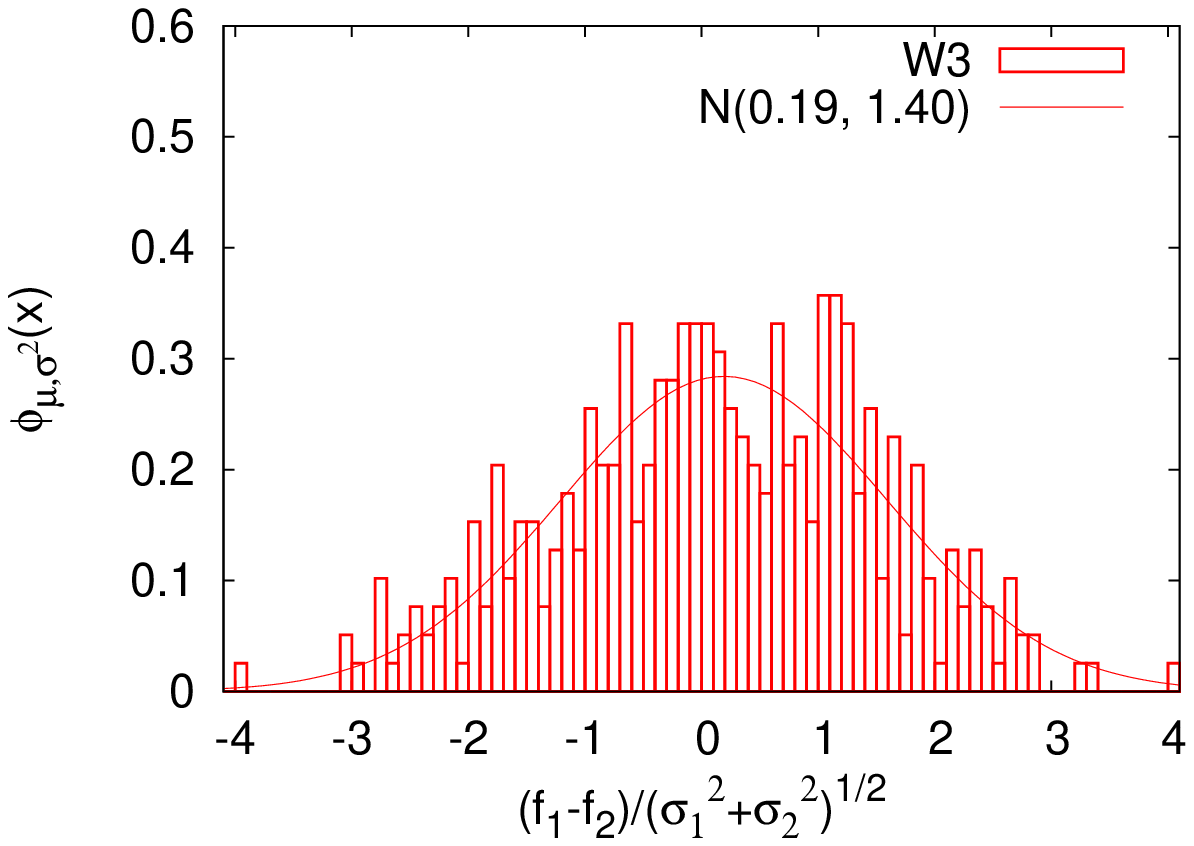}\includegraphics{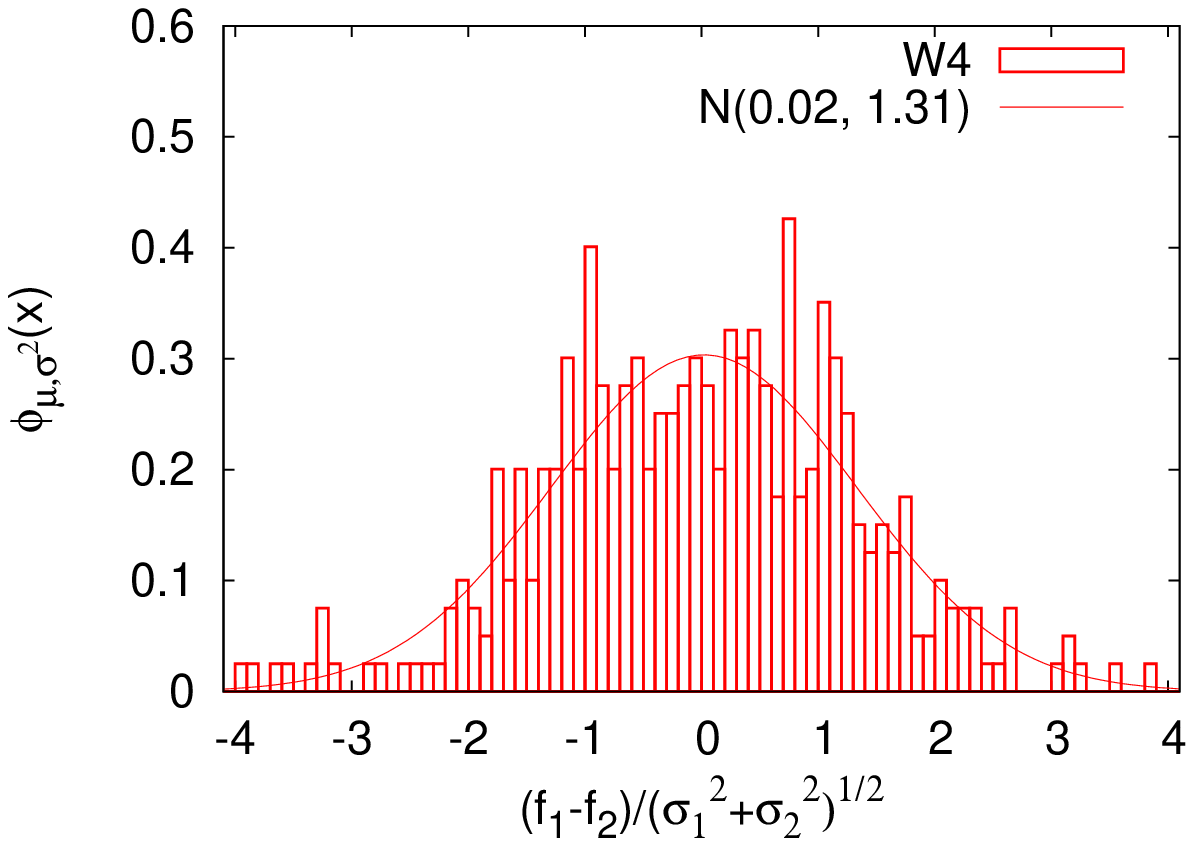}}\\
\end{center}
\caption{\label{img:sigma}Distribution of normalized differences $\bar f_{12}$ for all pairs of close points in our WISE thermal data set in filters W3 (left panel) and W4, respectively (right panel), and the best normal-fit approximations (lines).}
\end{figure*}

The purpose of this section is to describe the method we used to independently estimate the accuracy of the WISE thermal infrared data. 
Our estimate is based on the examination of WHAT we call {\em double detections} of asteroids. Due to an overlap of about 10\% between the areas of the field of view of two subsequently observed frames by WISE, we can sometimes find two flux measurements of an asteroid separated by the $\sim$11 seconds, which is the cadence between the WISE frames. We have identified $\sim$400 such double detections in the WISE data set for asteroids with convex models. The change in the thermal flux due to orbital and rotational evolution during this short time interval can be neglected, because it is smaller than the rotational spin-barrier period ($\sim$2.2 h) by a factor of $\sim$700. The flux of an asteroid with a double peaked thermal lightcurve and 2.2 hour rotational period changes only by $\sim$0.03\% in 11 s, which is significantly smaller than the typical flux error of $\sim$1--2\%. Thus we can use these measurements to estimate the errors.

Let us assume two measurements separated by 11 s with fluxes $f_1$ and $f_2$ ($f_1$ is always the first one observed) and reported uncertainties $\sigma_1$ and $\sigma_2$. We also assume that these independent points are drawn from normal distributions N($\mu$, $\sigma_1^2$) and N($\mu$, $\sigma_2^2$), respectively (the mean $\mu$ is unknown). The difference $f_1-f_2$ is also a random value from a normal distribution N(0, $\sigma_{12}^2$), where dispersion $\sigma_{12}$ corresponds to
\begin{equation}\label{eq:df12}
\sigma_{12} = \sqrt{\sigma_1^2 + \sigma_2^2}.
\end{equation}
By normalizing the difference $f_1-f_2\,$ by $\sigma_{12}$, we get a random value $\bar f_{12}$ from a distribution N(0,1). We calculate $\bar f_{12}$ for all double detections we find in filters W3 and W4 separately, and test the  hypothesis that the fluxes $f_1$ and $f_2$ are drawn from normal distributions N($\mu$, $\sigma_1^2$) and N($\mu$, $\sigma_2^2$). If we obtain values $\bar f_{12}$ following distribution N(0,1), the uncertainties $\sigma_1$ and $\sigma_2$ correspond to standard 1$\sigma$ values. However, by fitting these distributions of normalized differences $\bar f_{12}$, we determine that they correspond to N($0.19\pm0.07$, $(1.40\pm0.07)^2$) and N($0.02\pm0.06$, $(1.31\pm0.06)^2$), respectively (see Fig.~\ref{img:sigma}). The mean for W4 data is close to zero, and thus the data are not significantly offset. On the other hand, there is a small offset in the W3 filter consistent with the flux measurements from one position on the frame (i.e., $f_1$) to be systematically higher than the fluxes from the position on the opposite side of the frame (this is given by the particular scanning law). The reason could be vignetting and may not apply for the central parts of the frame. Our analysis shows that more realistic error bars in filters W3 and W4 are a factor $\sim$1.4 and $\sim$1.3 larger than the ones reported from the wise catalog. We also use the Kolmogorov-Smirnov test to test two zero hypotheses: (i)~values $\bar f_{12}$ are drawn from a normal distribution N(0,1), and (ii)~values $\bar f_{12}$ divided by the factors $\sim$1.4 and $\sim$1.3 are drawn from a normal distribution N(0,1). The hypothesis (i) is rejected at level $10^{-7}$ for W3 filter and 0.006 for W4 filter. On the other hand, hypothesis (ii) cannot be rejected at level 0.92 and 0.87.

%\textbf{As an alternative approach, we compare the number of pairs of points drawn from the same distribution N(0,1) whose uncertainty intervals intersect with the theoretical expectations. If we draw two random points from a distribution N(0,1), their uncertainty intervals intersect in $\sim84$\% cases. However, for observed pairs of points in both filters W3 and W4, we obtain lower presence of such points (68\% and 73\%). By enlarging the uncertainties of observed fluxes we can find the match between the number of expected and observed points which uncertainty intervals intersect, and thus determine again the true uncertainties of the thermal data. We find that errors of W3 and W4 data correspond to $\sim$1.4$\sigma$ and $\sim$1.25$\sigma$, respectively. We are aware that all the close points measurements are close to the edges of the frames and thus could be affected by systematic effects (such as vignetting) and their accuracy could be different (probably worse) than for the measurements closer to the middle of the frames. From this sense, our error estimates should be taken as upper limits. As we are determining the uncertainty intervals of modeled parameters such as thermal inertia or Bond albedo by applying the 1$\sigma$ criterion, we probably underestimate the errors by a factor of 1.3--1.4.}

\subsection*{A.2. Cross-calibration of the WISE data against Spitzer}\label{sec:WISE_systematics}

\begin{figure*}[!htbp]
\begin{center}
\resizebox{\hsize}{!}{\includegraphics{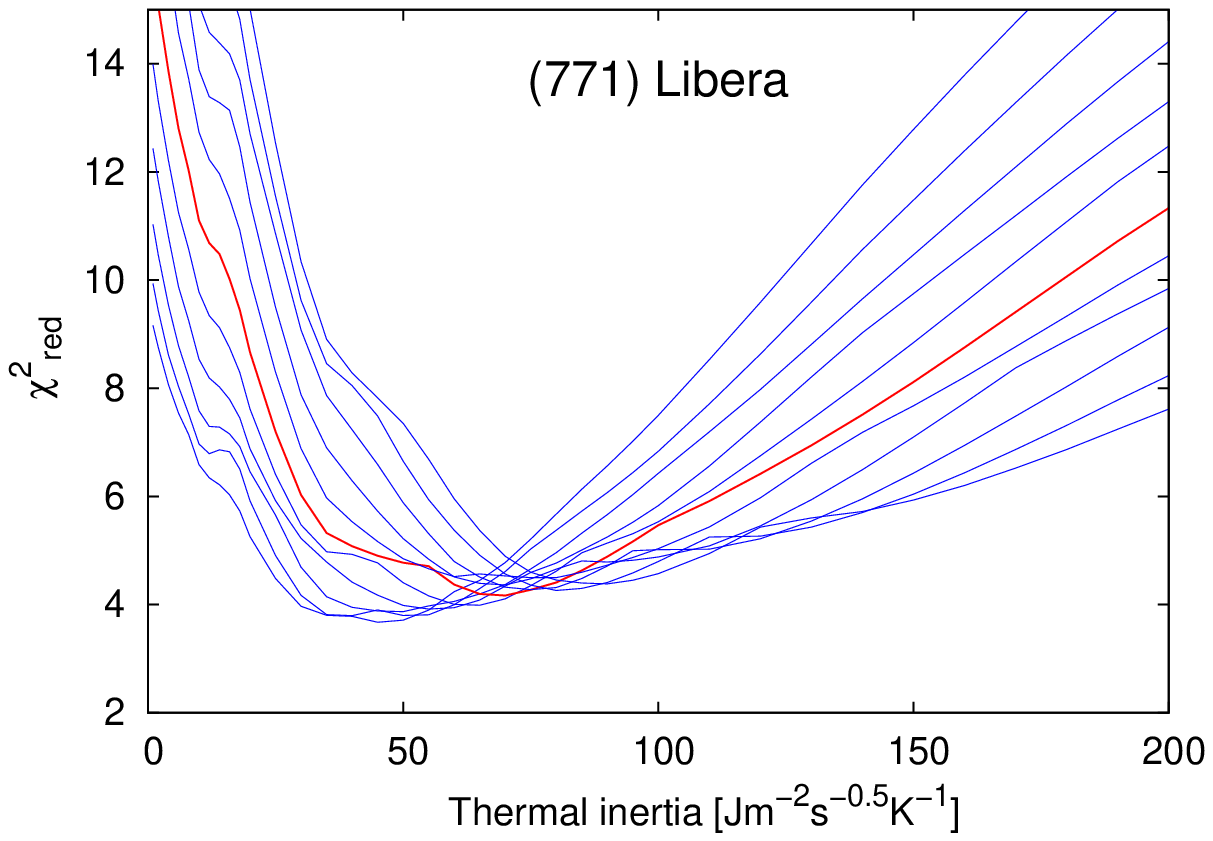}\includegraphics{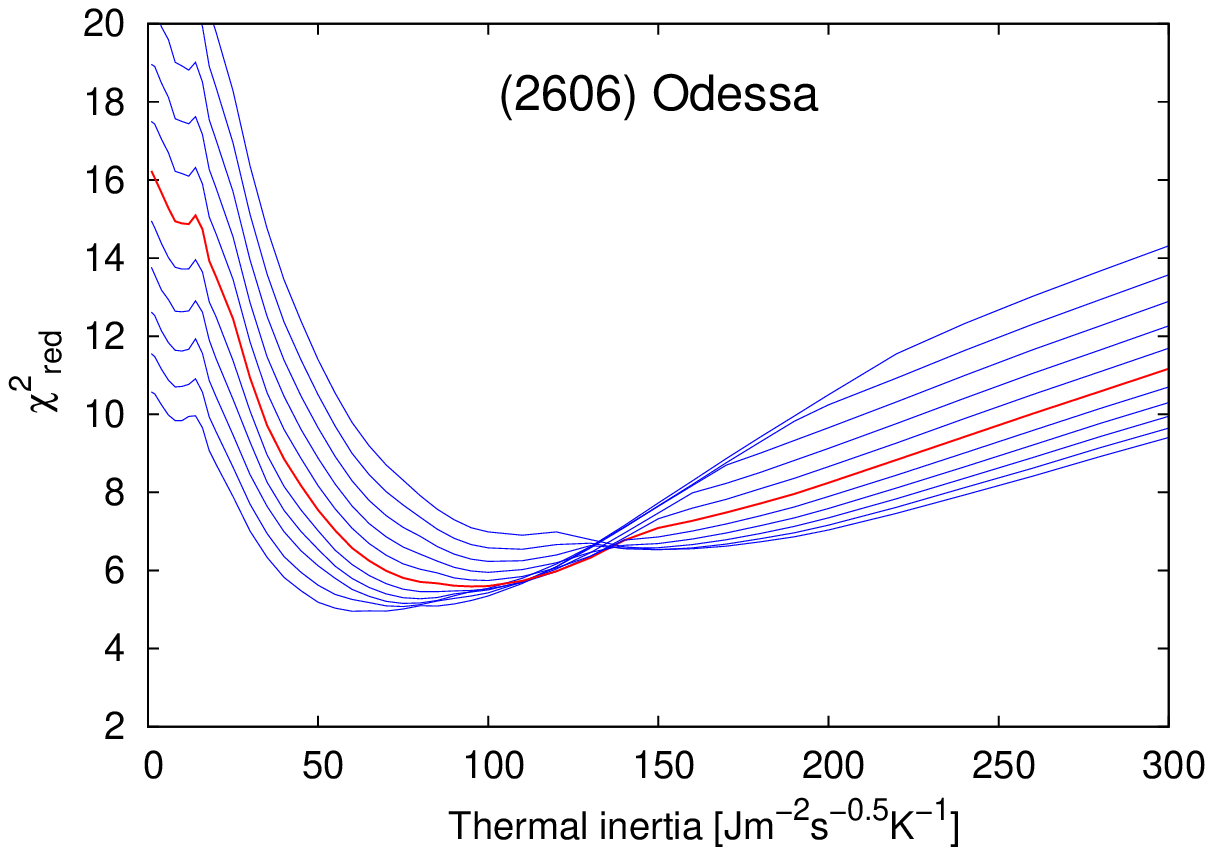}}\\
\end{center}
\caption{\label{img:WISE_syst}Chi-square recalculation for different offsets ($\pm$10\%) introduced to W4 thermal data (W3 data are fixed) of asteroids (771)~Libera and (2606)~Odessa. The red lines correspond to the curves obtained for the original thermal data. It shows the sensitivity of the TPM fits to the possible offsets observed by \citet{Jarrett2011} in their WISE-Spitzer cross-calibration study. Note that these offsets were characterized for stars and may not be applicable to asteroids.}
\end{figure*}

As already introduced in Sect~\ref{sec:thermal_data}, \citet{Jarrett2011} studied the accuracy of the absolute calibration of WISE data by performing an analysis of WISE-Spitzer flux cross-calibration of a number of calibration stars and one galaxy situated near the poles of the ecliptic. They found that (i)~the photometry for individual objects is stable for the whole cryogenic phase within less than 1\%, but (ii)~there is an rms scatter around the zero level of 4.5\% and 5.7\% of the WISE zero magnitudes in filters W3 and W4 when several objects are examined (the offset is different for each individual star).

Given that filters W3 and W4 are independent, we expect the offsets in these filters to be different for each individual asteroid. As a result, the fluxes will be changed by different multiplicative factors and the temperature will be different. To qualitatively estimate the corresponding change in the best-fitting values of parameters derived from the TPM, we perform the following test: we fix the observed fluxes in filter W3 and introduce various multiplicative factors in W4 data spanning $\pm$10\% with a step of 2\% and run the TPM scheme with the fixed shape models. We note that  $\pm$10\% is is a slightly overestimated range, because the rms scatter of $\approx$5\% in both filters should be summed in quadrature, which results in $\approx$7\%. 
In Fig.~\ref{img:WISE_syst}, we show the thermal inertia curves for different offsets introduced to W4 data of asteroids (771)~Libera and (2606)~Odessa. The effect on the thermal inertia is only apparent for the largest offsets, and, in general, the solutions remain well within the error bars given in Table~\ref{tab:TI}. Indeed, for the most extreme cases, the best-fitting thermal inertia $\Gamma$ could change by up to 50\% (which is not outside the limits for error bars reported in literature), the size $D$ by up to 10\% and the geometric visible albedo $p_\mathrm{V}$ by up to 20\%. This range of uncertainties represents the extreme cases that would be introduced by the discrepancy seen in the cross-calibration of WISE and Spitzer. 
Note that changing the fluxes by the same multiplicative factor will not affect the value of the thermal inertia or the surface roughness, and that it would only affect the diameter.

%In a less likely case of similar offsets in both filters, the resulting values of thermal inertia and surface roughness would not change because the offsets would be compensated only by the scale (thus diameter and Bond albedo).

\section*{Appendix B. Rotational phase of the shape model}\label{sec:rotational_phase}

In this section, we justify in more detail the necessity to optimize the rotational phase $\phi_0$ in the TPM. 

All shape models we use here are derived by the lightcurve inversion technique. This gradient-based method searches a large parameter space that includes sidereal rotational period, pole orientation, shape and scattering parameters, and converges to all local minima and essentially finds the deepest (global) minimum. The difference $\Delta P$ between two local minima in the parameter space of rotational periods corresponds to

\begin{equation}\label{eq:dP}
 \Delta P \approx \frac{P^2}{2T},
\end{equation}
where $P$ is the sidereal rotational period and $T=T_2-T_1$ is the timespan of the photometric data \citep{Kaasalainen2001b} and $T_1$ and $T_2$ are the epochs of the first and last observations of the asteroid within the photometric data set. The meaning of Eq.~(\ref{eq:dP}) is the following: the rotational phase shift during time $T$ due to a  change in period of $\Delta P$ is $180^{\circ}$ (assuming double-peaked sinusoidal lightcurve), thus the difference between the corresponding maxima of the lightcurves with $P$ and $P+\Delta P$ after time $T$ is exactly 180$^{\circ}$. 

Let us assume that the uncertainty of the initial shape model orientation $\delta\phi_0$ due to the noise in the optical lightcurves corresponds to $\sim$10 degrees, thus we can write for the uncertainty of the rotational period $\delta P$

\begin{equation}\label{eq:dp}
 \delta P = \frac{\Delta P}{20}.
\end{equation}
This means that the orientation of the shape model is only known to this level of certainty within the coverage of the optical observations. Because the WISE data were acquired outside the timespan $T$ ($T_{\mathrm{W}}>T_2$), the uncertainty in the shape model's initial orientation $\delta\phi_0$ is even larger:

\begin{equation}
% \delta\phi_0 [\mathrm{deg}]= 360\frac{T_{\mathrm{W}}-T_2}{P^2}\delta P.
% \delta\phi_0 [\mathrm{deg}]= \frac{1}{2}\frac{360}{20} + 360\frac{T_{\mathrm{W}}-T_2}{P^2}\delta P.
 \delta\phi_0 [\mathrm{deg}]= 10 + 360\frac{T_{\mathrm{W}}-T_2}{P^2}\delta P.
\end{equation}
So, values of the uncertainty $\delta\phi_0$ of the rotational phase are always at least 10$^{\circ}$ purely because of the uncertainty in $P$. Some of the shape models from the DAMIT database we use here were derived more than 10 years before WISE observations, thus the values of $\delta\phi_0$ are large. The most extreme case is the asteroid (1036)~Ganymed, for which $\delta\phi_0\sim540^{\circ}$, thus it is completely unconstrained (the model is based on the data from only one apparition in 1985). The expected uncertainties $\delta\phi_0$ for other studied asteroids are $\lesssim20^{\circ}$. Of course, $\delta\phi_0$ is dependent on the value of $\delta P$, for which we assume a reasonable value. 
\clearpage
}

\afterpage{%
\section*{Supplementary material}

\newpage

\begin{figure*}[!htbp]
	\begin{center}
	 \resizebox{1.0\hsize}{!}{\includegraphics{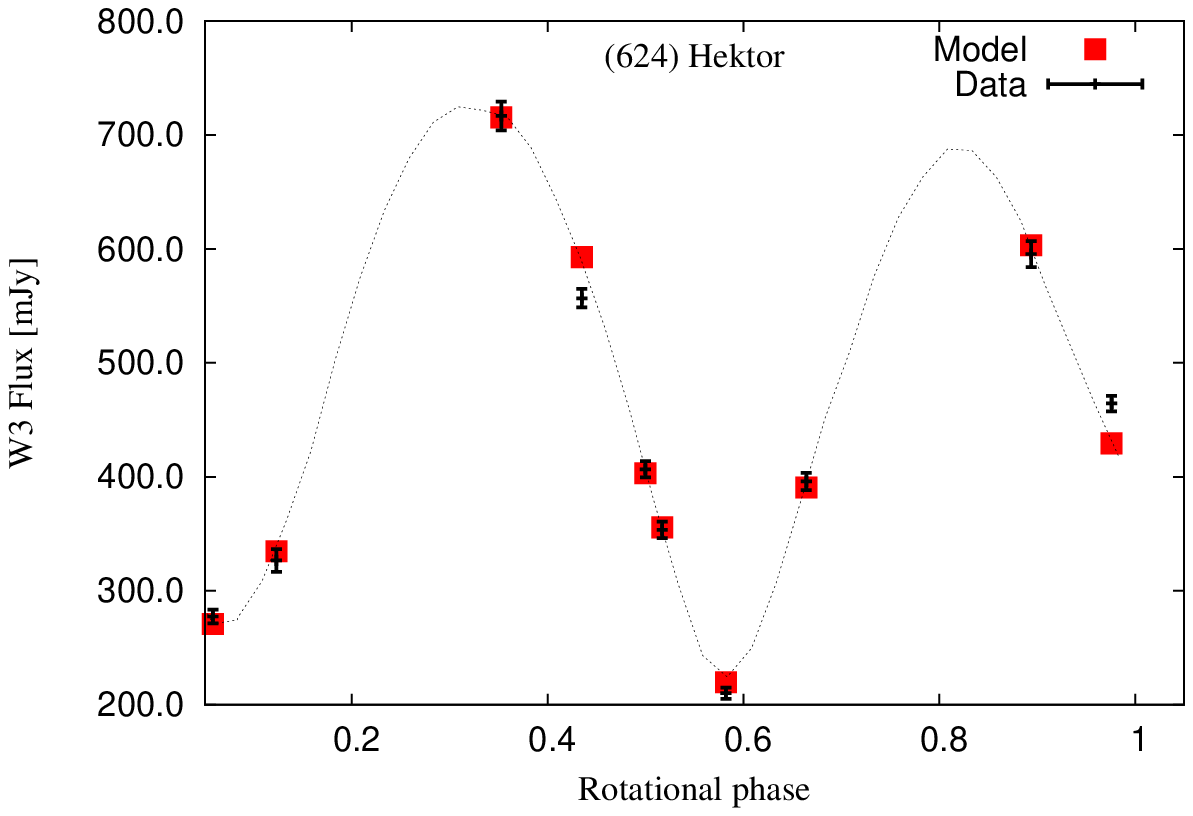}\,\includegraphics{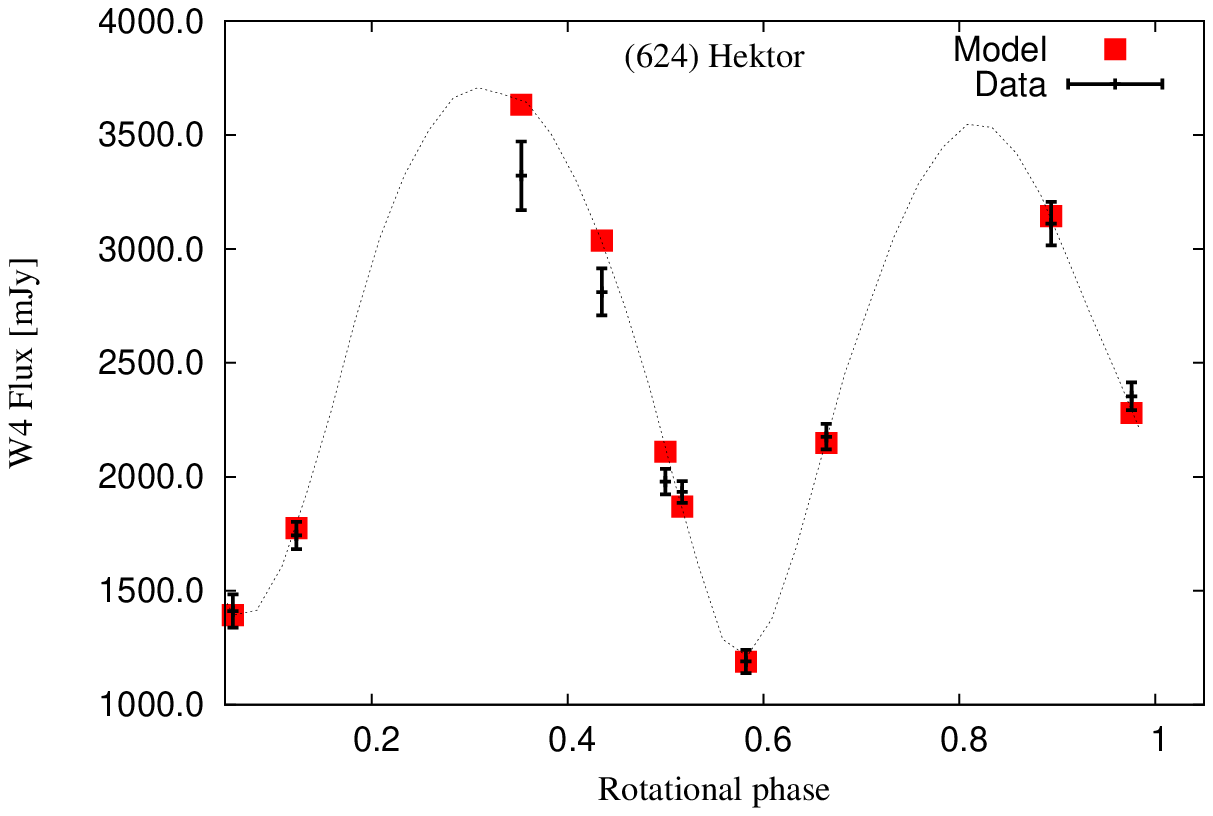}}\\
%	 \resizebox{1.0\hsize}{!}{\includegraphics{img/fluxW3_624_2_best.out.eps}\,\includegraphics{img/fluxW4_624_2_best.out.eps}}\\
	 \end{center}
	 \caption{\label{img:624}Comparison between the observed (WISE) and modeled (best TPM fit) thermal IR fluxes in filters W3 and W4 and synthetic thermal lightcurves (dotted lines) for asteroid (624)~Hektor (revised model).}
\end{figure*}

\newpage

\begin{figure*}[!htbp]
	\begin{center}
	 \resizebox{1.0\hsize}{!}{\includegraphics{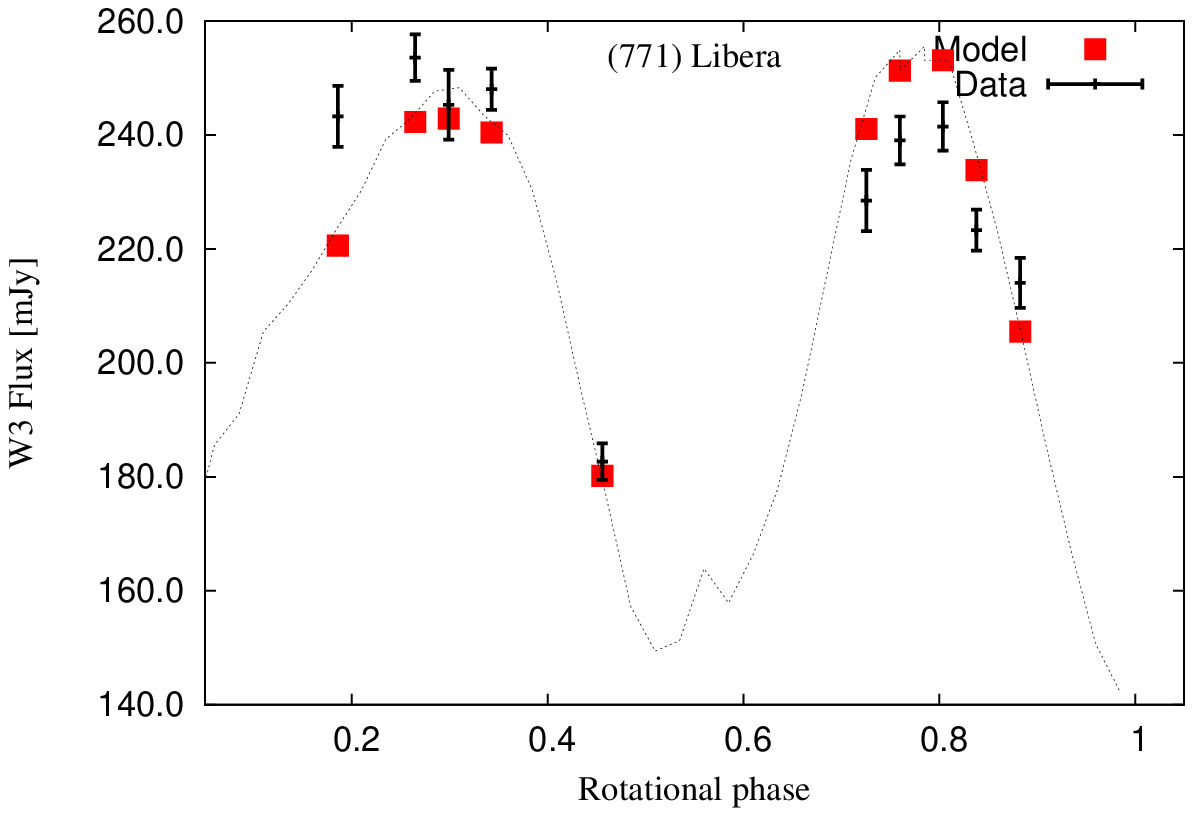}\,\includegraphics{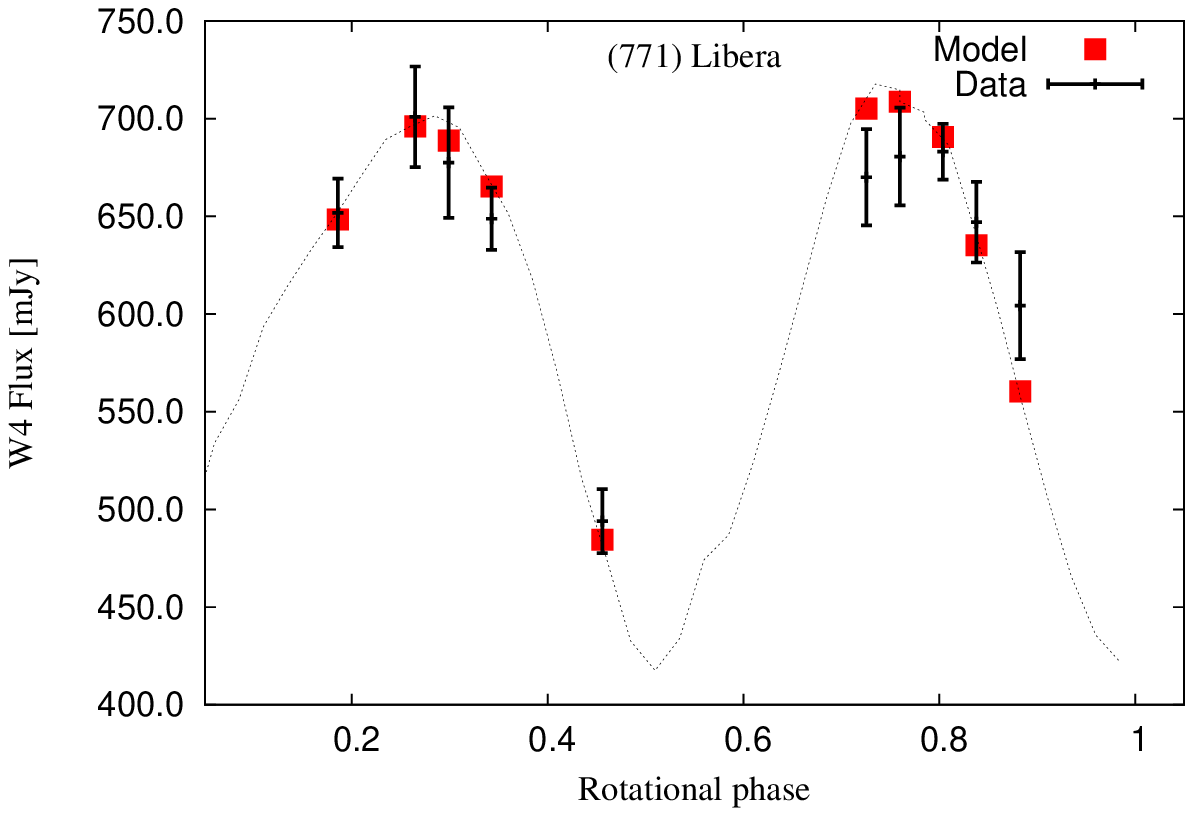}}\\
	 \resizebox{1.0\hsize}{!}{\includegraphics{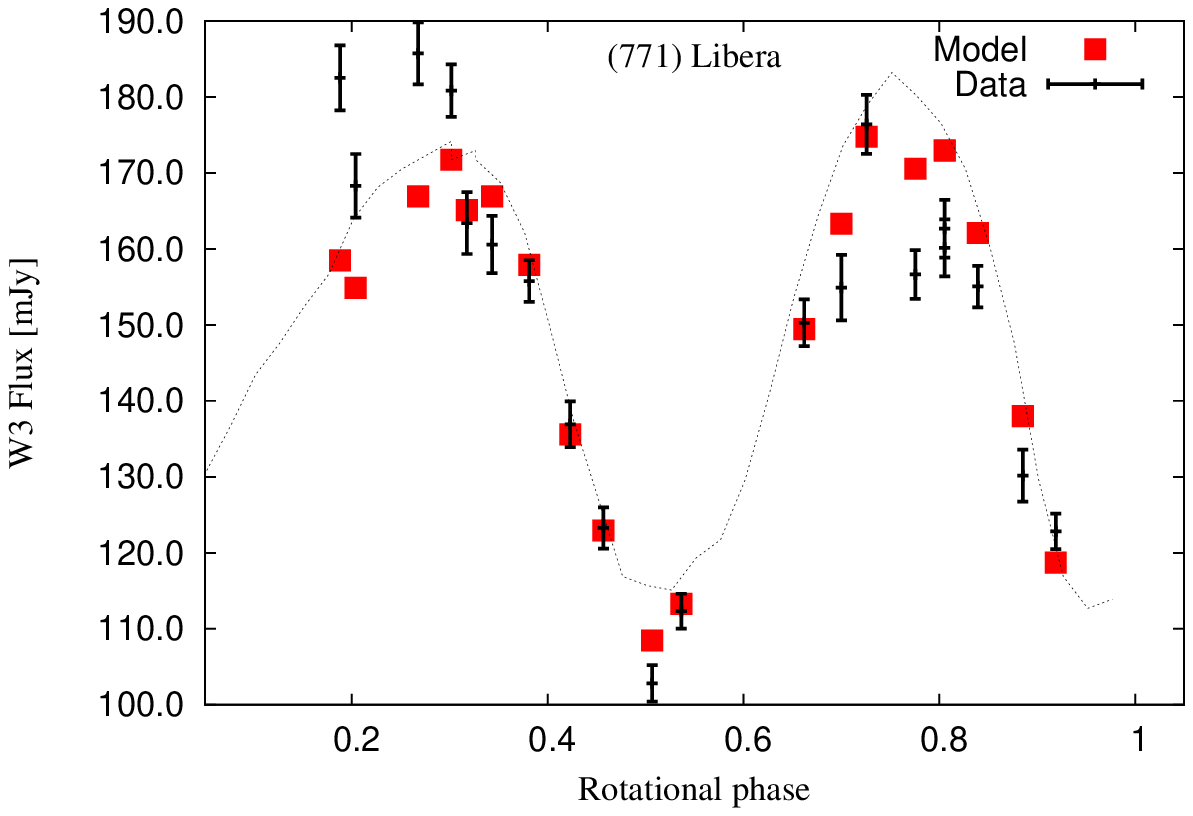}\,\includegraphics{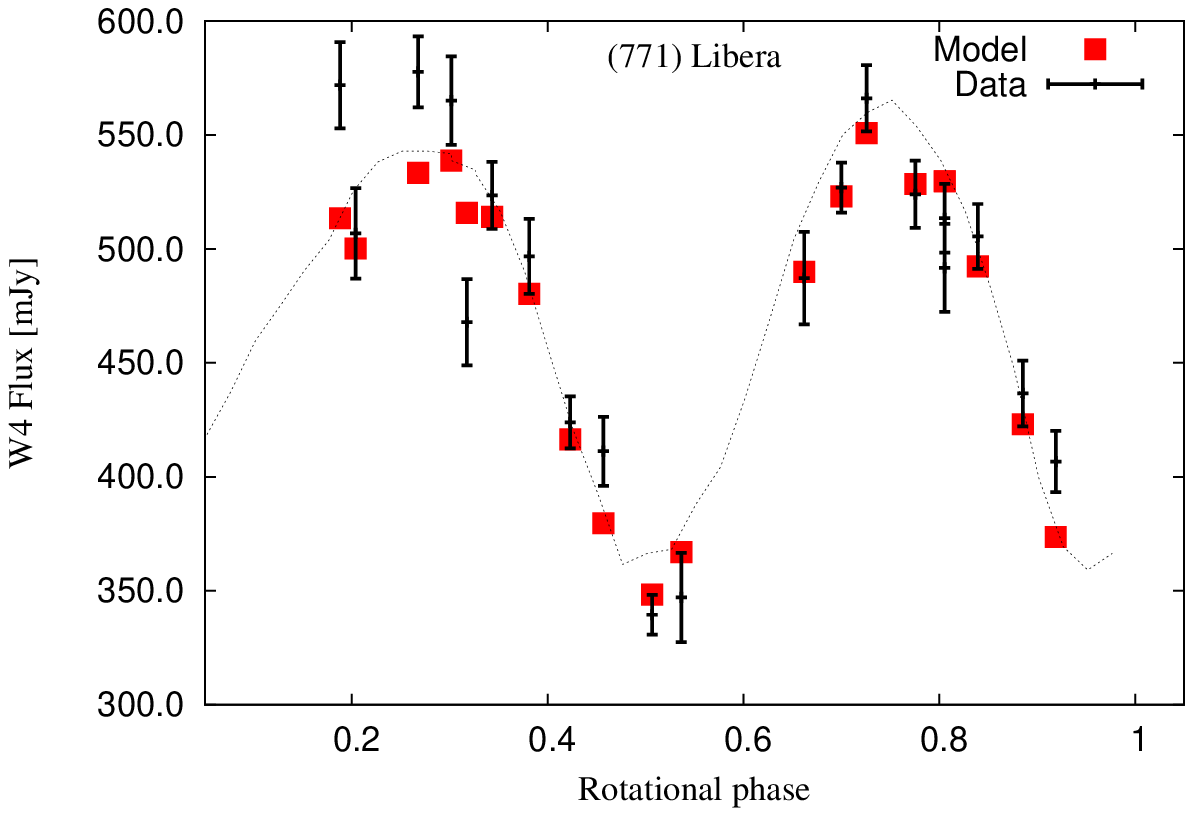}}\\
	 \end{center}
	 \caption{\label{img:771}Comparison between the observed (WISE) and modeled (best TPM fit) thermal IR fluxes in filters W3 and W4 and synthetic thermal lightcurves (dotted lines) for asteroid (771)~Libera (DAMIT model). There are two groups observations separated by few months (top and bottom panels). The offset between several model points and the synthetic lightcurve in the lower panels is caused by the fact that the observations span four days. In this time, the thermal lighcurve changes because of the variation in the observing geometry. However, the synthetic lightcurve corresponds to the beginning of this interval.}
\end{figure*}

\newpage

\begin{figure*}[!htbp]
	\begin{center}
	 \resizebox{1.0\hsize}{!}{\includegraphics{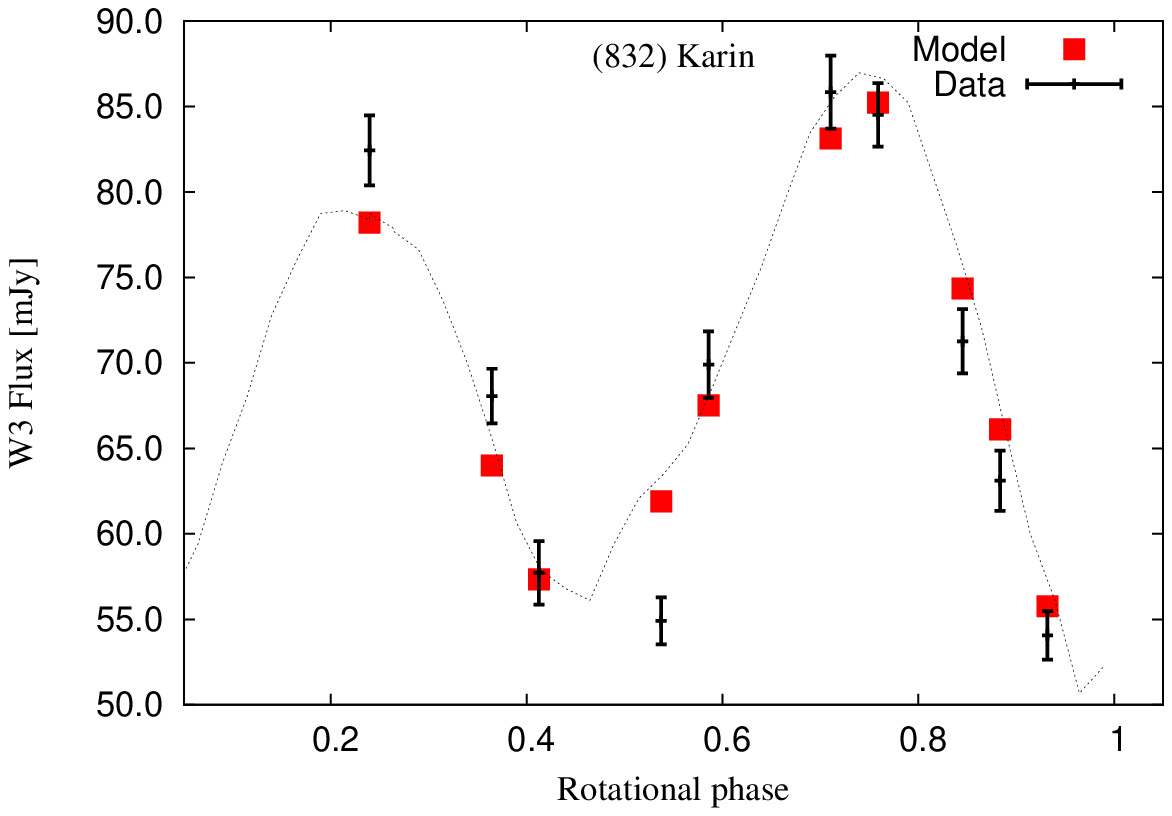}\,\includegraphics{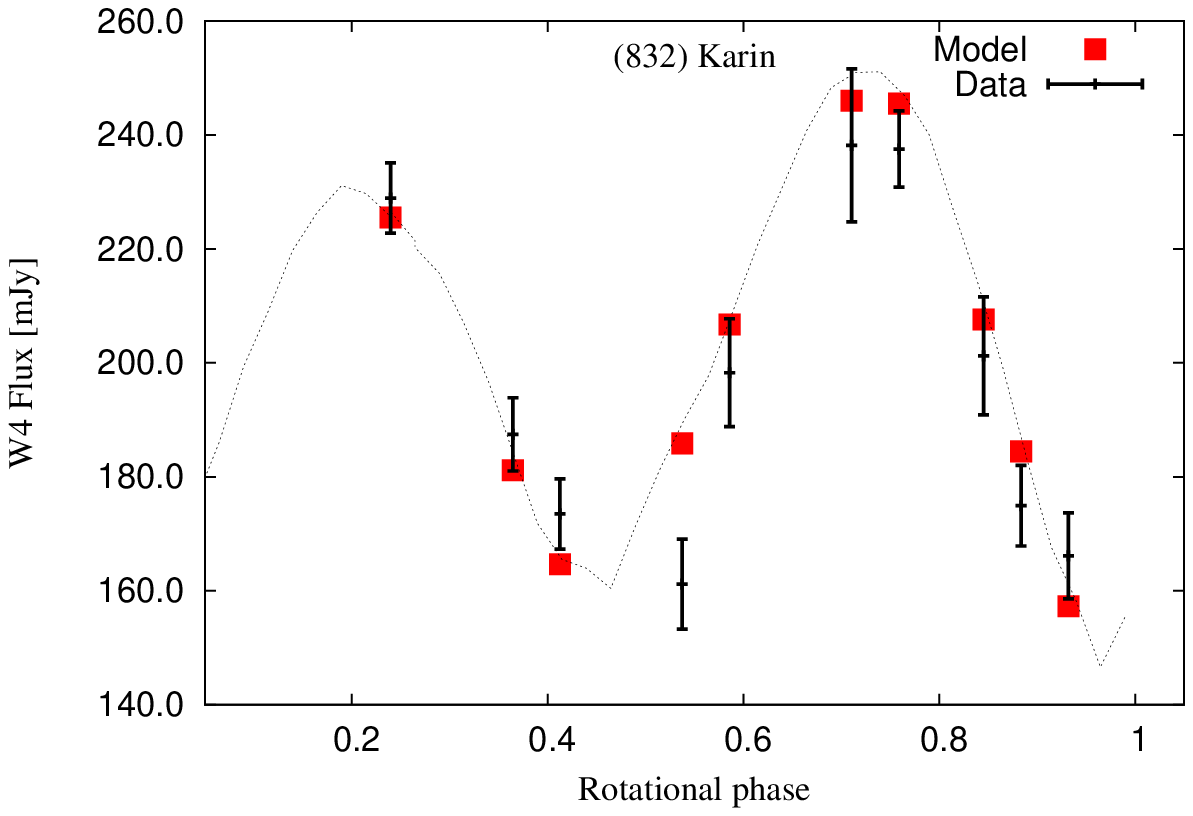}}\\
	 \resizebox{1.0\hsize}{!}{\includegraphics{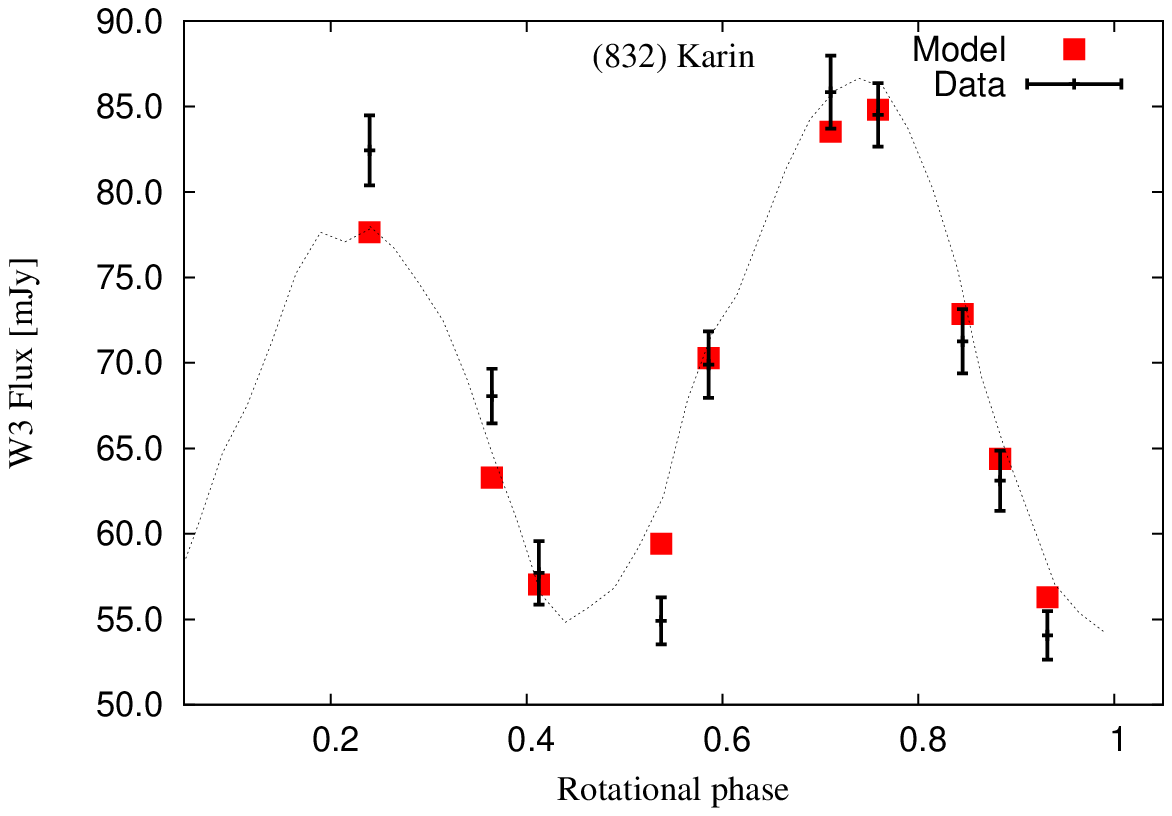}\,\includegraphics{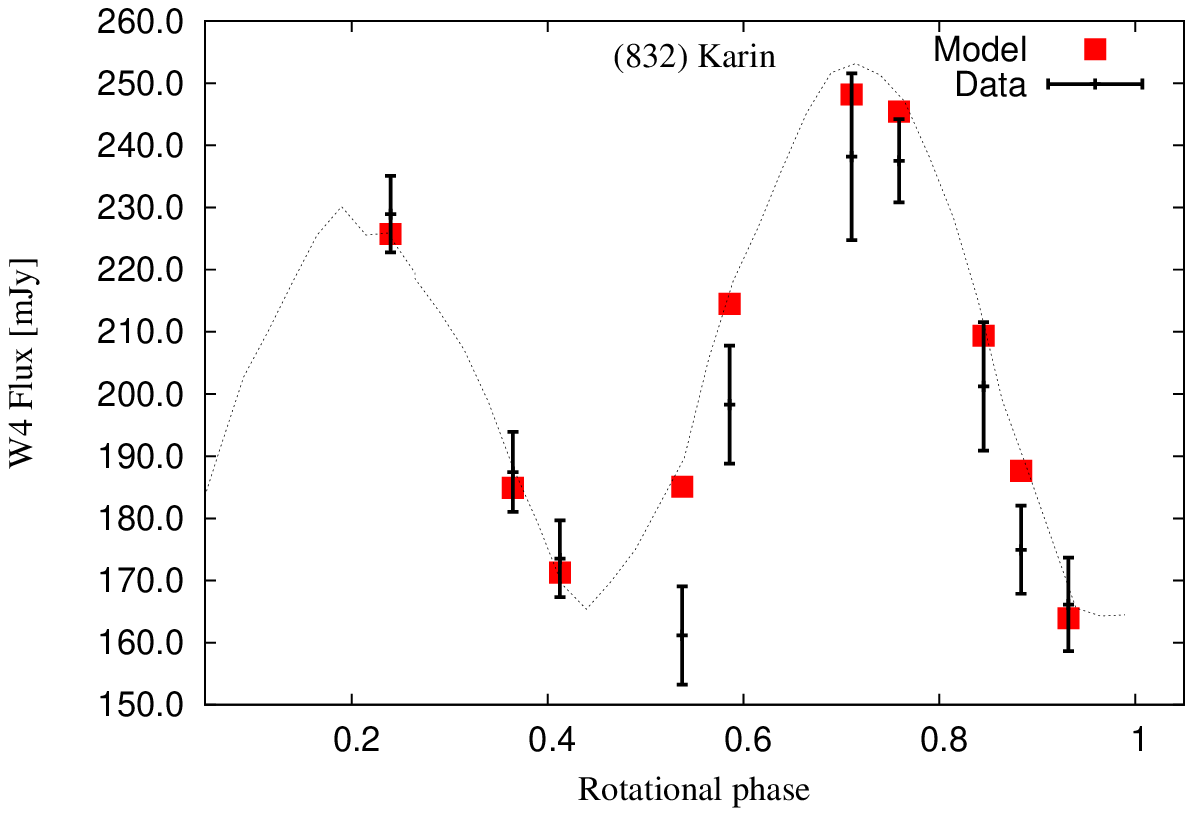}}\\
	 \end{center}
	 \caption{\label{img:832}Comparison between the observed (WISE) and modeled (best TPM fit) thermal IR fluxes in filters W3 and W4 and synthetic thermal lightcurves (dotted lines) for both pole solutions of asteroid (832)~Karin (DAMIT model). The small offset between several model points and the synthetic lightcurve is caused by the fact that the observations span several rotational periods. In this time, the thermal lighcurve changes because of the variation in the observing geometry. However, the synthetic lightcurve corresponds to the beginning of this interval.}
\end{figure*}

\newpage

\begin{figure*}[!htbp]
	\begin{center}
	 \resizebox{1.0\hsize}{!}{\includegraphics{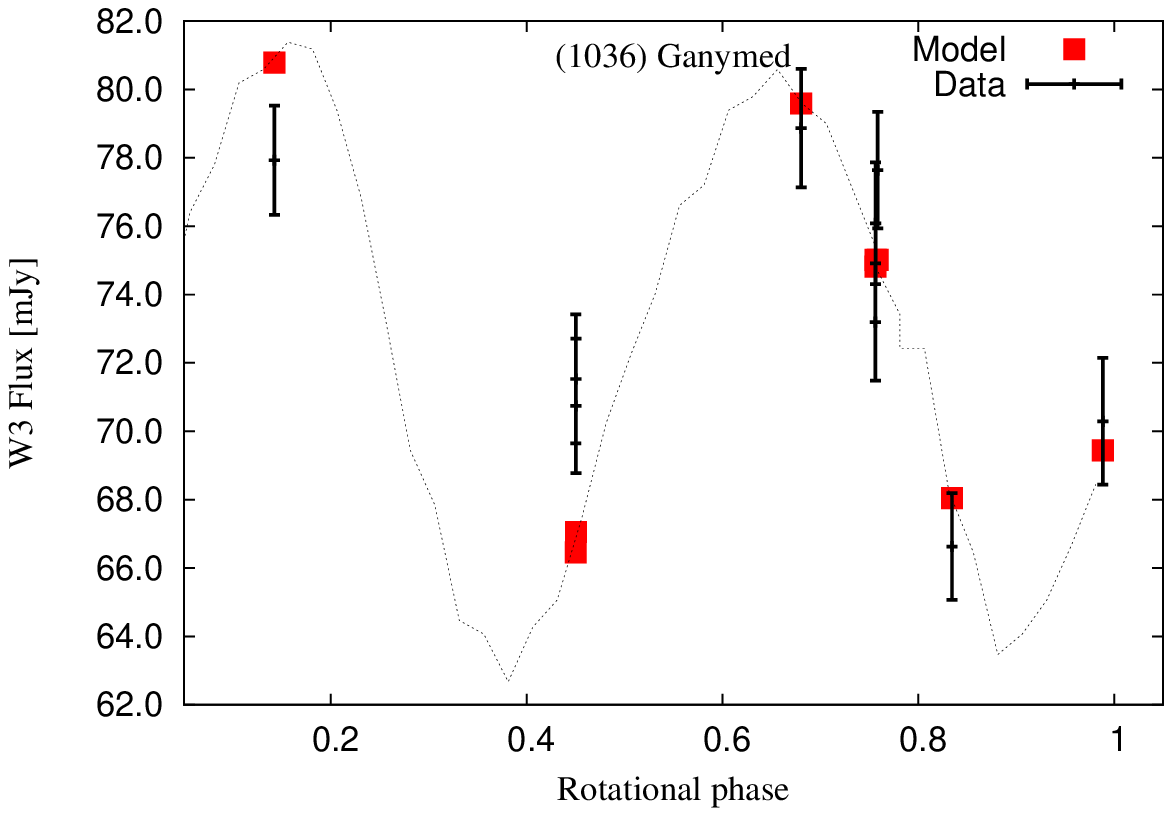}\,\includegraphics{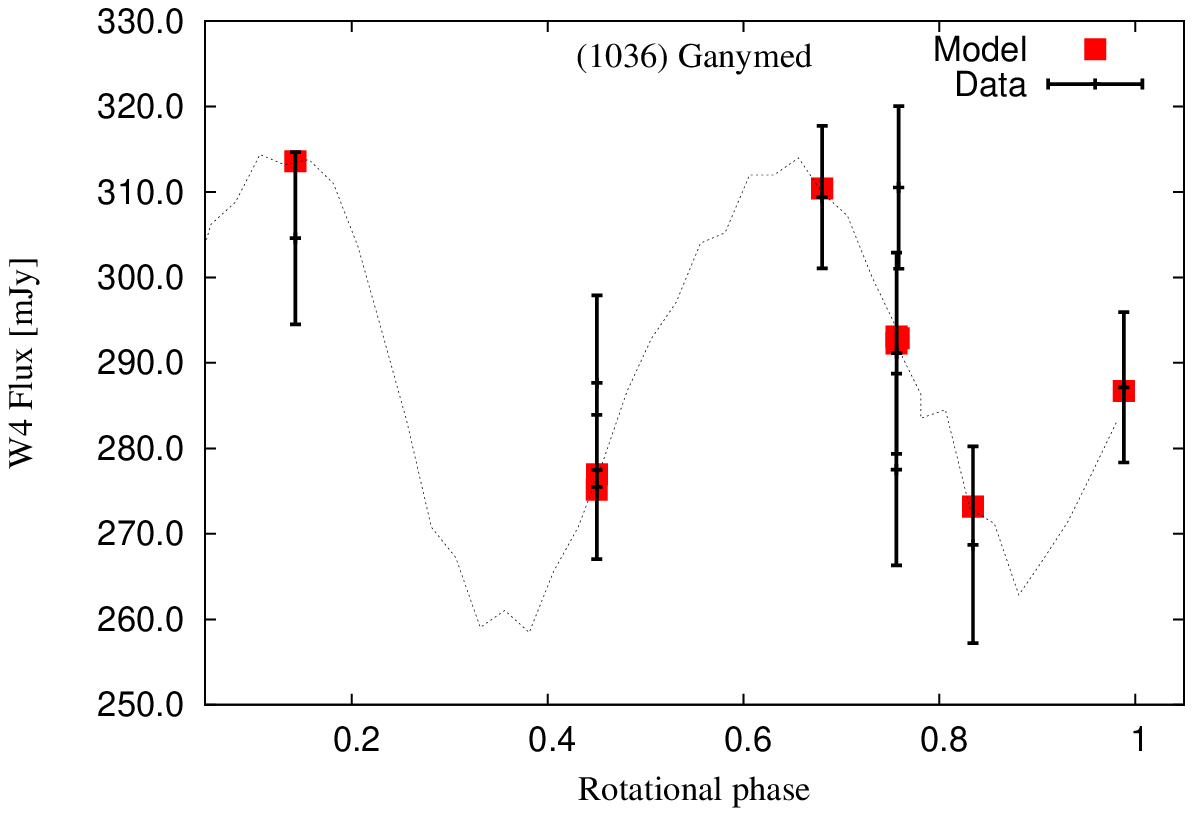}}\\
	 \resizebox{1.0\hsize}{!}{\includegraphics{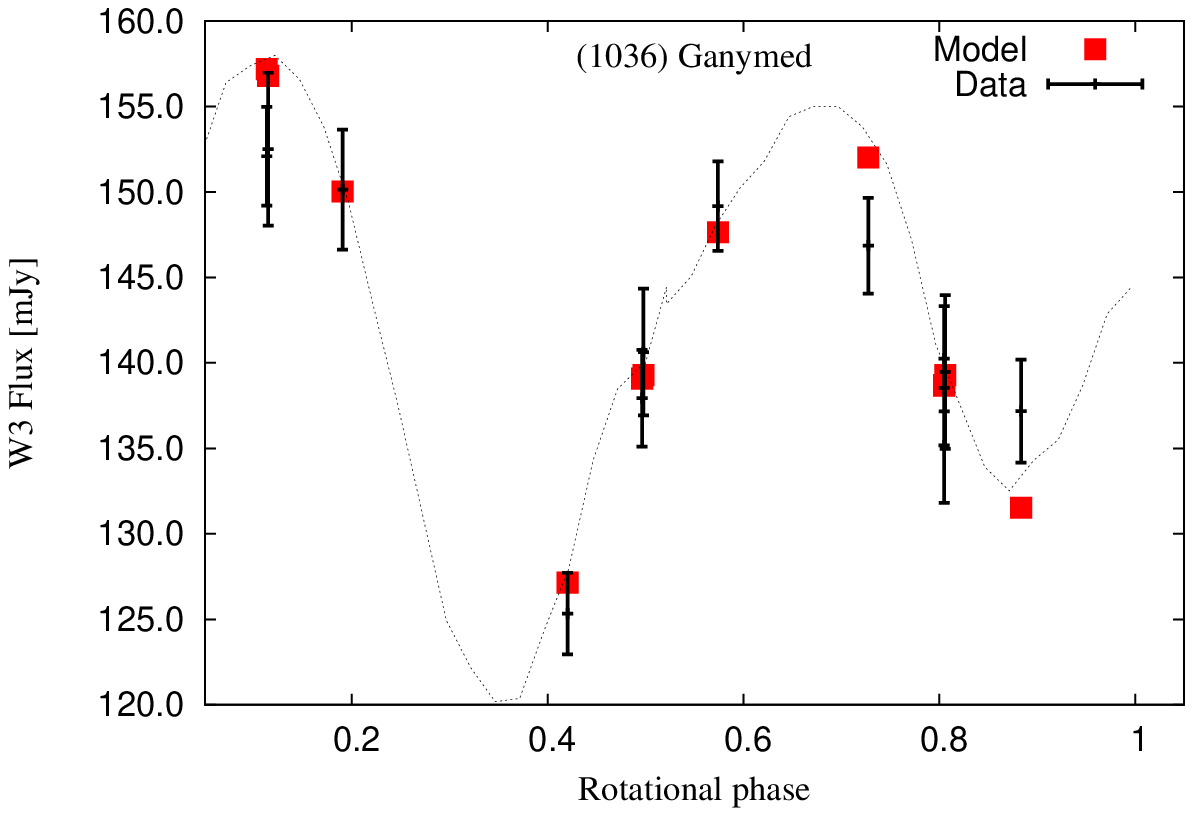}\,\includegraphics{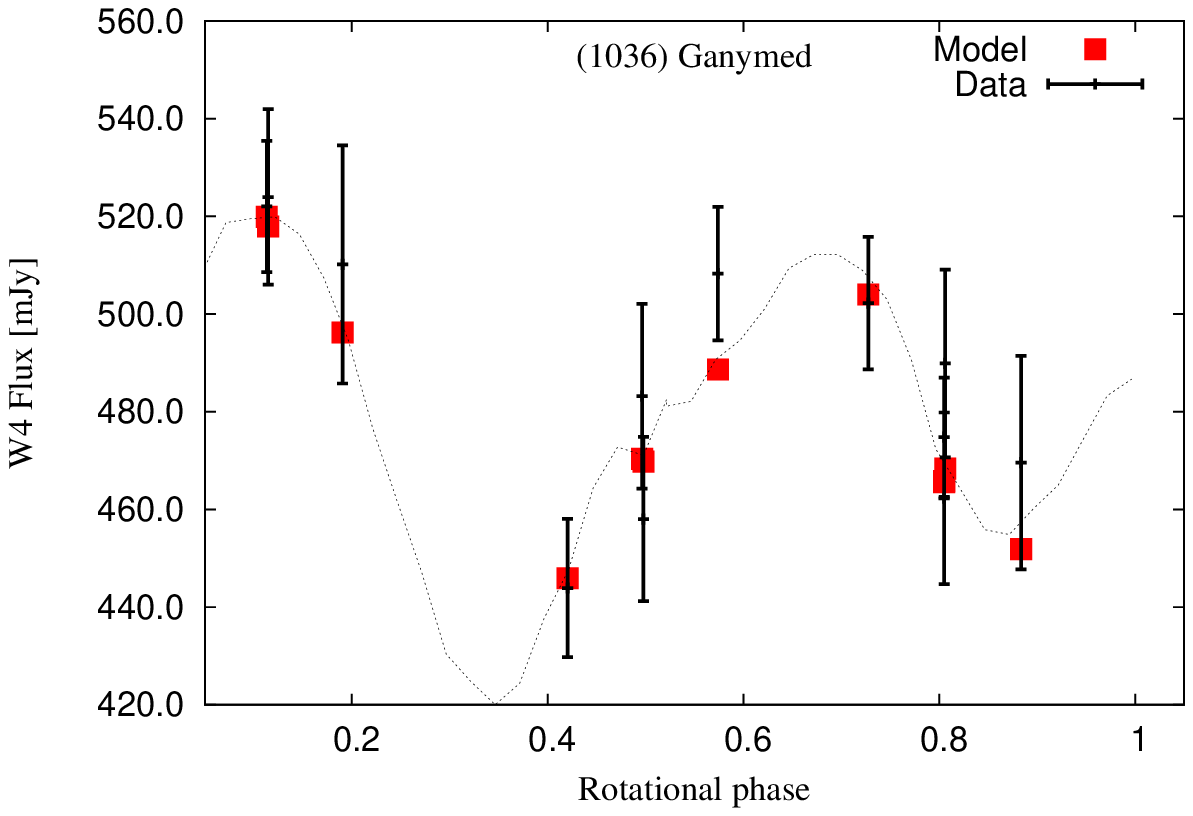}}\\
	 \end{center}
	 \caption{\label{img:1036}Comparison between the observed (WISE) and modeled (best TPM fit) thermal IR fluxes in filters W3 and W4 and synthetic thermal lightcurves (dotted lines) for asteroid (1036)~Ganymed (revised model). There are two groups observations separated by few months (top and bottom panels). The small offset between several model points and the synthetic lightcurve is caused by the fact that the observations span several rotational periods. In this time, the thermal lighcurve changes because of the variation in the observing geometry. However, the synthetic lightcurve corresponds to the beginning of this interval.}
\end{figure*}

\newpage

\begin{figure*}[!htbp]
	\begin{center}
	 \resizebox{1.0\hsize}{!}{\includegraphics{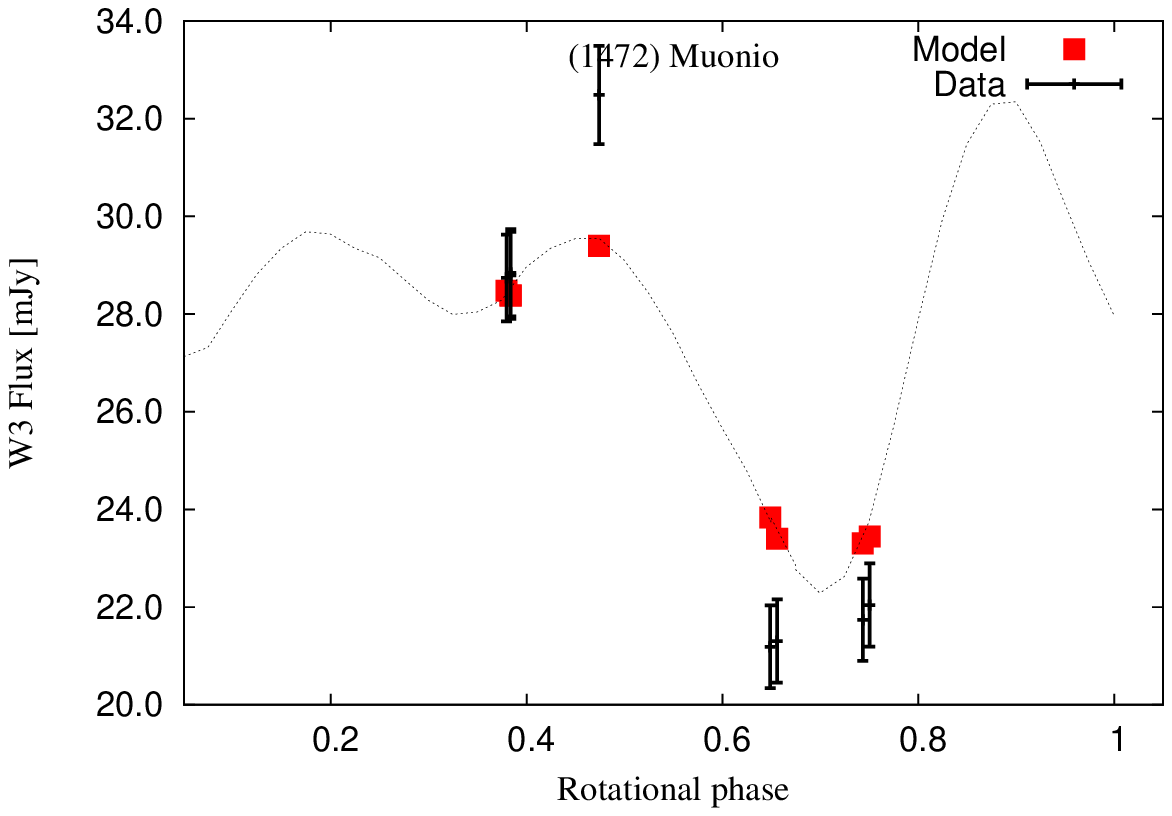}\,\includegraphics{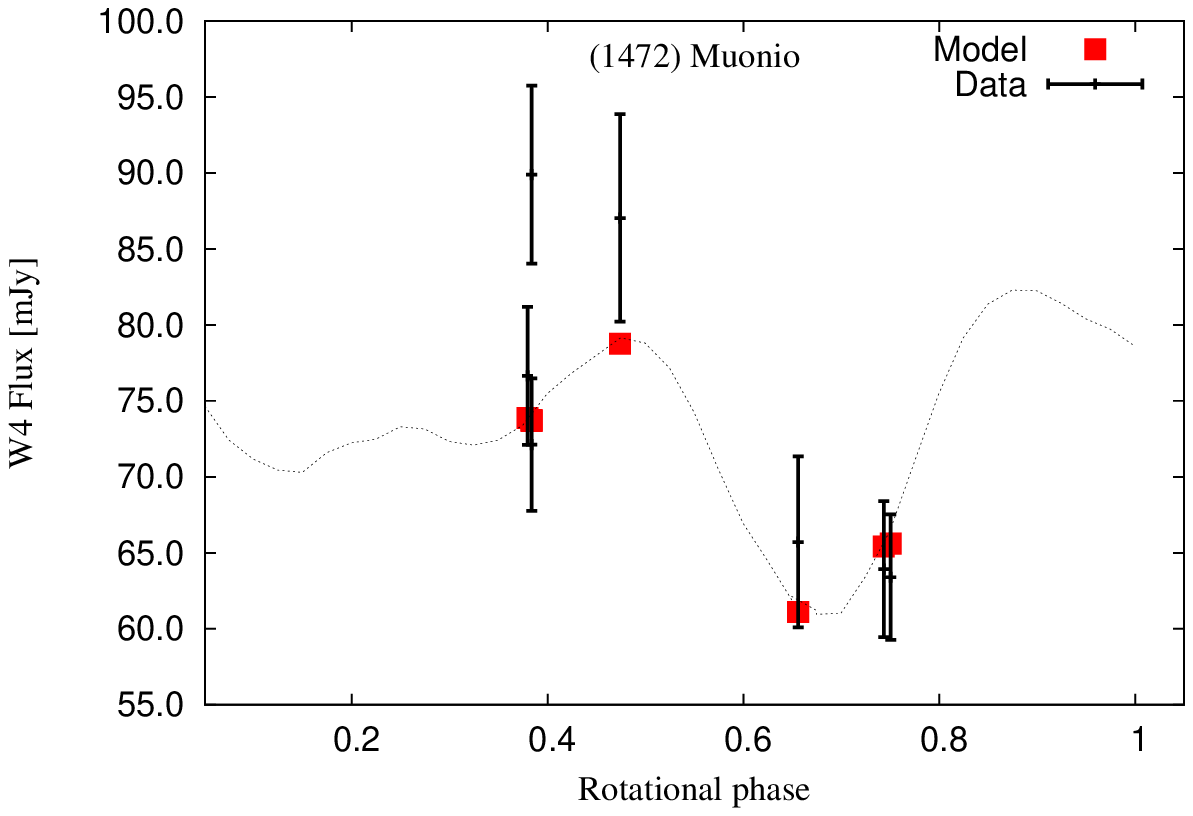}}\\
	 \resizebox{1.0\hsize}{!}{\includegraphics{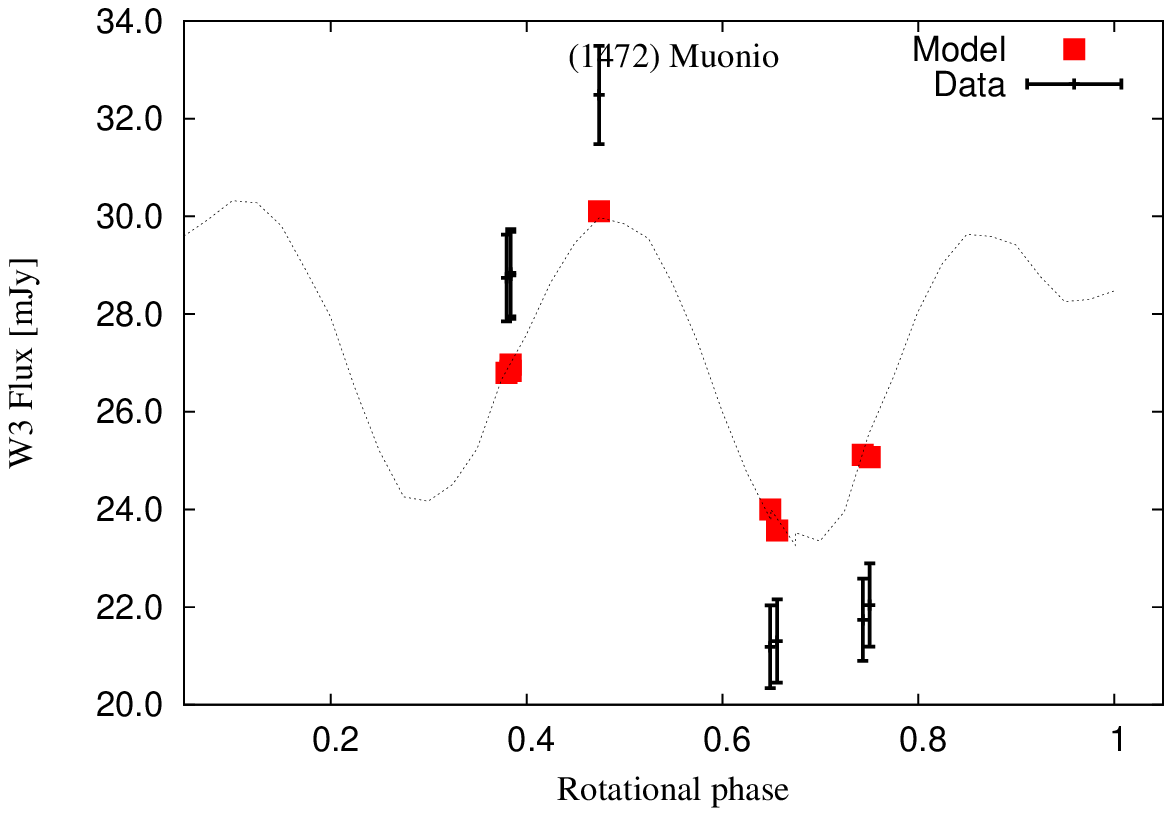}\,\includegraphics{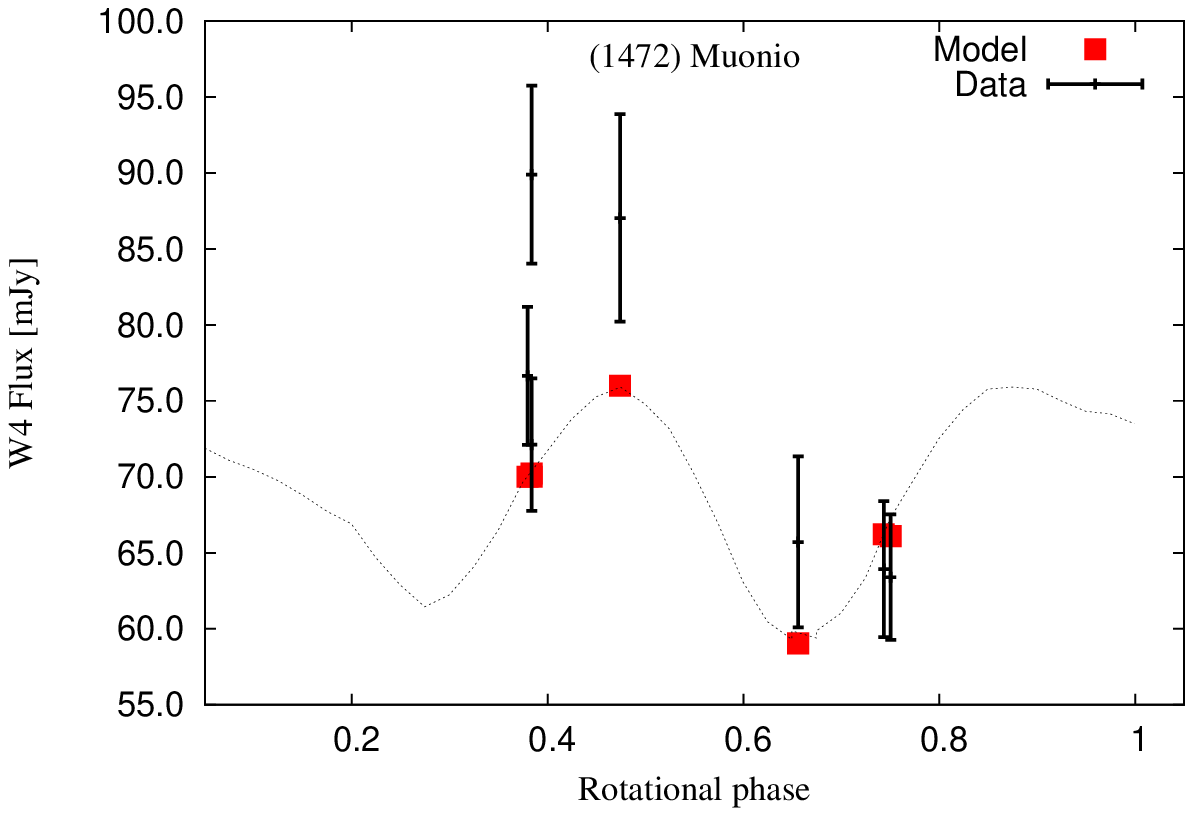}}\\
	 \end{center}
	 \caption{\label{img:1472}Comparison between the observed (WISE) and modeled (best TPM fit) thermal IR fluxes in filters W3 and W4 and synthetic thermal lightcurves (dotted lines) for both pole solutions of asteroid (1472)~Muonio (DAMIT model).}
\end{figure*}

\newpage

\begin{figure*}[!htbp]
	\begin{center}
	 \resizebox{1.0\hsize}{!}{\includegraphics{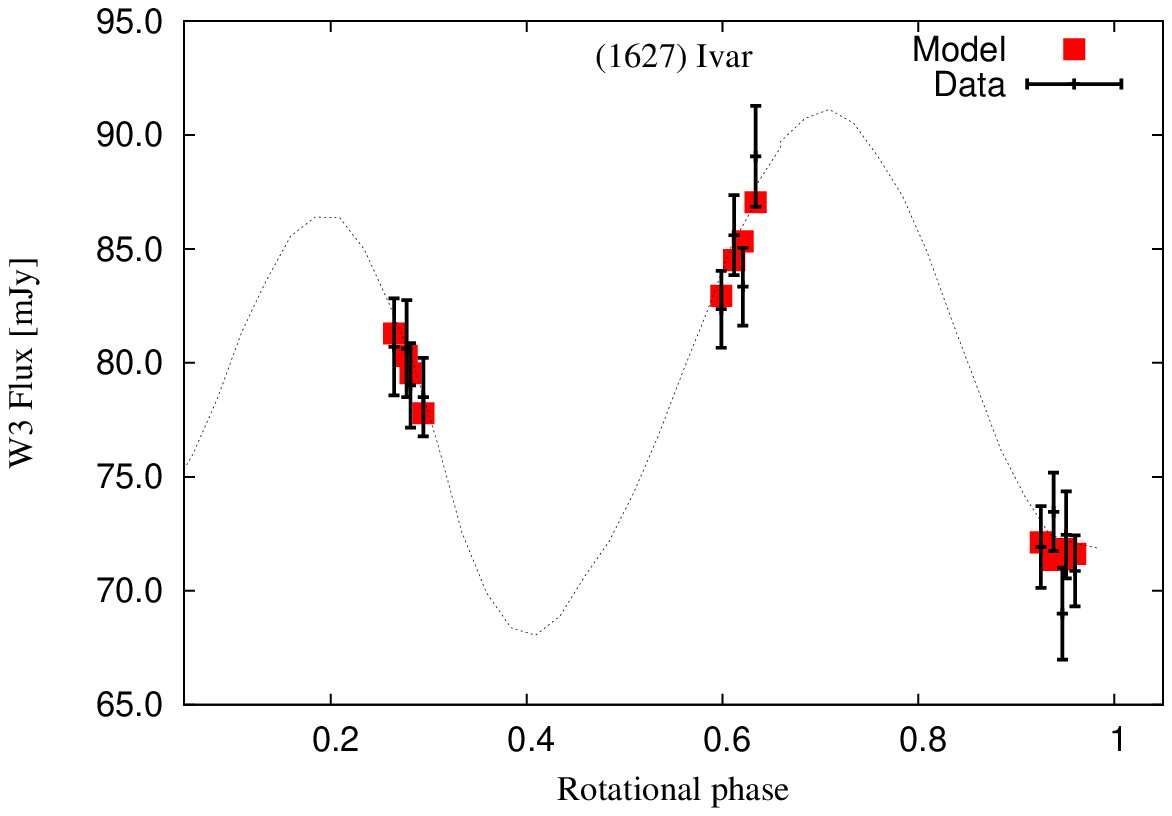}\,\includegraphics{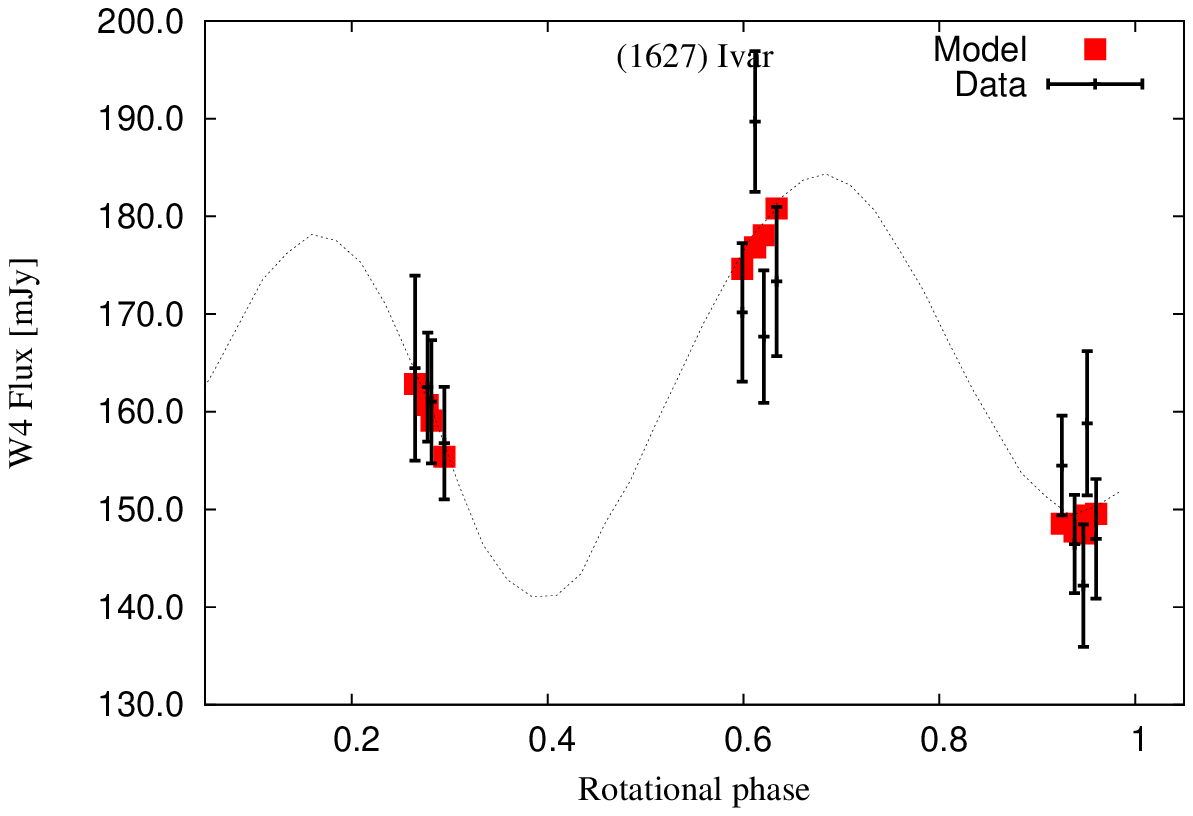}}\\
	 \end{center}
	 \caption{\label{img:1627}Comparison between the observed (WISE) and modeled (best TPM fit) thermal IR fluxes in filters W3 and W4 and synthetic thermal lightcurves (dotted lines) for asteroid (1627)~Ivar (revised model).}
\end{figure*}

\newpage

\begin{figure*}[!htbp]
	\begin{center}
%	 \resizebox{1.0\hsize}{!}{\includegraphics{img/fluxW3_1865_1_best.out.eps}\,\includegraphics{img/fluxW4_1865_1_best.out.eps}}\\
	 \resizebox{1.0\hsize}{!}{\includegraphics{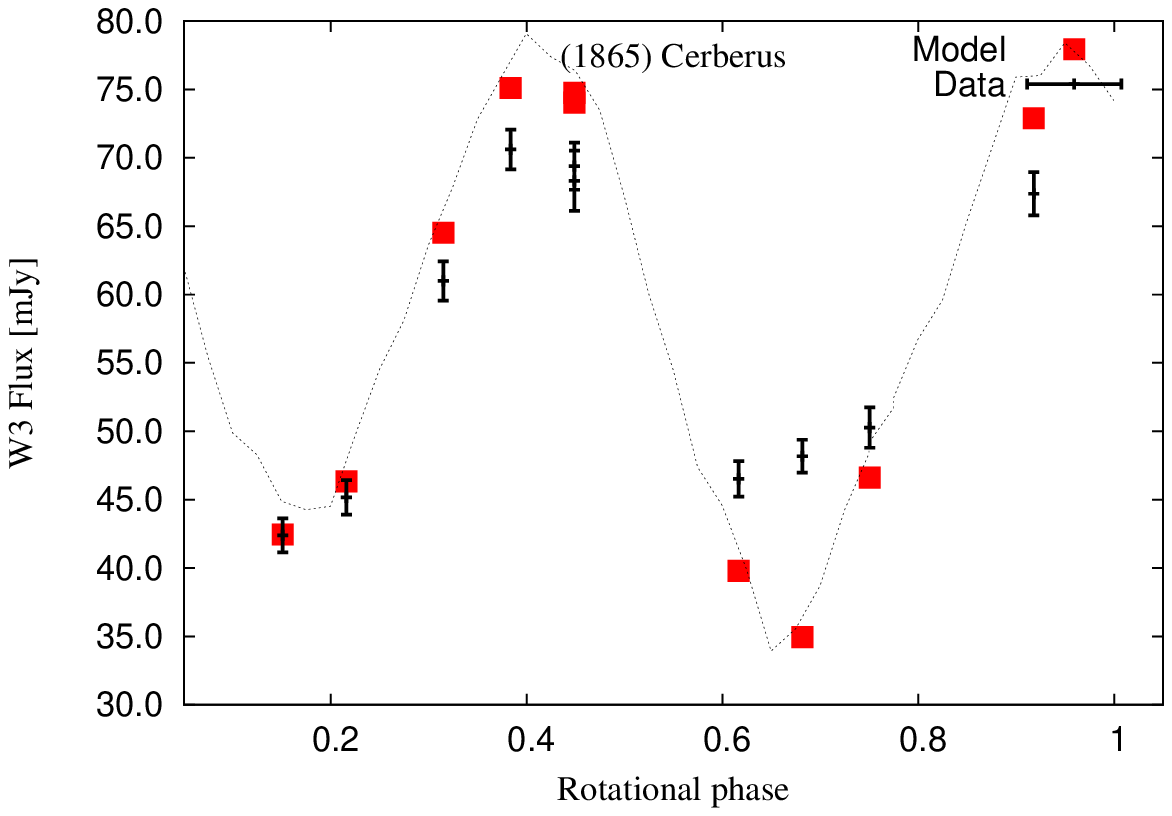}\,\includegraphics{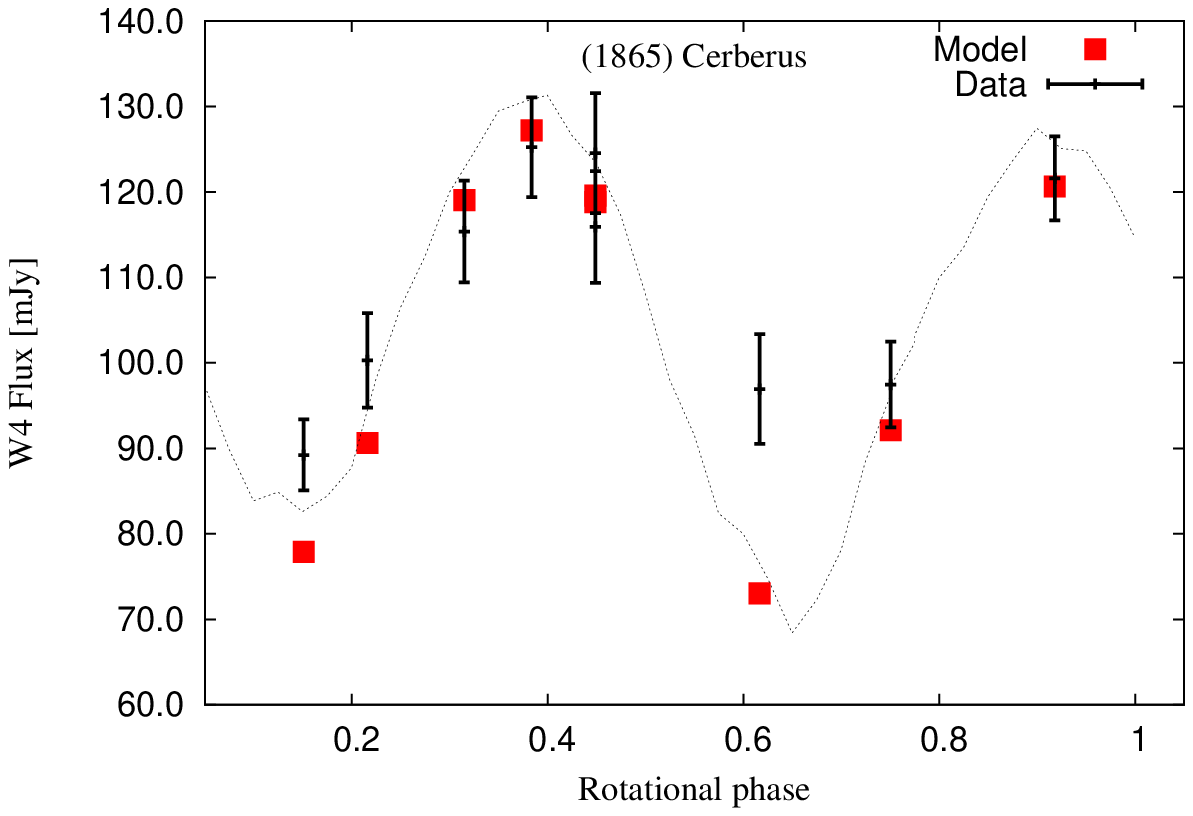}}\\
	 \end{center}
	 \caption{\label{img:1865}Comparison between the observed (WISE) and modeled (best TPM fit) thermal IR fluxes in filters W3 and W4 and synthetic thermal lightcurves (dotted lines) for asteroid (1865)~Cerberus (revised model). The small offset between several model points and the synthetic lightcurve is caused by the fact that the observations span several rotational periods. In this time, the thermal lighcurve changes because of the variation in the observing geometry. However, the synthetic lightcurve corresponds to the beginning of this interval.}
\end{figure*}

\newpage

\begin{figure*}[!htbp]
	\begin{center}
	 \resizebox{1.0\hsize}{!}{\includegraphics{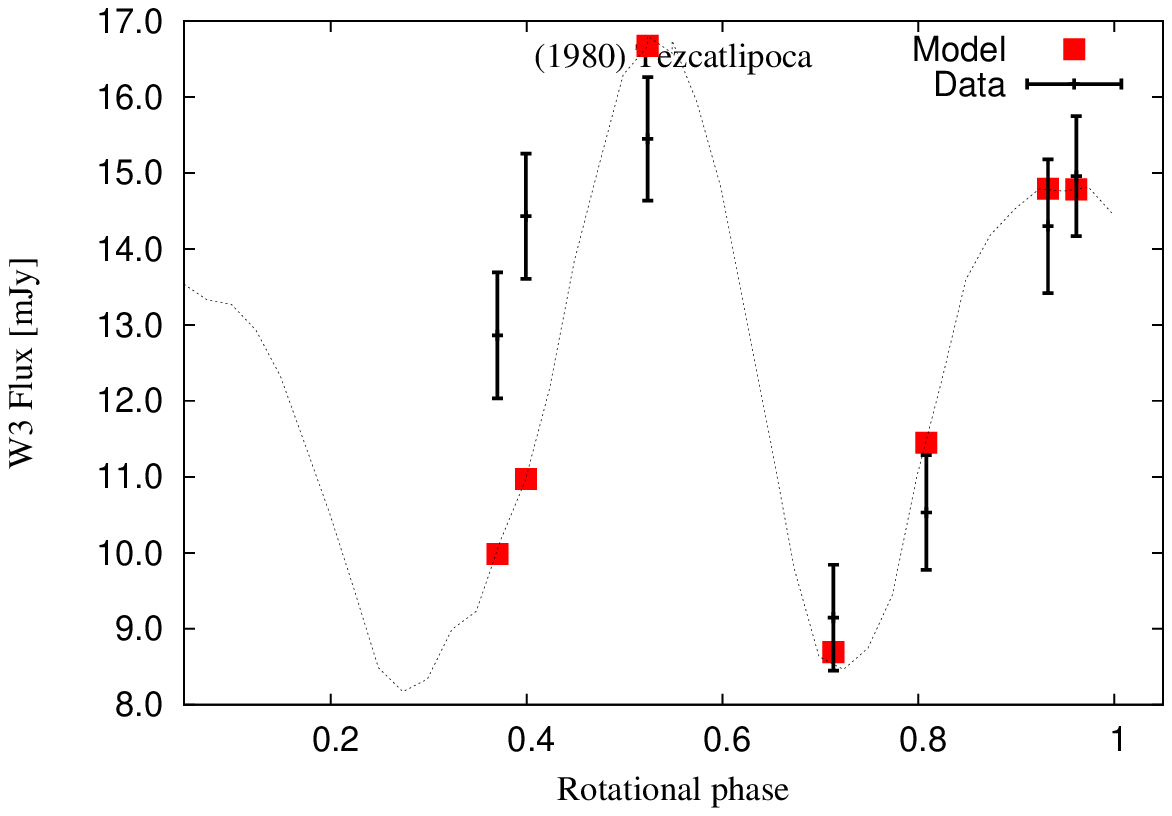}\,\includegraphics{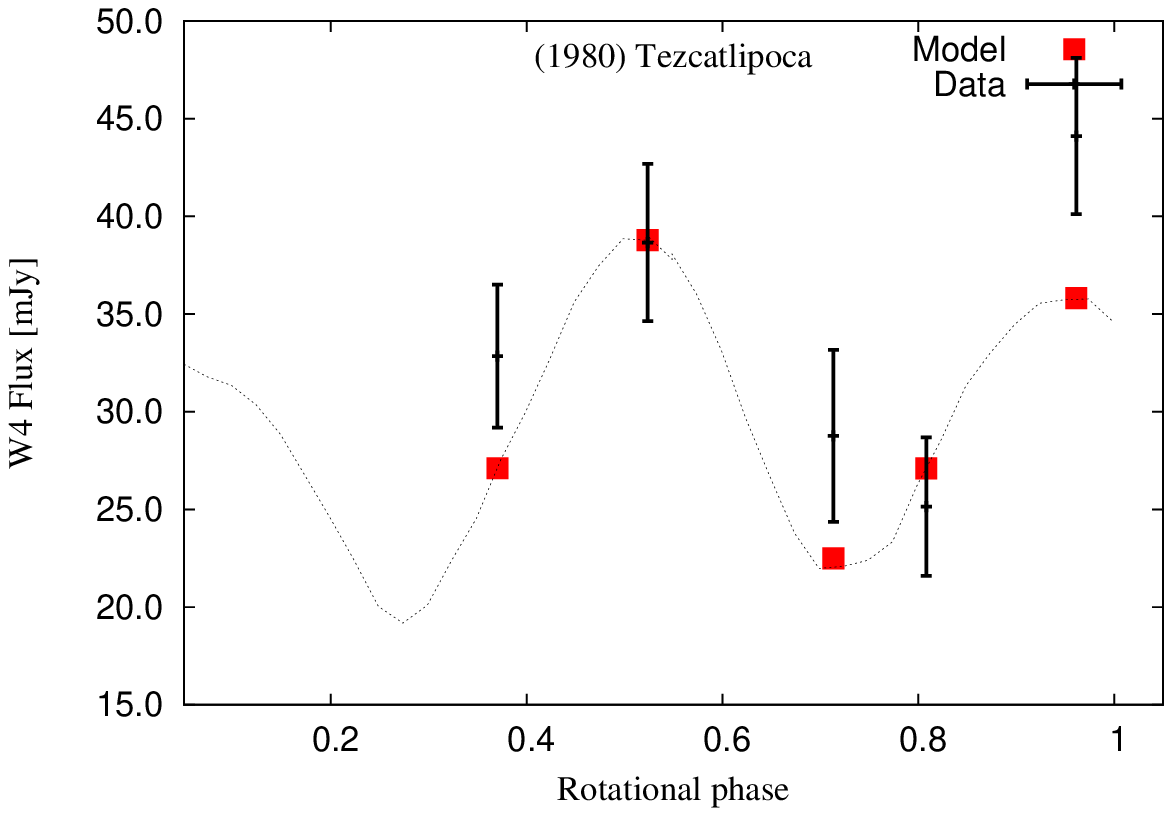}}\\
	 \resizebox{1.0\hsize}{!}{\includegraphics{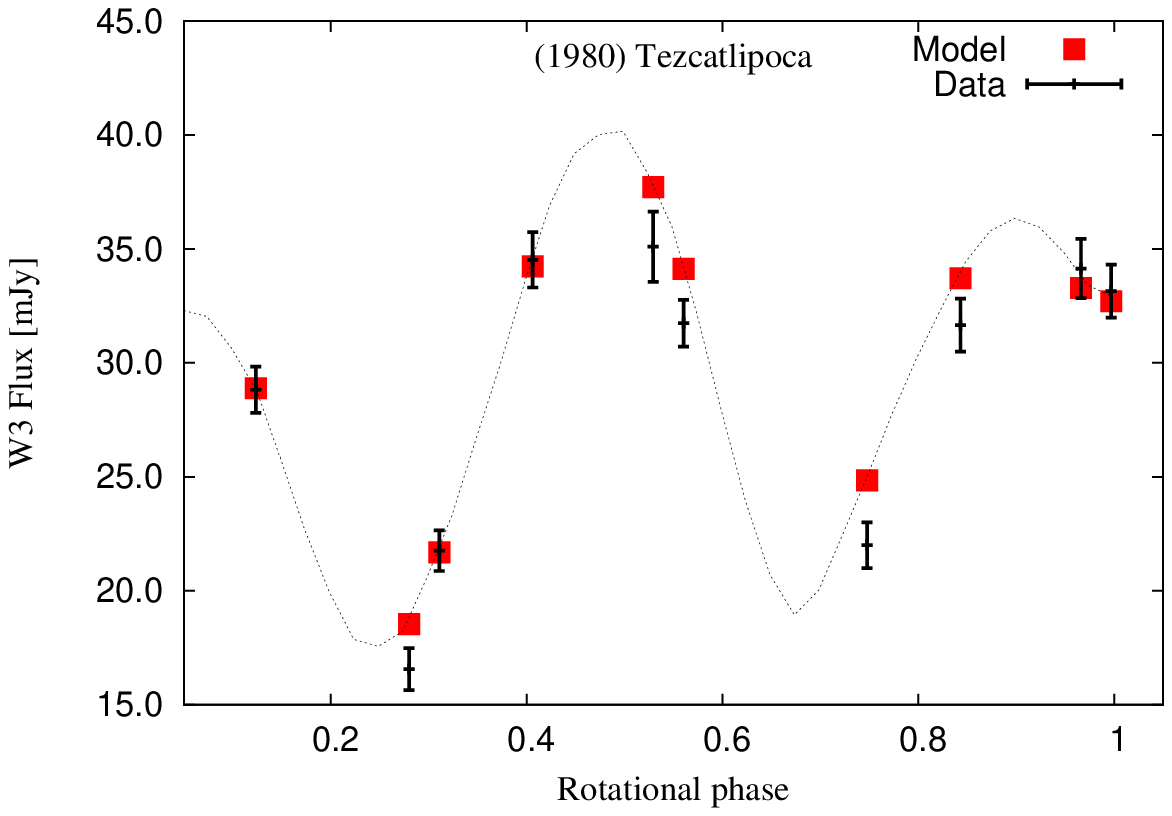}\,\includegraphics{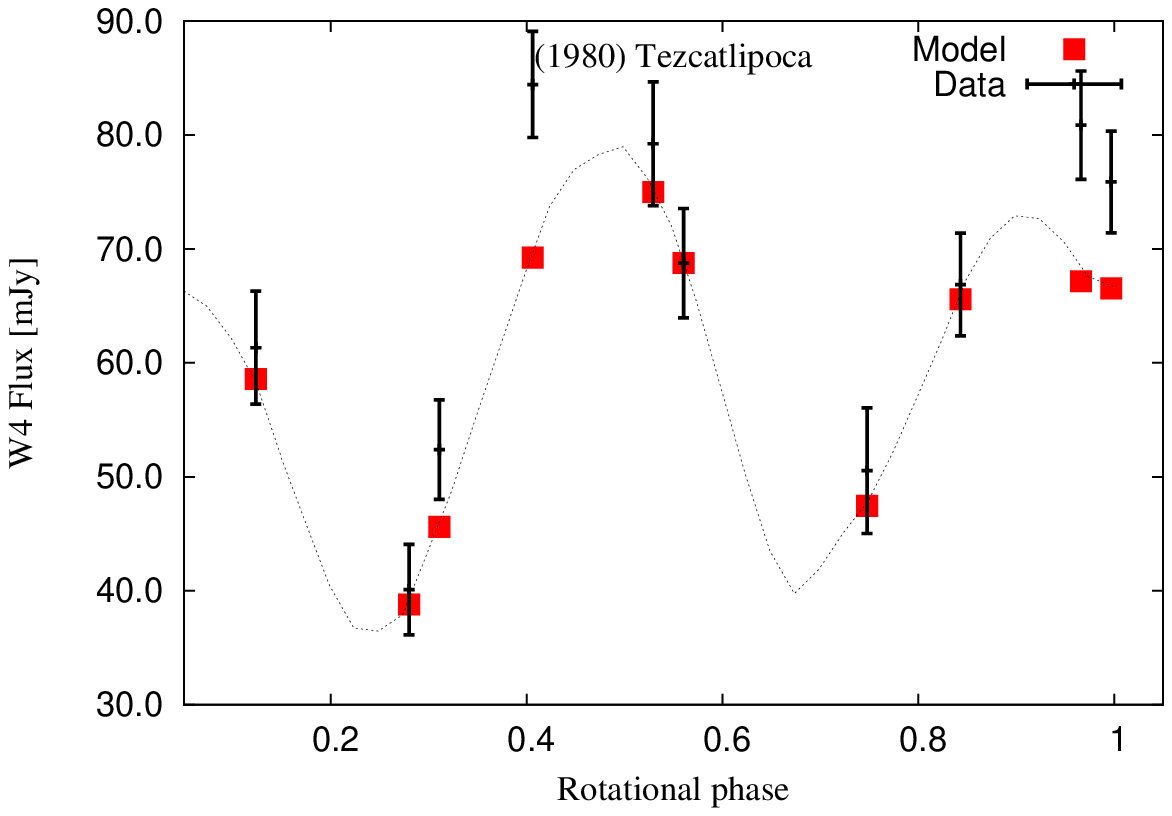}}\\
	 \end{center}
	 \caption{\label{img:1980}Comparison between the observed (WISE) and modeled (best TPM fit) thermal IR fluxes in filters W3 and W4 and synthetic thermal lightcurves (dotted lines) for asteroid (1980)~Tezcatlipoca (revised model). There are two groups observations separated by few months (top and bottom panels).}
\end{figure*}

\newpage

\begin{figure*}[!htbp]
	\begin{center}
	 \resizebox{1.0\hsize}{!}{\includegraphics{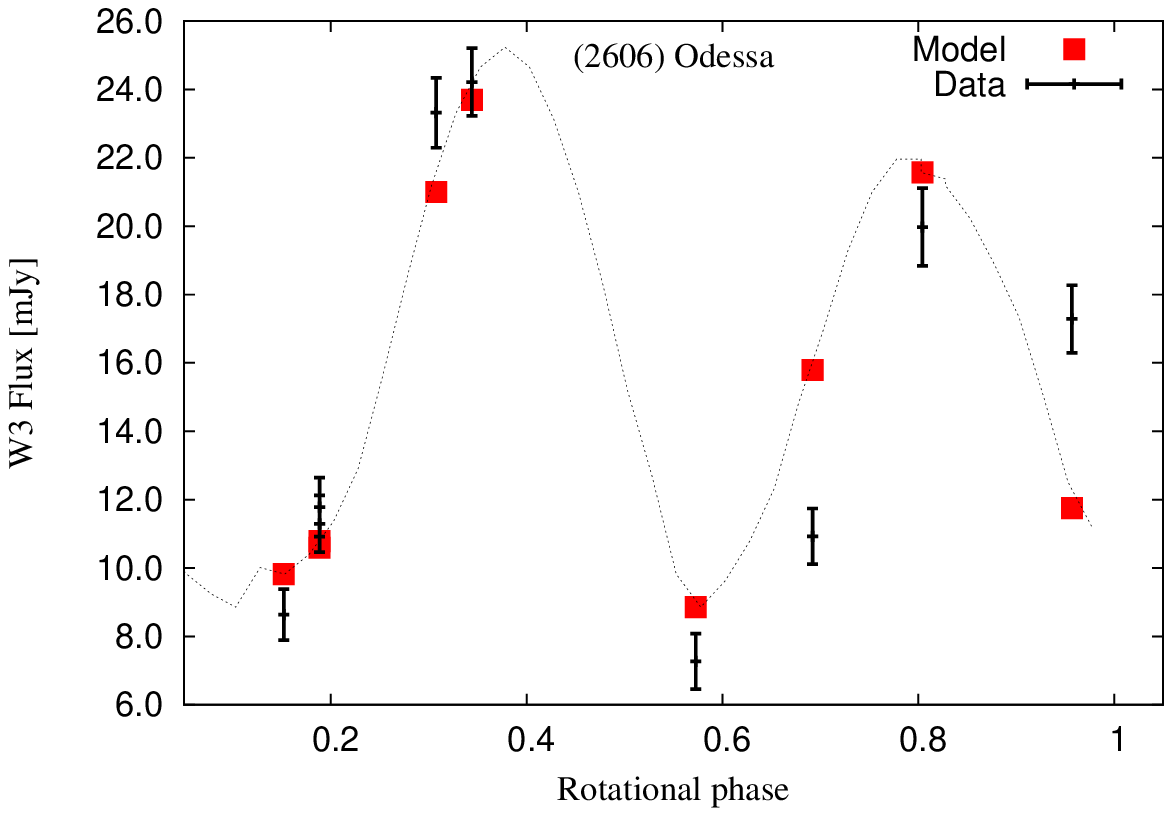}\,\includegraphics{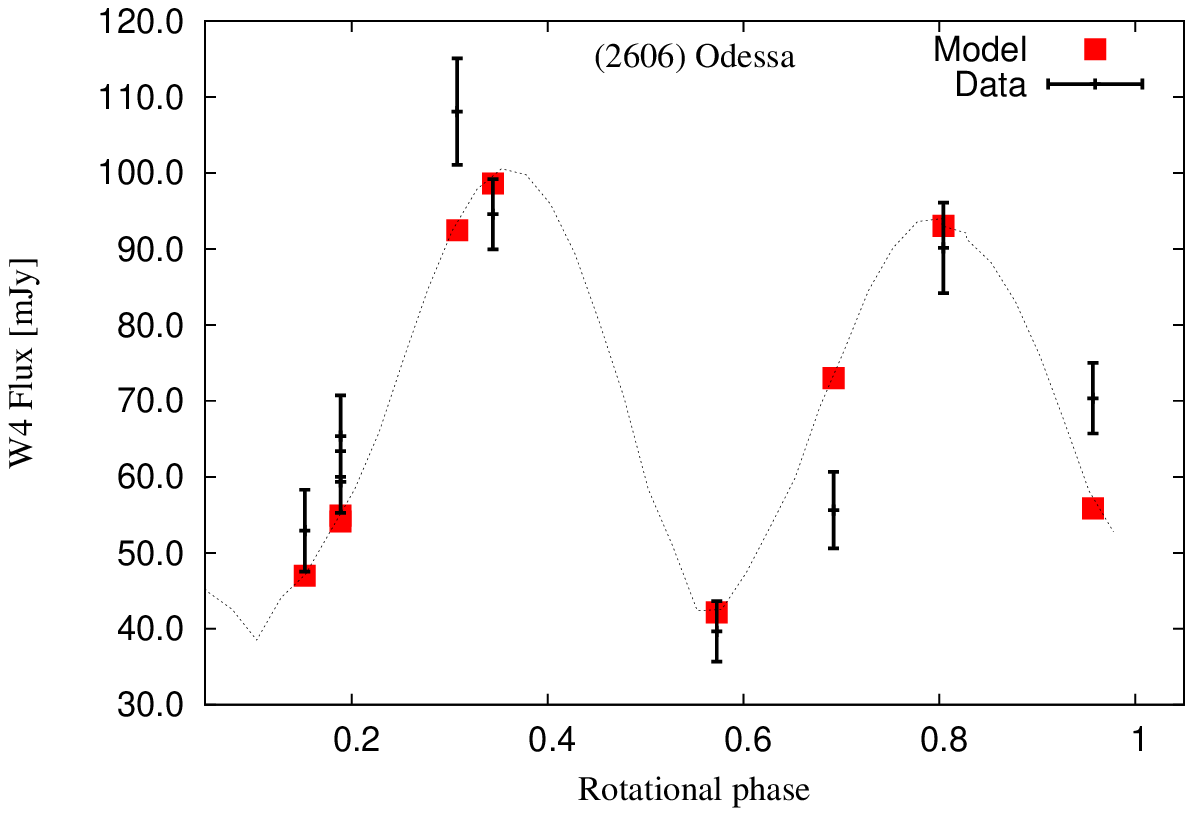}}\\
	 \resizebox{1.0\hsize}{!}{\includegraphics{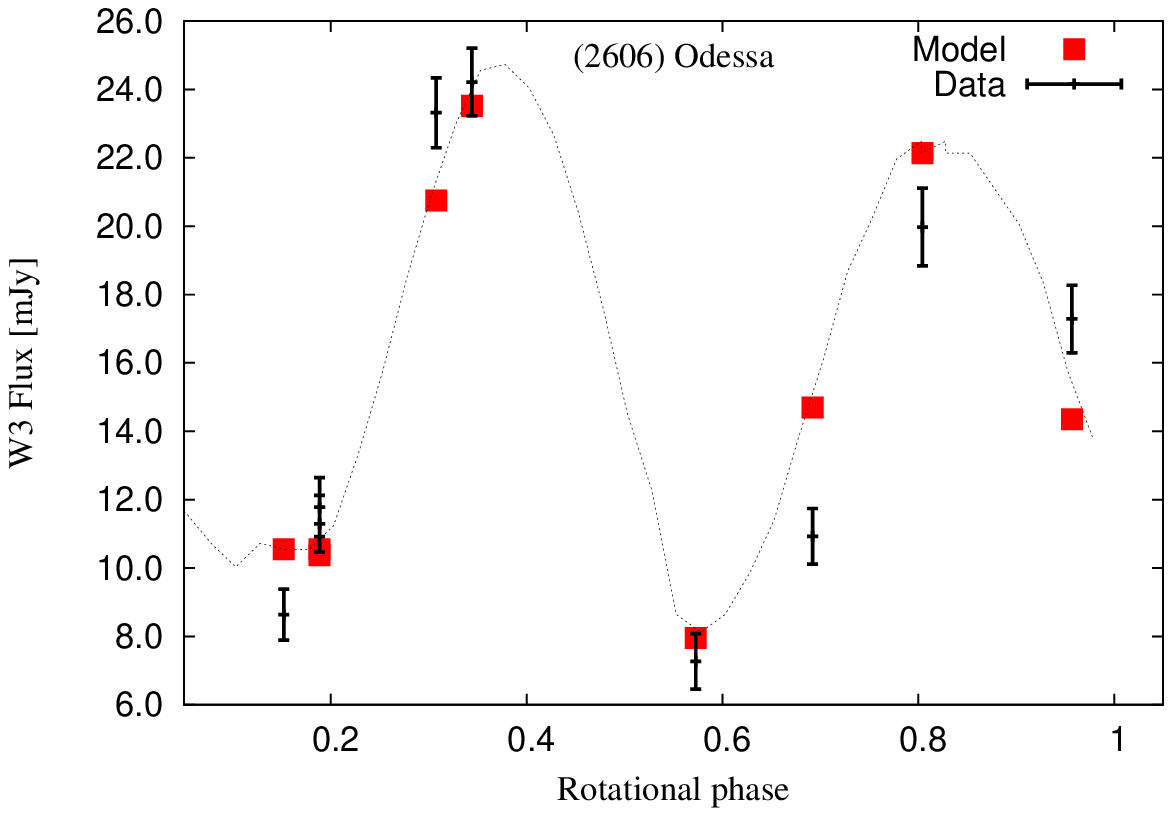}\,\includegraphics{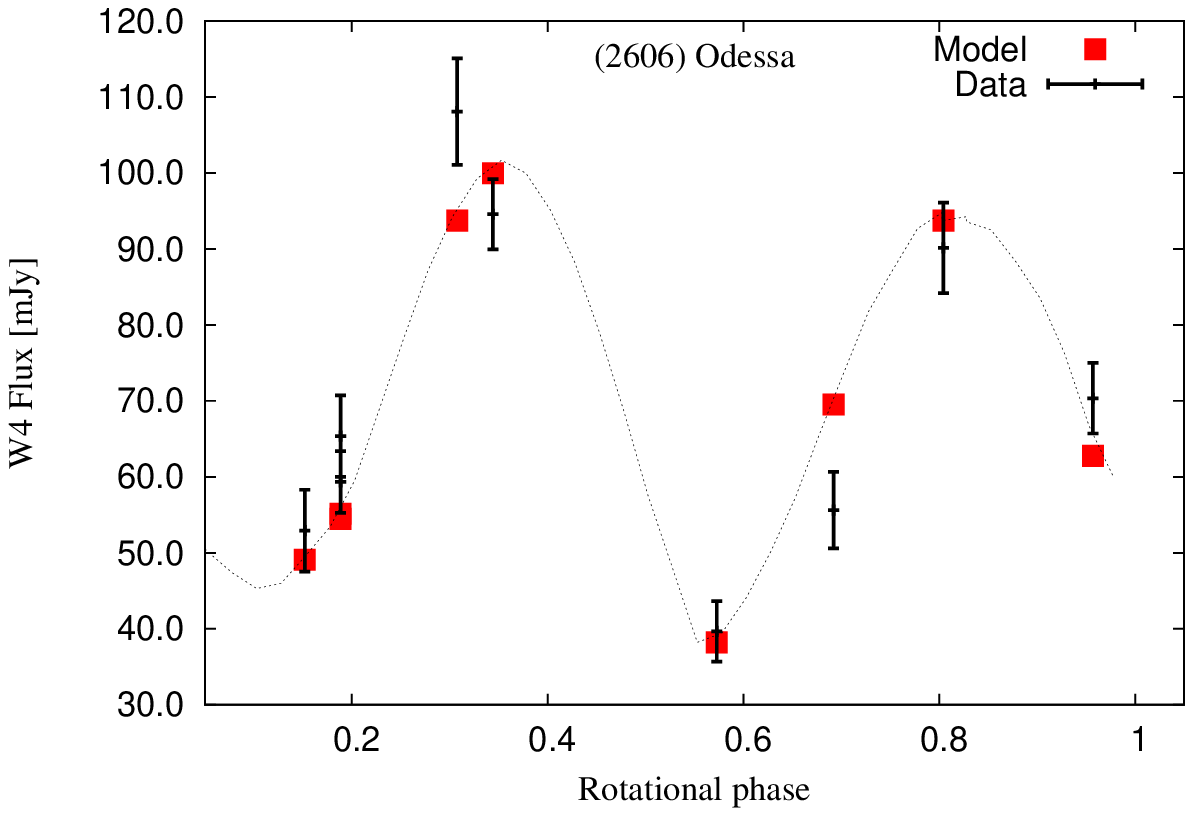}}\\
	 \end{center}
	 \caption{\label{img:2606}Comparison between the observed (WISE) and modeled (best TPM fit) thermal IR fluxes in filters W3 and W4 and synthetic thermal lightcurves (dotted lines) for both pole solutions of asteroid (2606)~Odessa (DAMIT model).}
\end{figure*}

%\newpage

\clearpage
}

\end{document}